%% file: main.tex
\documentclass[twocolumn,twocolappendix]{aastex63}
\usepackage[left]{lineno}
\usepackage{amsmath}
\usepackage{mathtools}
\usepackage[normalem]{ulem} 
\usepackage{natbib}
\usepackage{comment}
\usepackage{float}

\defcitealias{paper1}{S19}
\defcitealias{Sesar:2010}{S10}
\defcitealias{Martinez-Vazquez:2016b}{MV16}
\input{commands.tex}

\shorttitle{RR Lyrae in DES Year 6}
\shortauthors{Stringer, Drlica-Wagner, et al.}
\graphicspath{{./}{figures/}}
\reportnum{FERMILAB-PUB-20-610-AE}
\reportnum{DES-2020-0584}

\begin{document}
\title{Identifying RR Lyrae Variable Stars in Six Years of the Dark Energy Survey}
\correspondingauthor{Katelyn Stringer, Alex Drlica-Wagner}
\email{stringkm@gmail.com, kadrlica@fnal.gov}

\input{authors.tex}

\begin{abstract}

We present a search for RR Lyrae stars using the full six-year data set from the Dark Energy Survey (DES) covering $\roughly 5{,}000 \deg^2$ of the southern sky. Using a multi-stage multi-variate classification and light curve template-fitting scheme, we identify RR Lyrae candidates with a median of \CHECK{35} observations per candidate. We detect \Nrrab RR Lyrae candidates out to $\roughly 335 \kpc$, and we estimate that our sample is $>70\%$ complete at $\roughly 150\kpc$. We find excellent agreement with other wide-area RR Lyrae catalogs and RR Lyrae studies targeting the Magellanic Clouds and other Milky Way satellite galaxies. We fit the smooth stellar halo density profile using a broken-power-law model with fixed halo flattening ($q=0.7$), and we find strong evidence for a break at $R_0 = \Rbreak \kpc$ with an inner slope of $n_1 = \Ninner$ and an outer slope of $n_2 = \Nouter$. We use our catalog to perform a search for Milky Way satellite galaxies with large sizes and low luminosities. Using a set of simulated satellite galaxies, we find that our RR Lyrae-based search is more sensitive than those using resolved stellar populations in the regime of large ($\rhalf \gtrsim 500\pc$), low-surface-brightness dwarf galaxies. A blind search for large, diffuse satellites yields three candidate substructures. The first can be confidently associated with the dwarf galaxy Eridanus II. The second has a similar distance and proper motion to the ultra-faint dwarf galaxy Tucana II  but is separated by $\roughly 5\deg$. The third is close in projection to the globular cluster NGC 1851 but is $\roughly 10\kpc$ more distant and appears to differ in proper motion.

\end{abstract}


\section{Introduction} \label{sec:intro}

Studies of the Milky Way stellar halo provide unique insights into the formation and evolution of our Galaxy \citep[e.g.][]{Johnston:2002, Helmi:2008}. Over the past several decades, wide-area digital sky surveys have shown that the Galactic halo hosts a large population of stellar substructures that can be classified as satellite galaxies, star clusters, and stellar streams \citep[e.g.,][]{Belokurov:2006,McConnachie12,Shipp:2018,Simon20}. The abundance, distribution, and properties of halo substructures can be used to inform models for the assembly, chemical evolution, and star formation history of our Galaxy \citep[e.g.,][]{White:1991, Johnston:2008, Tolstoy:2009, Mo:2010,  Sharma:2011b, Gallart:2019}. In addition, halo structure and substructure are valuable tools for estimating the matter density profile and total mass of the Milky Way \citep[e.g.,][]{Deason:2012, Kafle:2012, Bonaca:2018}. A wide variety of luminous tracers have been used to map the structure and substructure of the Milky Way halo, including main-sequence turn-off stars \citep[e.g.,][]{Belokurov:2006,Shipp:2018}, blue horizontal branch stars \citep[e.g.,][]{Deason:2014}, and red giant branch stars \citep[e.g.,][]{Sharma:2011a,Sheffield:2014}. However, among these stellar tracers, pulsating variable RR Lyrae stars (RRL) are especially useful due to their distinct temporal signature and standardizable luminosity.

RRL are low-mass stars in the core helium burning phase of evolution that radially pulsate when they fall within the instability strip \citep[e.g.,][]{Walker:1989, Smith:1995, Bono:2011, Marconi:2012}. They are found in the horizontal branches of old stellar systems ($> 10 \Gyr$) and follow a well-understood period-luminosity-metallicity (PLZ) relation  \citep[e.g.,][]{Caceres:2008,Marconi:2015}.  Their age, relatively high luminosity \citep[$M_V = 0.59$ at $\feh = -1.5$;][]{Cacciari:2003}, and well-understood PLZ relation make them excellent distance indicators for old, low-metallicity stellar populations in the outer halo of the Milky Way \citep[e.g.,][]{Catelan:2004, Vivas:2004, Caceres:2008, Sesar:2010, Stetson:2014, Fiorentino:2015}. RRL are sufficiently luminous to be detected at large distances and are sufficiently numerous to trace the halo substructures with good spatial resolution \citep[e.g.,][]{Sesar:2010, Sesar:2014, Baker:2015, Martinez-Vazquez:2019, Torrealba:2015, Vivas:2006, Vivas:2020}.

The most numerous sub-type of RRL are RRab, which pulsate in the fundamental mode and have light curve shapes resembling a sawtooth curve with short periods ($0.4 \lesssim P \lesssim 0.9$ days) and large amplitudes ($0.5 \le A_g \le 1.5$ mag). In contrast, RRc pulsate in the first overtone and have smoother, more sinusoidal-shaped light curves with shorter periods ($0.2 \lesssim P \lesssim 0.5$ days) and smaller amplitudes ($0.2 \le A_{g} \le 0.5$ mag). The detection and classification of RRL that contain additional pulsation modes require very well-sampled light curves over a long and continuous baseline. For example, RRd pulsate simultaneously in the fundamental and first overtone \citep[e.g.,][]{Jerzykiewicz:1977}, while RRL may be subject to the Blazhko effect \citep{Blazhko:1907,Buchler:2011}, i.e., a modulation of period and amplitude of unknown origin that can span several to hundreds of days. 

Thanks to their distinct light curves and well-defined PLZ relation, RRL overdensities have been shown to be a good tracer of halo substructure \citep[e.g.,][]{Vivas:2001, Ivezic:2004, Sesar:2014, Baker:2015, Sanderson:2017}. Indeed, RRL have been detected in nearly all Milky Way satellite galaxies (e.g., see the recent compilation in \citealt{Martinez-Vazquez:2019}) and are abundant in Milky Way stellar streams \citep[e.g.,][]{Mateu:2018,Price-Whelan:2019,Ramos:2020,Koposov:2019}. While \citet{Martinez-Vazquez:2019} and \citet{Vivas:2020} showed that galaxies fainter than $M_{V}\sim-4.5$ are expected to contain fewer than three of these variables, even a few tightly clustered RRL in the outer halo could indicate the presence of an ultra-faint galaxy \citep{Baker:2015}. Such a technique was recently used to aid in the discovery of the ultra-diffuse satellite Antlia 2 \citep{Torrealba:2019}.

As the sensitivity of wide-field optical imaging surveys has increased, it has become possible to use RRL to probe the outer halo of our Milky Way at increasingly large distances. However, the temporal coverage of these surveys can be sparse and non-uniform, requiring the development of statistical algorithms to detect and measure RRL \citep{Hernitschek:2016,Sesar:2017,Medina:2018}. In \citet[hereafter S19]{paper1} we showed that a substantial number of RRab can be detected even in extremely sparsely sampled multi-band light curves from the first three years of the Dark Energy Survey \citep[DES;][]{DES:2005, DES:2016}. Here, we extend this work to use the full six-year DES data set (DES Y6) to assemble a catalog of RRL over $5{,}000 \deg^2$ in the southern Galactic cap with sensitivity out to a heliocentric distance of $\sim 500$ \kpc.

On average, DES Y6 has approximately twice as many observations of each astronomical source as the three-year data (DES Y3) explored in \citetalias{paper1}. This larger data set allows us to perform better identification and characterization of RRL candidates. While we show that our catalog agrees well with other overlapping surveys, DES Y6 only provides a median of \CHECK{35} observations (combining all filters) per RRL candidate. High-cadence follow-up observations will be able to confirm and better characterize candidates in our sample. The catalog resulting from our analysis of the DES Y6 data consists of the locations, periods, and estimated distances of \Nrrab RRL, with the most distant candidate residing at \CHECK{$\roughly 335 \kpc$}. We clearly resolve RRL structures associated with classical Milky Way satellite galaxies (i.e., the Large Magellanic Cloud, Small Magellanic Cloud, Fornax, and Sculptor), we detect previously known RRL associated with Milky Way ultra-faint satellite galaxies (i.e., Tucana II, Phoenix II, and Grus I), and we report the first candidate RRL associated with the ultra-faint satellites Eridanus II, Cetus III, and Tucana IV. Based on the successful detection of RRL associated with known ultra-faint satellites, we use our catalog to perform a search for previously undiscovered satellite galaxies in the DES footprint. No high-confidence satellite galaxy candidates are discovered, and we interpret the sensitivity of our search in the context of a suite of satellite galaxy simulations from \citet{Drlica-Wagner:2020}.

This paper is organized as follows. In \secref{data}, we present the DES Y6 single-epoch catalog data, our criteria for selecting stellar objects, and our calibration of the photometric uncertainties of steady sources. In \secref{selection}, we describe the color and variability criteria used to select a set of objects for further analysis. In \secref{temp_classification}, we describe the RRL light-curve template-fitting procedure, which yields our catalog of candidate RRL. 
In \secref{results}, we discuss our resulting catalog of candidate RRab. 
We estimate the total efficiency of our identification techniques (\secref{sim}), and we associate our RRab catalog with known classical dwarf satellite galaxies, ultra-faint satellites, and globular clusters residing in the DES footprint (\secref{MC}--\secref{ultra}).
In \secref{field} we use our catalog to estimate the halo density profile. In \secref{satsearch}, we perform a search for low-surface-brightness substructures using our RRab catalog. We state our conclusions in \secref{conclusion}.
The catalog of DES Y6 RRab candidates is available online.\footnote{\url{https://des.ncsa.illinois.edu/releases/other/y6-rrl}}

\section{Data Preparation} \label{sec:data}

\subsection{DES Y6 Quick Catalog} \label{sec:y6q}

DES \citep{DES:2005,DES:2016} was a six-year optical/near-infrared imaging survey covering $\roughly 5000 \deg^2$ of the southern Galactic cap using the Dark Energy Camera \citep[DECam;][]{Flaugher:2015} mounted at the prime focus of the 4\,m Blanco telescope at the Cerro Tololo Inter-American Observatory (CTIO). Observations were completed in 2019 January. DES obtained $\roughly 10 \times 90 \second$ exposures in five broadband filters, $grizY$ \citep{Neilsen:2019}.\footnote{DES took 45s exposures in Y-band in the first three years.} DES observed with $gri$ in dark time, $iz$ in gray time, and $zY$ in bright time, with each field of the footprint being observed 2--3 times per year \citep{Diehl:2016,Diehl:2019,Neilsen:2019}.  The $5\sigma$ point-source depth of the DES exposures is estimated to be $grizY \sim (24.3, 24.1, 23.5, 22.9, 21.5)$ \citep{Morganson:2018}.

As in \citetalias{paper1}, the light curves for this work were assembled using the internal DES ``Quick" release pipeline. The DES Y6 Quick Release catalog (hereafter Y6Q) was constructed using survey exposures processed with the ``Final Cut'' pipeline from the DES Data Management system \citep[DESDM,][]{Morganson:2018}. This pipeline applies instrumental calibrations and detrending corrections to the images, then creates photometric source catalogs for each exposure using \SExtractor \citep{Bertin:1996}. The full details of the DESDM image-reduction and catalog-creation pipelines are summarized in \citet{Morganson:2018}; we note that the Y6 processing used a lower source detection threshold (Y6 single-epoch catalogs have a detection threshold of $\roughly 3\sigma$ compared to a Y3 threshold of $\roughly 5\sigma$), and it has an improved astrometric calibration based on {\it Gaia} DR2. All coordinates used in this paper are in the (J2000) equinox. The photometric calibration is performed with the Forward Global Calibration Module \citep[FGCM,][]{Burke:2018}. The relative photometric calibration accuracy in $griz$ is estimated to be better than 3 \mmag across the footprint \citep[][]{Sevilla:2020}, while the absolute photometric calibration in these bands is estimated to be $\roughly 3 \mmag$ based on a comparison with the {\it HST} standard star C26202 \citep{2018ApJS..239...18A}. Complete details of the Y6 data processing, calibration, and validation will be released in forthcoming publications by the DES Collaboration. 

After the images were reduced through the Final Cut pipeline, several quality cuts were applied to select exposures for the single-epoch catalog. Any images with insufficient depth, poor seeing, poor sky subtraction, or astrometric errors, or which contained artifacts such as ghosts, bleed trails, and airplane streaks were excluded. Additionally, only exposures with FGCM zeropoint solutions \citep{Burke:2018} were included in the catalog. These selections were applied to images from years one through six of survey operations (Y6), yielding a total of 78,364 exposures.\footnote{The DES Y6Q catalog does not include observations from the DES Science Verification period.}

We assembled a Y6Q unique catalog of astronomical objects by matching sources in a given image to the nearest neighboring detections (within $1\arcsec$ in radius) in all other exposures using \code{cKDTree} as implemented in \code{scipy.spatial} \citep{Scipy:2020}. When multiple detections from one exposure were located within $1\arcsec$, these were split into multiple objects in the Y6Q catalog. All objects with at least one detection in any band were included in the Y6Q catalog to ensure  the inclusion of transient and moving objects. 
Additionally, the Y6Q catalog was cross-matched with objects detected in the \code{Y6A1\_COADD} images to provide easy reference to quantities only available from the DES processing of the coadded images.

The resulting Y6Q catalog contained $\roughly610$ million objects distributed across the DES survey footprint. 
A large number of these objects possess only a single detection, which can occur due to spurious background fluctuations or transient objects. 
Overall,  objects in the Y6Q catalog had a median of 6 observations spread over the $grizY$ bands.
If we require that objects be detected at least once in every band, then the catalog is reduced to \CHECK{$\roughly 109$ million} objects and the median total number of observations for all objects across all bands is \CHECK{34}.
This nearly doubles the median total number of observations for RRL identified in the DES Y3 release \citepalias{paper1}. 
Because of the lower signal-to-noise threshold for detections, this catalog is deeper than the DES Y3 one by $\roughly 0.75 \magn$ in each band.

Unless explicitly stated otherwise, all magnitudes in this paper are point-spread function (PSF) magnitudes derived by \SExtractor and corrected for interstellar extinction.
Interstellar extinction is calculated following the prescription in \citet{2018ApJS..239...18A} using the \citet{Schlegel:1998} dust maps with the normalization correction from \citet{Schlafly:2011} and the \citet{Fitzpatrick:1999} reddening law.

\subsection{Rescaling Photometric Uncertainties}\label{sec:error_rescale}

\begin{figure*}[!bht]
  \includegraphics[width=1\textwidth]{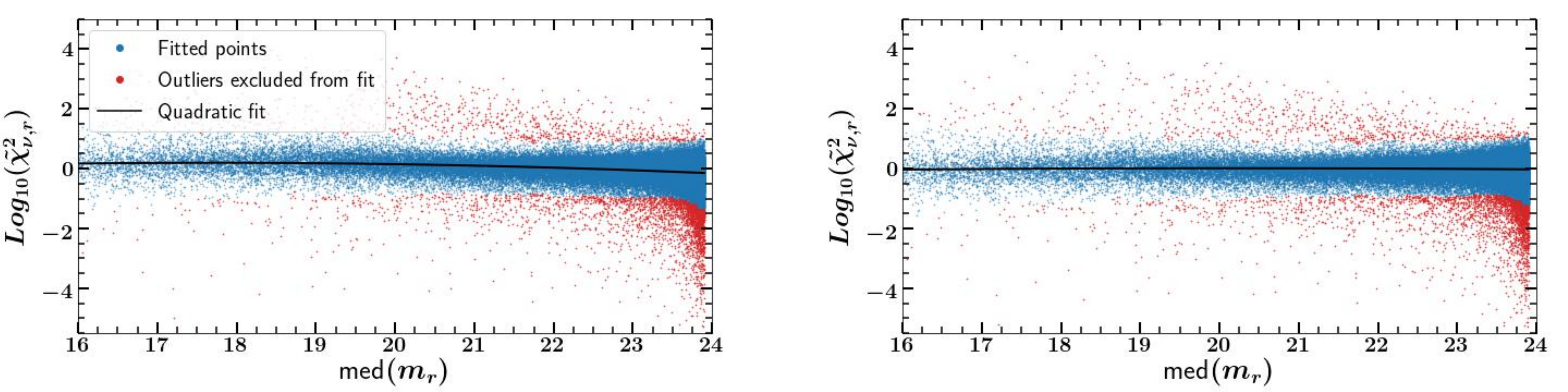}
  \caption{\textit{Left:} Reduced median chi-squared in the $r$ band, $\log_{10}(\tilde{\chi}_{\nu,r}^{2})$, vs.\ median $r$-band magnitude, $\overline{m_{r}}$, for objects classified as stars in a single DES \healpix pixel ($l,b \sim 302,-52$). Unlike DES Y3, the photometric uncertainties are found to be slightly underestimated for brighter objects in the DES Y6 data. Blue points show objects that were used to fit the quadratic curve (black line), while red points were objects excluded by a 3$\sigma$ outlier clipping. \textit{Right:} We correct this trend by rescaling the photometric errors using the quadratic fit so that the remaining trend in $\log_{10}(\tilde{\chi}_{\nu,r}^{2})$ is flat. Although this trend is very minor, it is necessary to correct because the trends differ as a function of band and sky position.}
  \label{fig:error_rescale}
\end{figure*}

Similar to \citetalias{paper1}, we found magnitude-dependent residuals in the reduced chi-squared of the photometric measurements of objects classified as stars using the criterion described in the following section. The residuals found in DES Y6 were significantly smaller than those found in DES Y3, but they were still large enough to bias the identification of variable sources if left uncorrected. Thus, we followed the same procedure as in \citetalias{paper1} to rescale the magnitude errors of each observation according to trends in the reduced chi-squared. Within each region of the survey (\healpix pixel with $\nside=32$), we define the reduced median chi-squared in band $b$ as
\begin{equation}
   \medchisq = \frac{1}{N_{b}-1}\sum_{1}^{N_{b}}\frac{\left(m_{i,b}-\mathrm{med}(m_{b})\right)^{2}}{\sigma_{i,b}^{2}},
\label{eqn:medchisq}
\end{equation}
\noindent where $m_{i,b}$ is the observed PSF magnitude of an object in observation $i$, $\sigma_{i,b}$ is the reported magnitude uncertainty on that observation, and $N_b$ is the total number of observations of that object in that band.
We fit a quadratic function of the form
\begin{equation}
\begin{aligned}
\mathrm{log}_{10}(\medchisq)& =\ c_{0,b}\\ 
                                   &+ c_{1,b} (\mathrm{med}(m_{b})-20) \\ 
                                   &+ c_{2,b} (\mathrm{med}(m_{b})-20)^{2}.
\end{aligned}
\label{eq:quadfit}
\end{equation}

\noindent \figref{error_rescale} shows a noticeable trend in log$_{10}(\tilde{\chi}_{\nu,r}^{2})$, similar to those seen in \citetalias{paper1}; however, we find that the uncertainties are slightly underestimated for bright objects in DES Y6, as shown by the negative slope in \figref{error_rescale}. Although this trend is far less pronounced than the trend in DES Y3, we perform this correction since the trends vary slightly over the wide-field footprint. We independently fit the coefficient for each \healpix region in each band and corrected the uncertainties of the observed \var{MAG\_PSF} quantities using the appropriate scale factor calculated from these relations. This process effectively rescaled the errors and flattened the trends in \medchisq.


\section{Selection for Template Fitting}\label{sec:selection}
\subsection{Stellar Source Selection}\label{sec:star_selection}

Since the faint end of the DES object sample is dominated by galaxies, we perform an initial star--galaxy separation to select stellar sources. We use the \var{SPREAD\_MODEL\_I} parameter from the exposure with the largest effective exposure time \citep{Neilsen:2016}. The $i$ band is preferred for star--galaxy separation because it typically has the best seeing of the $gri$ bands observed during dark time (see \S2.3 in \citealt{2018ApJS..239...18A} and Figure~8 in \citealt{Diehl:2019}). This procedure differs from \citetalias{paper1}, which considered all objects that passed this \var{SPREAD\_MODEL} criterion in any of the $griz$ bands to avoid omitting objects that were missing observations in a single band. While this enabled the catalog in \citetalias{paper1} to be more complete, many extended sources leaked into the sample and had to be removed through visual inspection. In comparison, a much larger fraction of DES Y6 sources have at least one measurement in the $i$ band. Any object for which $\left|\var{SPREAD\_MODEL\_I}\right| < (0.003 + \var{SPREADERR\_MODEL\_I})$ was considered ($\roughly 310$ million sources from the Y6Q catalog pass this cut). Additionally, only objects that were associated with an object detected in the Y6A1 coadded images were considered (0\farcs7 match radius).  

\subsection{External Catalogs and Simulated Data Used to Define the Sample}
\label{sec:extcat_select}

After rescaling the photometric uncertainties and applying the star--galaxy separation, we remove any objects with fewer than 10 total observations. Such a small number of observations skews their variability statistics and makes template fitting challenging. 
We then further reduce the size of our catalog by selecting objects that match the colors and temporal variability that are characteristic of RRL. 

To determine optimal color and variability cuts to select RRL, we cross-match objects in Y6Q with RRL identified by external surveys, variable objects that are not RRL, and stars used for the photometric calibration that are largely non-variable. Our sample of external RRL includes objects from Sloan Digital Sky Survey (SDSS) Stripe 82 \citep[hereafter S10;][]{Sesar:2010}, the Catalina Sky Surveys DR2 \citep{Drake:2013a,Drake:2013b,Drake:2014,Torrealba:2015,Drake:2017}, Pan-STARRS PS1 \citep{Sesar:2017}, variables from the Sculptor dSph \citep{Martinez-Vazquez:2016b}, RRL from the Fornax dSph \citep{Bersier:2002}, and RRL from {\it Gaia} DR2 with measured periods \citep{Holl:2018, Clementini:2019, Rimoldini:2019}. As a likely contaminant class, we also select $\roughly16{,}000$ quasars (QSOs) from the KiDS DR3 survey \citep{Nakoneczny:2019,deJong:2017} and the SDSS-POSS southern sample \citep{MacLeod:2012}. Finally, to compare against other (likely non-variable) contaminants, we match to an internal DES catalog of $\roughly17$ million stars that were used in the FGCM zeropoint calibration \citep{Burke:2018}. We remove calibration stars located less than 10 arcmin from the centers of the Fornax and Sculptor dwarf galaxies since the DES photometry suffers from crowding in these regions. We randomly downsample the QSO and calibration star catalogs to match the size of our external RRab sample. The resulting comparison samples contain 5055 RRab, 472 RRc, 46 RRd, 4 Blazkho RRL, 5055 QSOs, and 5055 calibration stars. 

To further guide our selection criteria, specifically at faint magnitudes, we simulate a set of mock RRL light curves. This process follows the procedure described in \citetalias{paper1}, and we only provide a brief summary below.
Our simulated light curves are based on well-sampled light curves of RRL from \citetalias{Sesar:2010}. First, we construct smoothed light-curve shapes using the best-fitting templates and observational parameters for each of the 483 RRL (379 RRab and 104 RRc) identified by \citetalias{Sesar:2010} and convert their magnitudes into the DES filter system.\footnote{The SDSS-DES filter transformation equations can be found in Appendix F of \citetalias{paper1}.} We then subtract the \citetalias{Sesar:2010} estimated distance moduli from each light curve to transform to absolute magnitude. For a set of magnitude bins in the range from $15.5 \le g \le 24.5$ with a bin width of 0.5, we shift the light curve for each of the 483 light-curve shapes to an average $g$ magnitude randomly drawn from a uniform distribution within that magnitude bin. As each light curve is shifted to a random distance, any simulated measurement fainter than the Y6 single-epoch limiting magnitude in that band is removed.\footnote{The single-epoch limiting magnitude was estimated from Table 1 of \citet{2018ApJS..239...18A} with an adjustment to account for the lower object detection threshold in Y6.} We then assign photometric uncertainties to each observation using the rescaled values from \secref{error_rescale} and use them to introduce scatter into the observations. The light curves are then downsampled to the DES observing cadence as determined from a random set of bright stars in DES. The total simulated sample includes 8211 light curves, of which 6443 were RRab and 1768 were RRc. We do not specifically search for RRc in this work, but include their simulated curves to assess the RRc contamination in our final sample. 

\subsection{Color Selections}\label{sec:colors}

\begin{figure}[!hbt]
  \includegraphics[width=0.5\textwidth]{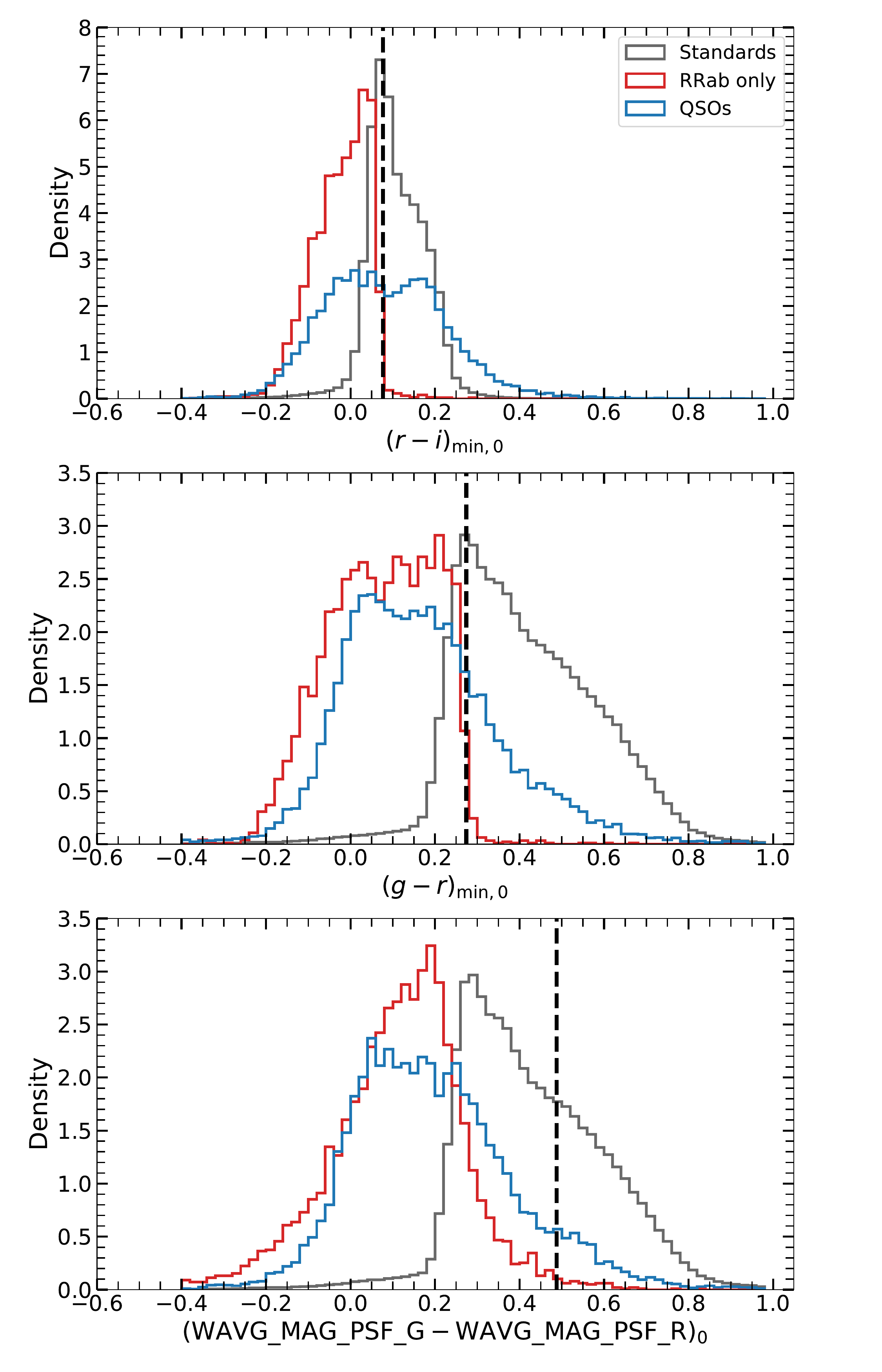}
  \caption{Multi-stage color cuts applied to DES stars to remove non-RRL. Each panel shows the distribution of colors for RRab in red, QSOs in blue, and stars used for photometric calibration in gray. The black dashed line indicates the 99th percentile of the RRab distribution and was set as the threshold for inclusion in our sample. \textit{Top:} Distribution of instantaneous $(r-i)_{{\rm min},0}$. \textit{Middle:} Distribution of instantaneous $(g-r)_{{\rm min},0}$.  \textit{Bottom:} Distribution of the extinction-corrected $(g-r)_{0}$ using weighted-average magnitudes.  
}
  \label{fig:color_cuts}
\end{figure}

RRL inhabit a well-defined region of color-color space \citep[e.g.,][]{Ivezic:2005}. We therefore apply a color selection to further reduce the number of objects that are passed to the light-curve template-fitting stage. This selection specifically excludes variable stars that are too red to be RRL (e.g., low-mass main-sequence stars).
Since the colors of RRL change over the course of their pulsation cycle \citep[e.g.,][]{Guldenschuh:2005,Vivas:2017}, calculations based on observations obtained at multiple phases will degrade the separation power of this method. Thus, we take advantage of the DES observing cadence to measure ``instantaneous colors'', based on observations taken within one hour of each other. To reduce the time spent slewing between fields, DES often took sequences of 2--3 consecutive exposures of the same field in different filters, with the filters chosen according to the seeing, lunar phase, and number of previous observations (see Fig.~3 in \citealt{Diehl:2016}). To select these sequences, we group together any observations taken within one hour of each other and, depending on the filters used, calculate $g-r$, $r-i$, or $i-z$ colors. Due to the observing cadence of DES, the median separation time for each color combination is $\roughly 120\second$. If there are multiple instantaneous colors for an object, we store only the maximum and minimum values. 

We develop a multistage color selection to remove objects based on their instantaneous colors (when available), while retaining any objects that did not have a particular instantaneous color available. 
Using our sample of previously identified RRab, we defined selections as the 99\% percentile value of the RRL population (\figref{color_cuts}). 
The first two selections use the extinction-corrected minimum instantaneous colors, $(r-i)_{\min,0}$ and $(g-r)_{\min,0}$, to select objects lying near the blue end of the stellar locus. 
The first cut selects objects with $(r-i)_{\min,0} - \sigma(r-i)_{\min} \le 0.068$. Any objects that do not have an instantaneous measurement of $r-i$ but have one in  $g-r$ are passed into the next step, which retains all objects with $(g-r)_{\min,0} - \sigma(g-r)_{\min} \le 0.268$. 
Any objects that do not have instantaneous measurements of $r-i$ or $g-r$ are retained if they satisfy $(\mathrm{WAVG\_MAG\_PSF\_G}-\mathrm{WAVG\_MAG\_PSF\_R})_{0} \le 0.488$. As can be seen from \figref{color_cuts}, this is the least restrictive of the three cuts. In total, $\roughly 51$ million stellar sources pass our color cuts, including 98.65\% of the sample of previously known RRab.

\subsection{Variability Selection}\label{sec:var_select}

We select temporally variable objects by calculating several variability statistics. 
Many of these quantities, which are summarized in \tabref{stats}, are based on the analysis of \citet{Sokolovsky:2017} and are described in detail in \appref{varmetrics}. We calculate these variability statistics from the MAG\_PSF measurements and their rescaled uncertainties as described in \secref{error_rescale}.

Since numerous color, magnitude, and variability measurements are calculated (many of which are correlated), we use the random forest algorithm to ``learn'' the optimal boundaries in feature space to separate RRab from non-RRab. Random forests use a collection of decision trees to predict an object's type. 
Each decision tree is trained on a random subset of the training population by repeatedly subdividing the sample based on feature values until a user-defined maximum depth or other specified stopping condition is reached. 
The specific features that are used in each split are chosen randomly and can influence the characteristics of the population that a tree learns to detect. A random forest classifier takes advantage of the learning differences between individual trees by averaging the predictions from a large number of trees to produce an aggregate score \citep{Amit:1997,Breiman:2001}. Random forests are a good choice for this task because they are largely insensitive to uninformative features, so the inclusion of a feature that does not separate the object types well will not harm the overall results. For our classifiers, we use the \code{RandomForest} implementation in \code{scikit-learn} \citep{Pedegrosa:2012}. To reduce our sample of light curves down to a manageable size for template fitting, we divide our pretemplate selection criteria into two phases: (1) remove non-variable sources, and (2) remove common variable contaminants (i.e., QSOs). 

For the first classifier, our training set consists of equal numbers of simulated RRab and calibration stars that are largely non-variable. We choose to use simulated RRab instead of RRab cross-matched from external catalogs because the simulated RRab cover the entire magnitude range of DES Y6 and thus do a better job of including the decreasing sensitivity to variability for more distant objects with larger photometric uncertainties. Hyperparameters for the random forest classifiers are determined using the \code{GridSearchCV} function of \code{scikit-learn}. This first classifier has 35 trees with a depth of eight splits and eight features allowed at each split.  

This stage of the selection is intended to remove as many non-variable sources as possible, so we choose the cutoff classifier score using the $F_{\beta}$ score \citep{Baeza-Yates:1999},

\begin{equation}
F_{\beta} = \frac{(1+\beta)^{2}\cdot\mathrm{TP}}{(1+\beta^2)\cdot\mathrm{TP} + \beta^2\cdot\mathrm{FN} + \mathrm{FP}}.
\end{equation}
\noindent where TP (true positives) and FN (false negatives) reflect how many true RRab have classifier scores above or below the cutoff threshold, respectively. Similarly, the FP (false positives) and TN (true negatives) show the number of non-RRab with classifier scores above or below the cutoff threshold. We choose this particular score over other popular metrics like the ``informedness'' or ``$F_1$-score'' because we wish to prioritize purity at this stage in the analysis \citep{Powers:2008}. 
The $F_{\beta}$ score with $\beta$=0.5 allows us to weigh the precision twice as heavily as the recall. 
The classifier value that maximizes $F_{\beta}$ is 0.755. 
This value yields a sample with an RRab precision of 99.71\% and a recall (completeness) of 97.96\% when applied to the training set.
Approximately 20\% of the 51 million input objects pass this classification. 

As a second step, we train and apply a classifier optimized to remove variable objects that do not show the strong variability pattern of RRL. 
In particular, due to the sparse temporal sampling of DES, it can be difficult to distinguish the variability of QSOs from that of RRL \citepalias[e.g.,][]{paper1}. 
In addition, QSOs can be unresolved, have blue colors similar to RRL, and are abundant at the faint magnitudes reached by DES \citep[e.g.,][]{Tie:2017}. 
Thus, for our training set, we use an equal number of simulated RRab and real QSOs cross-matched from the KiDS DR3 survey \citep{Nakoneczny:2019,deJong:2017} and the SDSS-POSS southern sample \citep{MacLeod:2012}.
For this classifier, we use a random forest with 18 deeper trees with nine features allowed for consideration at a total of 16 splits. 
To prioritize RRL completeness, we use the $F_{\beta}$ score with $\beta=1.5$ to choose a cutoff score of 0.34. 
This cutoff score returns a purity of 86.28\% and completeness of 96.04\% for the training set.  

After applying both of these classifiers, $\roughly 1.1$ million objects remain for light-curve template fitting. The performance curves for both of these initial variability classifiers and their top features are included in \appref{roc_features}.

\section{Template Fitting and Classification}\label{sec:temp_classification}

\begin{deluxetable*}{lrrrrcrrrrr}[!thb]
	\tabletypesize{\footnotesize}
	\tablecaption{DES Y6 RRab Candidates}
	\tablehead{\colhead{DES Y6 ID} & \colhead{$\alpha$} & \colhead{$\delta$} & \colhead{$\langle g \rangle$} & \colhead{$\langle r \rangle$} & \colhead{$\langle i \rangle$} & \colhead{$\langle z \rangle$} & \colhead{$\langle Y \rangle$} & \colhead{$P$} & \colhead{$A_{g}$} & \colhead{$\mu$} \\
    \colhead{} & \multicolumn{2}{c}{(deg, J2000)} & \multicolumn{5}{c}{(mag)} & \colhead{(days)} & \multicolumn{2}{c}{(mag)}}
    \decimals
    \startdata
    871765223 & 300.6754 & -53.9670 & 16.647 & 16.495 & 16.489 & 16.512 & 16.561 & 0.5956 & 0.881 & 15.98 \\
    871820547 & 300.4591 & -50.7457 & 15.352 & 15.201 & 15.221 & 15.225 & 15.039 & 0.5441 & 1.107 & 14.55 \\
    872060389 & 301.1574 & -54.3232 & 19.217 & 18.935 & 18.841 & 18.647 & 18.647 & 0.6158 & 1.181 & 18.14 \\
    872212127 & 301.4859 & -57.0505 & 17.941 & 17.671 & 17.770 & 17.715 & 17.810 & 0.5638 & 1.059 & 17.13 \\
    872268895 & 301.5402 & -52.5811 & 14.797 & 14.721 & --- & 14.637 & 14.741 & 0.5807 & 0.670 & 14.17 \\
    872444946 & 300.5736 & -56.2696 & 16.417 & 16.272 & 16.270 & 16.223 & 16.241 & 0.6750 & 0.974 & 15.68 \\
    \enddata
	\tablecomments{DES Y6 ID: DES Y6A1 {\sc coadd\_object\_id} number. $\alpha$: Right Ascension. $\delta$: Declination. $\langle grizY \rangle$: Mean extinction-corrected magnitude. $P$: Best-fit period. $A_{g}$: Best-fit amplitude in DES $g$. $\mu$: Best-fit distance modulus. The full version of this catalog, including feature values and cross-matching information, is available in the online data products at \href{https://des.ncsa.illinois.edu/releases/other/y6-rrl}{this URL}.}
    \label{tab:candidates}
\end{deluxetable*}

\subsection{Template Fitting}\label{sec:template}

Following \citetalias{paper1}, we fit a multiband RRab light-curve template to the extinction-corrected light curves of each object passing our aforementioned selection criteria. Our template is empirically derived from the well-sampled RRab light curves of \citetalias{Sesar:2010}. Our template-fitting procedure is particularly effective when applied to sparsely sampled multiband light curves because it solves for only four independent parameters: the period ($P$), phase ($\phi$), $g$-band amplitude ($A_{g}$), and distance modulus ($\mu$). In \citetalias{paper1}, we showed that these parameters can be effectively constrained with a small number of observations, and we refer the reader to that paper for a thorough discussion of this method. 

We apply the same template-fitting procedure as \citetalias{paper1} with only minor modifications. In particular, we expanded the period range from 0.44--0.89 days in \citetalias{paper1} to  0.2--1 days. As in \citetalias{paper1}, the periods, amplitudes, phases, distance moduli, and residual sum of squares (RSS) per degree of freedom of the fits are kept for the three best-fitting templates. Fitting one light curve requires $\sim$2--6 minutes, depending on the CPU load at the time of processing. 

Although both RRab and RRc can be used to estimate distances \citep[][and references therein]{Caceres:2008,Marconi:2015}, RRc are more easily confused with other types of variable objects due to their lower amplitude of variation and more sinusoidal light curves. This, combined with their lower rates of occurrence in halo RRL populations \citep{2017ApJ...850..137M}, makes them less attractive targets for our analysis of Galactic structures in \secref{results}. Thus, our template fit and subsequent classification are optimized to identify and fit RRab light curves and properties. Some RRc do pass all of the steps of our identification and fitting procedure and are misidentified as RRab. We explore the recovery rate for RRc with simulated light curves in \secref{sim}.

\subsection{Classification}\label{sec:rf2}

Even though our color and variability selections decrease by $>\!100\!\times$ the number of objects that require template fitting, there are still too many for individual visual inspection. Thus, we train another random forest classifier to select potential RRL. We input the RSS, amplitudes, and von Mises--Fisher concentration parameter $\kappa$ (see \S~4.3 in \citetalias{paper1} for more details) from the top three best template fits for each candidate. As in \citetalias{paper1}, we also calculate the distances of each set of periods and amplitudes to the scaled Oosterhoff relations \citep{Oosterhoff:1939} parameterized by \citet{Fabrizio:2019} and scaled to the $g$ band by \citet[][see their Fig.~7]{Vivas:2020b}. 

For this training set, we used the previously known cross-matched RRab described in \secref{data}, instead of the simulated light curves.  Although the former do not span the full magnitude range of DES Y6, they provide a more realistic representation of the performance of template fits to RRab beyond those from \citetalias{Sesar:2010}.  The template-fitting procedure performs artificially too well on simulated light curves because the templates and simulations are based on the same data \citepalias{paper1}.  For the non-RRab set, we include objects likely to contaminate our sample, such as the cross-matched QSOs from \secref{extcat_select} as well as objects that were visually rejected as extended sources and artifacts in a previous iteration of the catalog.

The cutoff score is selected using Matthew's correlation coefficient (MCC), which balances true and false positives and negatives for binary classification problems, even in the case of imbalanced class sizes \citep{Matthews:1975}. The MCC is defined as
\begin{align}\label{eq:mcc}
    &{\rm MCC} = \\ \nonumber
    &{\rm \frac{TP\times TN - FP\times FN}{\sqrt{(TP + FP)(TP+FN)(TN+FP)(TN+FN)}}},
\end{align}
\noindent where TP is the true positives, TN is the true negatives, FP is the false positives, and FN is the false negatives. We choose a classifier score cutoff of 0.605 that maximizes the MCC for the training sample. For the training set, this cutoff score recovers 96.07\% of the input RRab with 98.92\% purity.  When we apply this classifier to the full results of the template fitting, 10,812 objects pass the cutoff as potential candidates. The performance curves and top features are shown in \appref{roc_features}.  We discuss the recovery fraction for a broader range of RRL distances in \secref{sim}.

\begin{figure*}[!t]
  \includegraphics[width=1\textwidth]{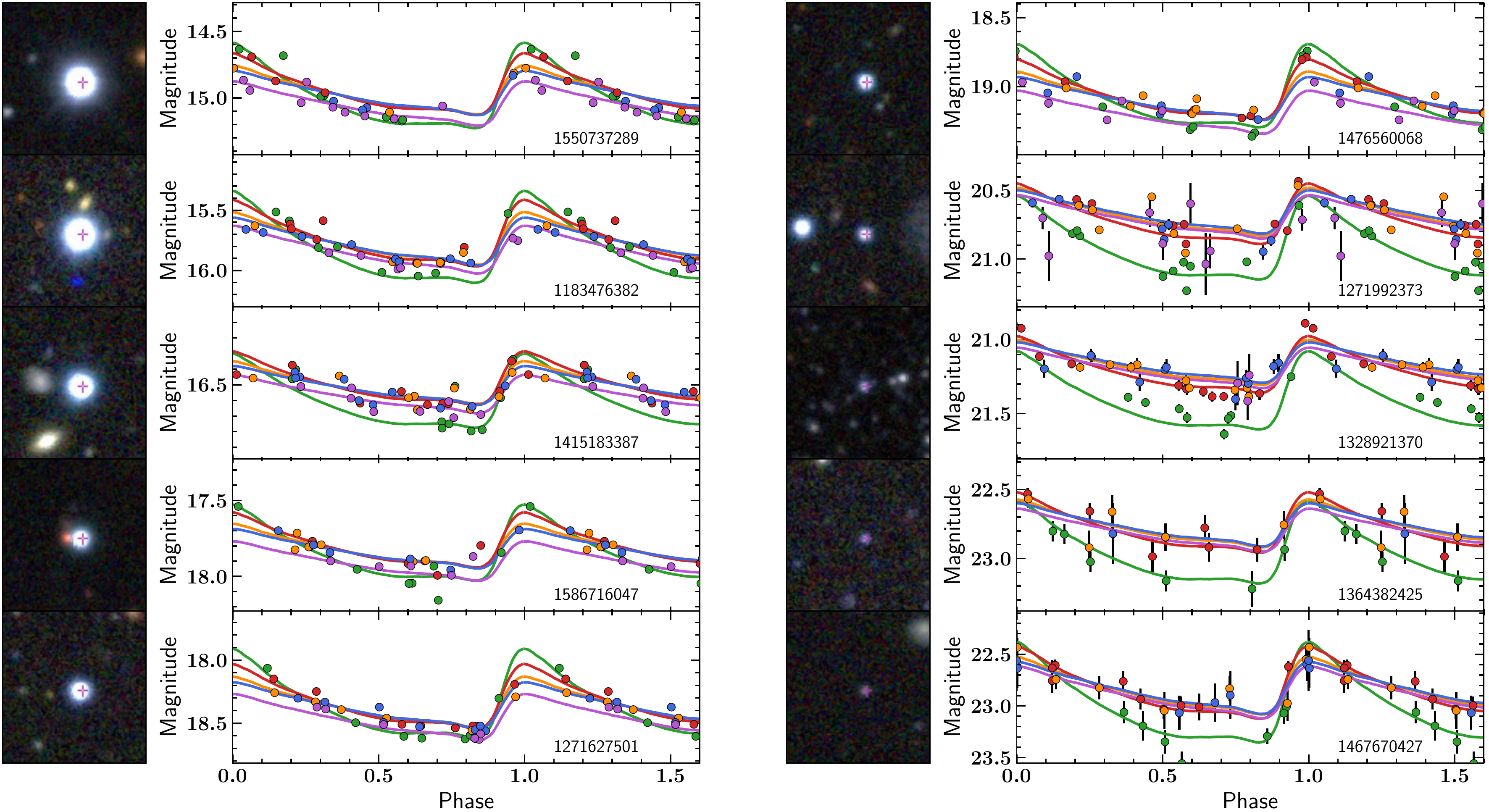}
  \caption{Sample of RRab candidates selected to represent the magnitude range of our catalog. Light curves are arranged vertically by increasing magnitude with best-fit heliocentric distances ranging from 7\kpc (top left) to 280\kpc (lower right). Observed magnitudes and RRab template light curves are colored by band ($g$: green, $r$: red, $i$: orange, $z$: blue, $Y$: purple). If not visible, photometric errors are smaller than the plotting symbols.}
  \label{fig:lc_grid}
\end{figure*}

Since the purity of the DES star--galaxy separation decreases dramatically at fainter magnitudes \citep[e.g.,][]{Shipp:2018}, we visually inspect every candidate that had not been previously identified as an RRab by another survey (see \figref{lc_grid}). During this visual inspection, we remove any obvious extended sources, bright oversaturated objects, and extremely poor-fitting light curves. 
After this visual inspection, our final sample consists of \Nrrab objects. 

\ \par

\begin{figure*}[!t]
  \includegraphics[width=\textwidth]{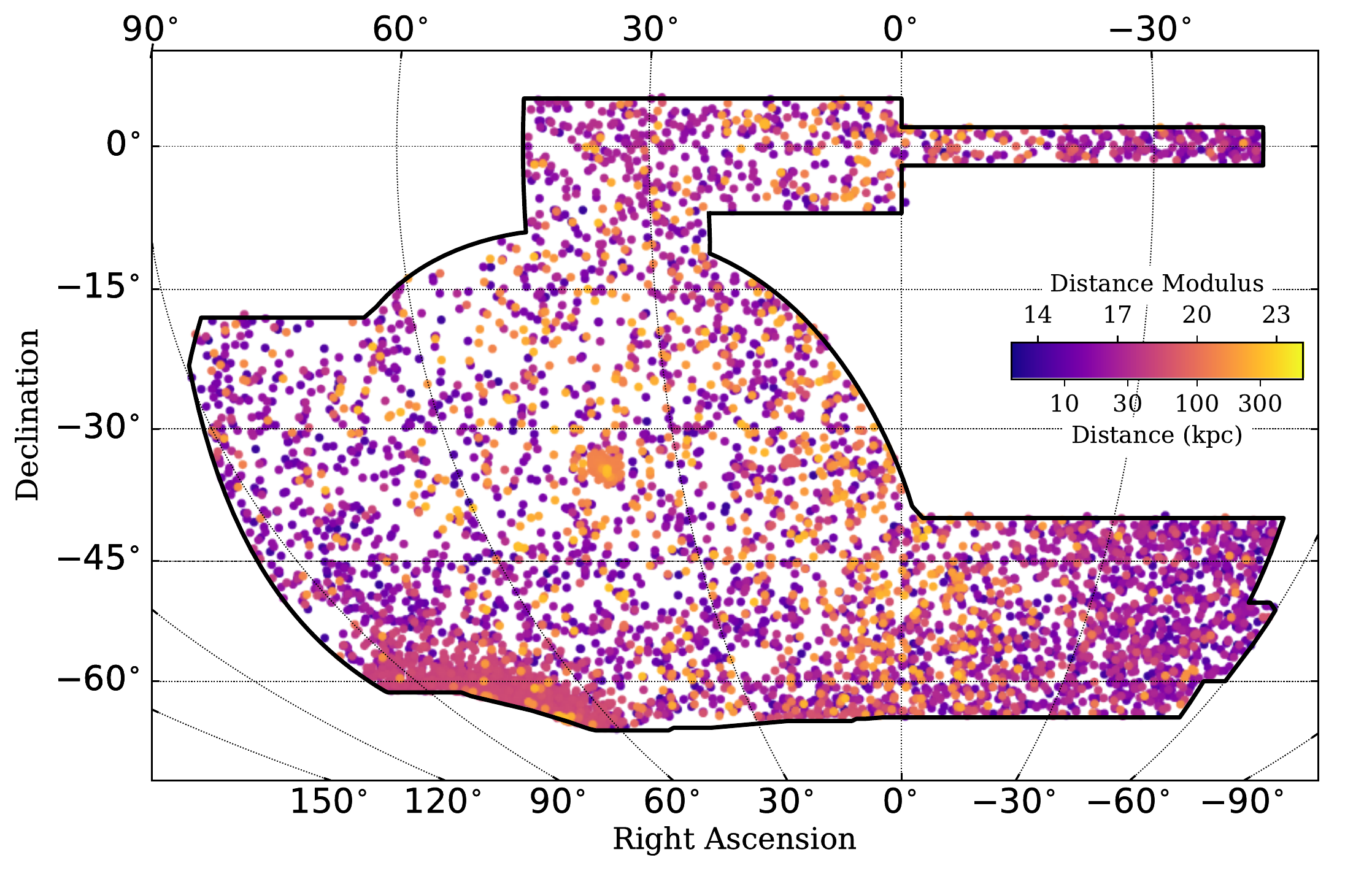}
  \includegraphics[width=\textwidth]{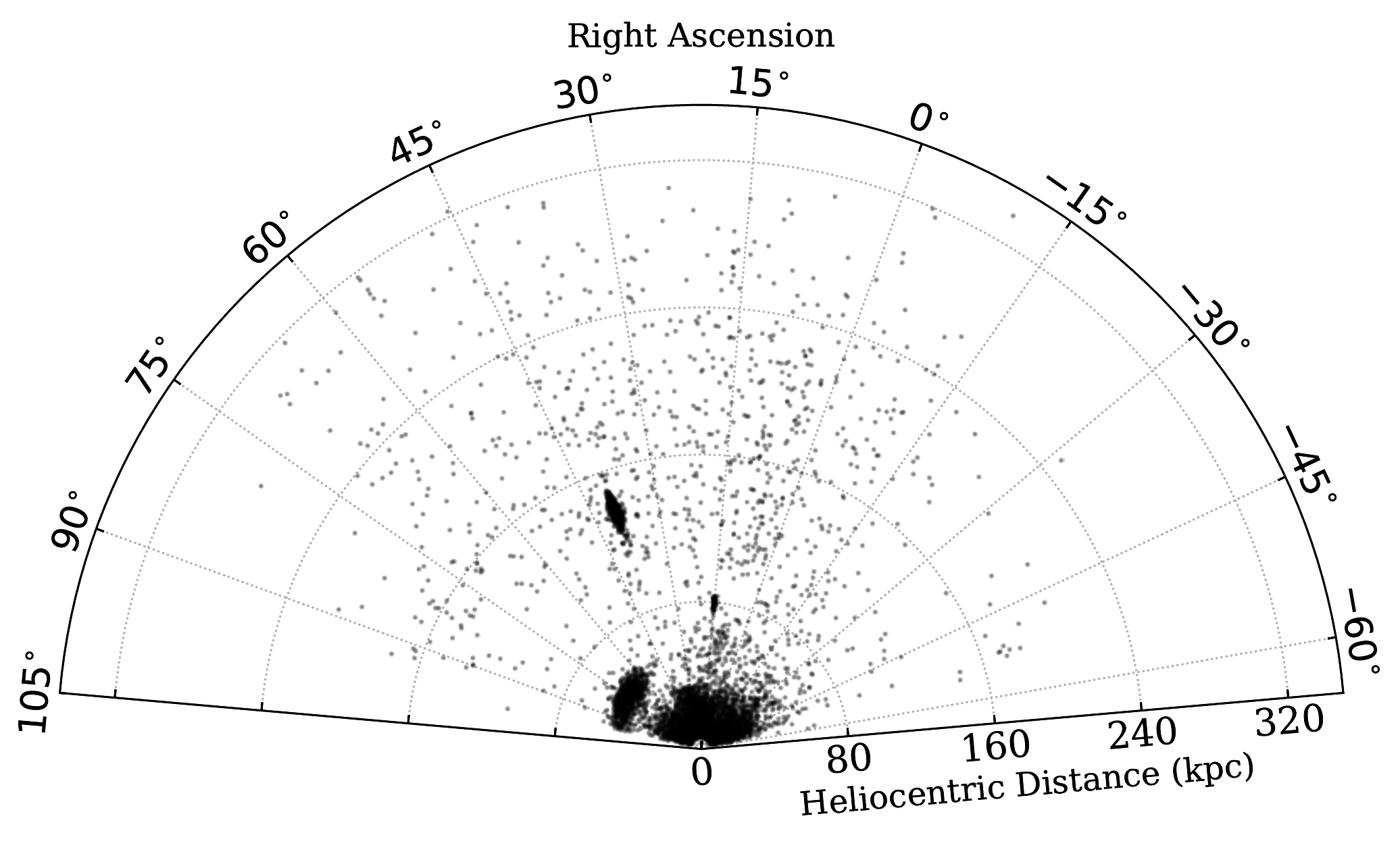}
  \caption{{\it Top:} Sky location of DES Y6 RRab candidates colored by distance modulus. This figure uses a McBryde--Thomas flat polar quartic projection with the DES footprint shown in black. {\it Bottom:} Heliocentric distance distribution of the DES Y6 RRab candidates as a function of right ascension. Overdensities corresponding to the LMC ($\alpha \sim 80 \degree$), Fornax ($\alpha \sim 39\degree$), and Sculptor ($\alpha \sim 15\degree$) can be seen in both panels. Stars associated with Eridanus II can be seen in the bottom panel at $\alpha \sim 56\degree$ and $D > 300\kpc$.}
  \label{fig:skyplots}
\end{figure*}

\begin{figure*}[!t]
    \includegraphics[width=1\textwidth]{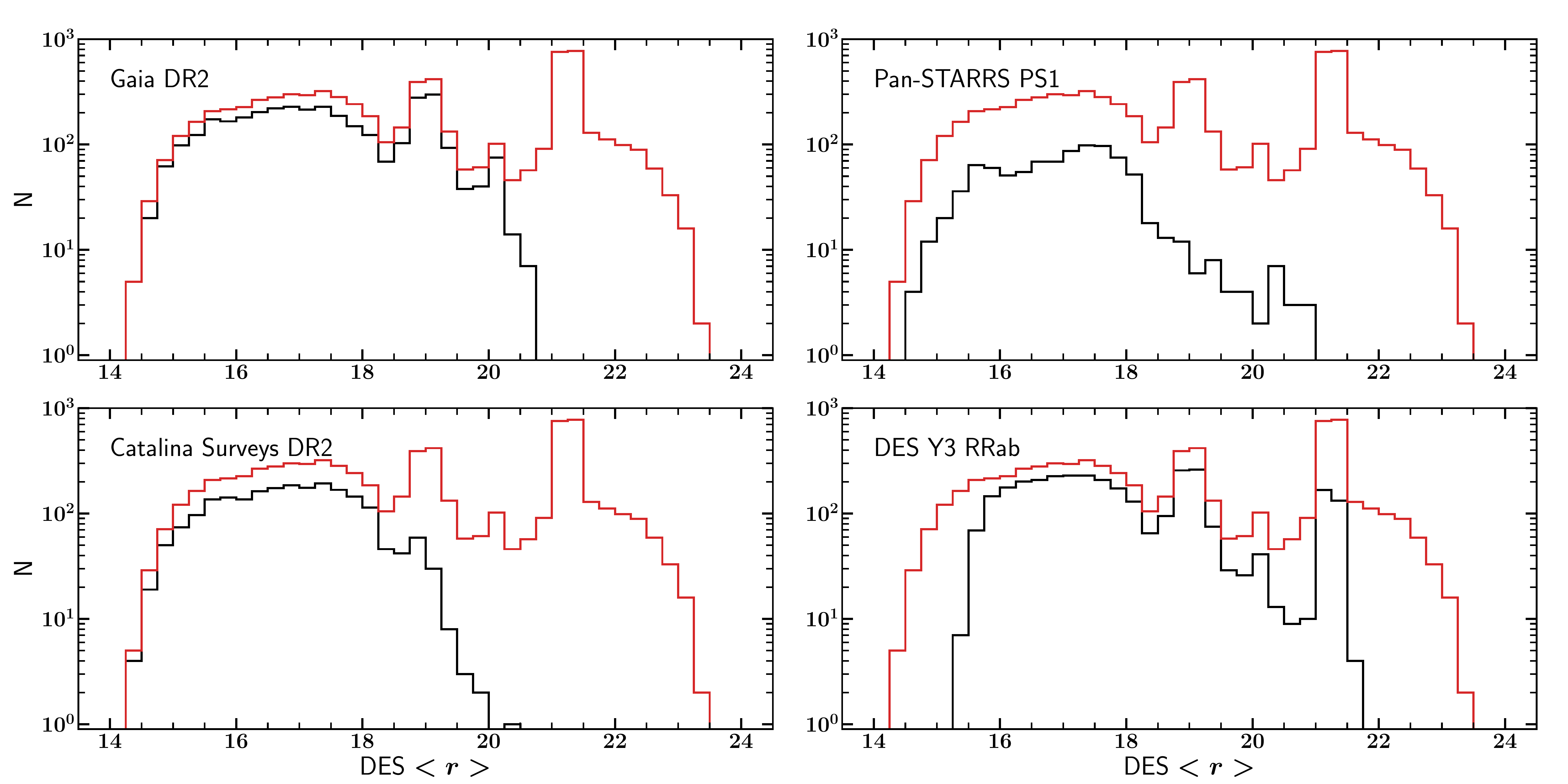}
    \caption{Overlap with external RRab catalogs. The distribution of this RRab catalog as a function of the average extinction-corrected $r$ magnitudes, $\langle r\rangle$, are plotted in red. The distributions of previously identified RRab from external catalogs that were recovered in our catalog are plotted in black. \textit{Bottom right:} We note that the combination of the increased number of observations, decreased signal-to-noise threshold, and alternative selection cuts in Y6 increased the depth of our catalog by $\sim 1$ magnitude, but also extended the detection range to brighter magnitudes as well.}
    \label{fig:external_overlap}
\end{figure*}

\section{Results} \label{sec:results}

Our selection process results in a catalog of \Nrrab objects that are consistent with the colors, variability, and light curves of RRab. We hereafter refer to objects in this sample as DES Y6 ``RRab candidates''. 
We show a selection of light curves covering the full magnitude range of our sample in \figref{lc_grid}. 
DES Y6 provides a median of \CHECK{35} observations per RRab candidate.
The equatorial positions and heliocentric distances of DES Y6 RRab candidates are shown in the top and bottom panels of \figref{skyplots}, respectively. The RRab candidates that were previously identified by other studies, including the DES Y3 RRab search of \citetalias{paper1}, are shown in \figref{external_overlap}. Because the boundaries of the DES Y3 and Y6 catalogs differ (most significantly around the LMC outskirts), this is not a one-to-one comparison. 

Although image cutouts and light-curve fits are visually inspected for each of these candidates, they still require additional observations to confirm that they are RRL. This is especially true of objects with sparsely sampled light curves and/or located beyond 150 kpc. The properties of visually-accepted and previously-identified RRab candidates are summarized in \tabref{candidates} (full version available online).

\subsection{Completeness for faint objects} \label{sec:sim}

\begin{figure*}[t]
    \includegraphics[width=1\textwidth]{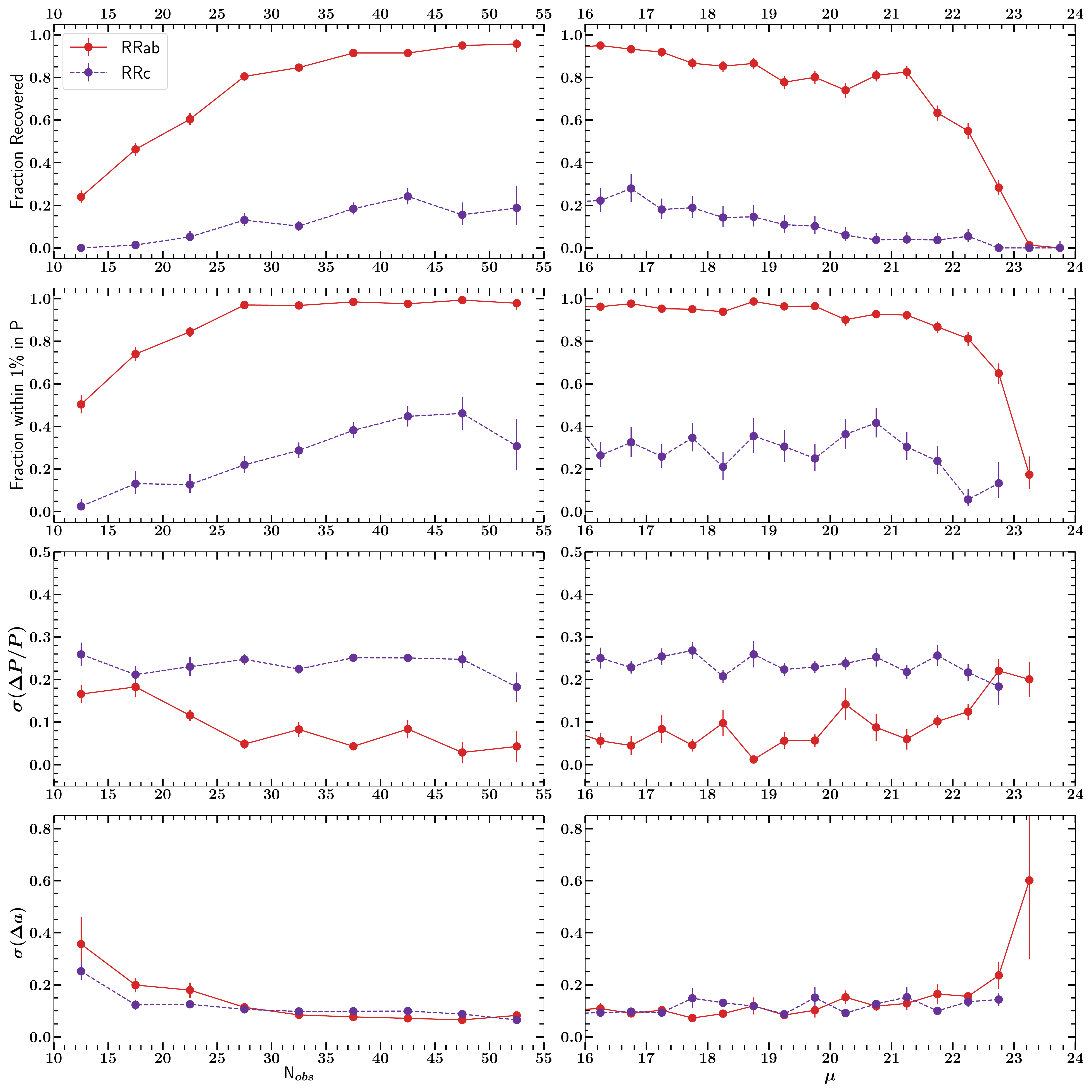}
    \caption{\label{fig:sim_results} Recovery of simulated RRL light curves. RRab results are plotted in red, while RRc are plotted in purple. All quantities in the left column are plotted as functions of the number of light-curve observations, $N_{\rm obs}$, while the right column shows these same quantities plotted as a function of distance modulus, $\mu$. We do not explicitly search for RRc in this work, and these values can be used to assess the RRc contamination in our catalog. The quantities shown are as follows. 
\textit{First row:} Fraction of simulated curves identified as RRab by the full classification pipeline. 
\textit{Second row:} Fraction of periods correctly estimated within 1\% of their input values. The period accuracy approaches 100\% near $N_{\rm obs} \ge 30$ and rapidly decreases beyond $\mu \sim 21$. \textit{Third row:} Standard deviation of period estimate precision. \textit{Bottom row:} Standard deviation of amplitude estimate precision. We see a degradation in precision of the amplitude measurement for light curves with $N_{\rm obs} \lesssim 30$ and $\mu \gtrsim 21$, similar to that of the period estimates.
Uncertainties on the recovery efficiency in the first two rows are assessed through the Bayesian technique of \citet{Paterno:2004}, while the bottom two rows show uncertainties from bootstrap resampling \citep[e.g.,][]{Efron:1982}.
}
\end{figure*}

Due to the limited depth and area of external RRL catalogs, we are required to use simulated RRL to assess the completeness of our selection for faint RRL ($D \gtrsim 100 \kpc$). 
To assess the recovery of our template fitting and classification process and to assess how frequently RRc are misclassified as RRab, we follow the procedure described in \secref{extcat_select} to simulate an independent set of light curves for $\roughly 3{,}100$ RRab and $\roughly 900$ RRc.
We subject these simulated light curves to the same minimum number of observation, color cuts, variability classifier, template fitting, and final classifier that are applied to the real data. 
The recovery rate and precision of the parameter estimates are shown in \figref{sim_results}.

We find that our ability to identify RRab and recover periods within 1\% of the input value improves dramatically as the number of observations increases and degrades dramatically at distances $\gtrsim 200\kpc$. 
The uncertainties in period and amplitude for both RRab and RRc using the template-fitting technique improve when there are more observations available and decline for objects at larger distances since their photometric uncertainties are larger. The period recovery for RRab approaches 100\% when the light curve has $\ge 30$ observations. The raction of RRc misidentified as RRab by the classification pipeline is overall very low ($\le 10\%$ for light curves with $\ge 25$ observations and $\less 20\%$ at all distances). Given that RRc make up $\lesssim 30\%$ of RRL \citep{Soszynski19,2017ApJ...850..137M}, the low efficiency suggests that RRc are unlikely to contribute substantially to our final catalog.

Our recovery rate for distant RRab ($\geq 50\%$ for $\mu \leq 22$) opens an exciting new discovery space for stellar structures beyond 100 kpc. In the following sections, we compare the DES Y6 RRab candidate sample to external catalogs in classical and ultra-faint dwarf galaxies. In addition, we use this catalog to fit the Milky Way stellar halo profile and search for previously undiscovered Milky Way satellites.


\begin{figure*}[t]
\center
\includegraphics[trim={0 0 2.0cm 0},clip,width=0.52\textwidth]{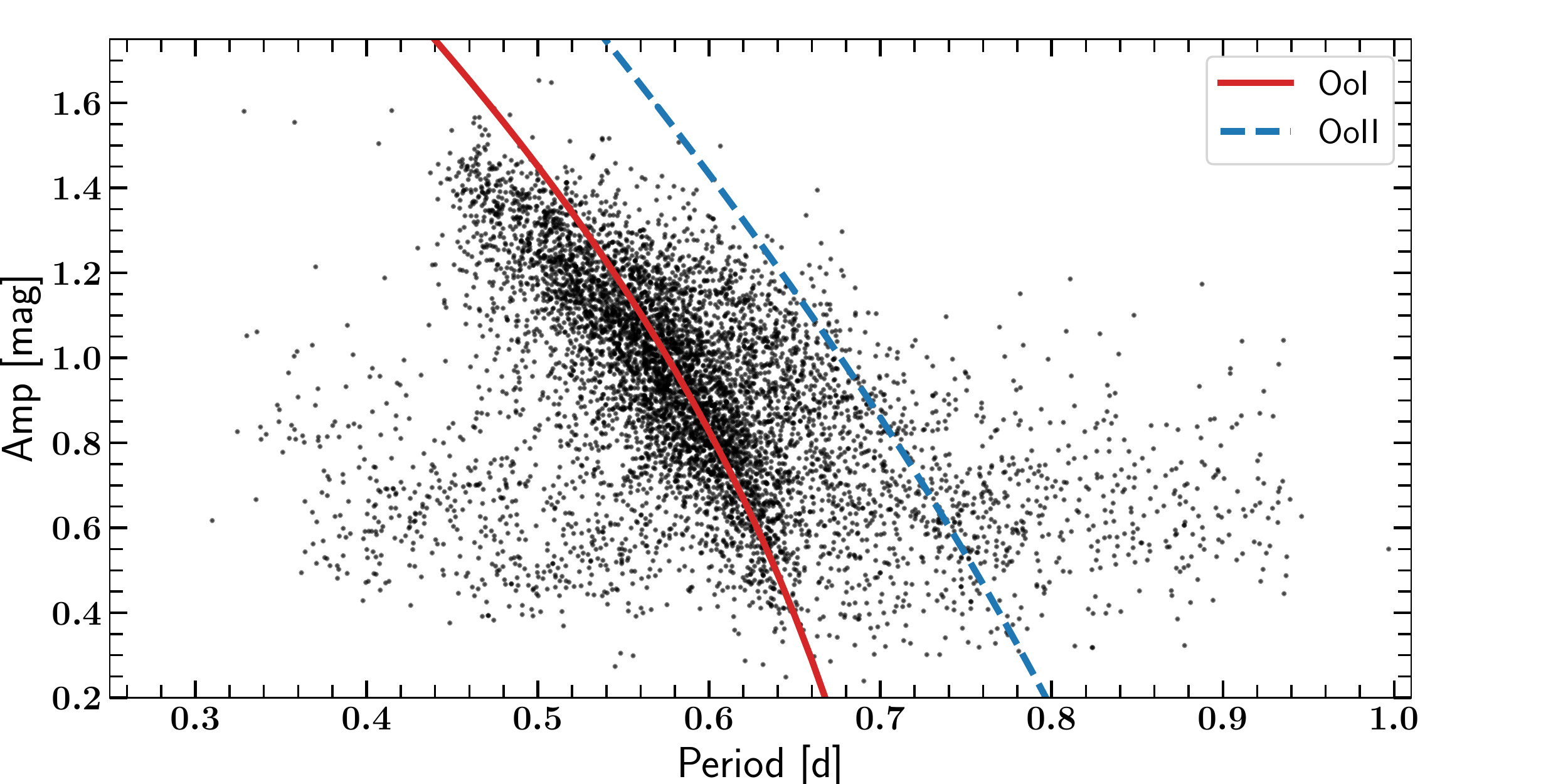}
\includegraphics[width=0.43\textwidth]{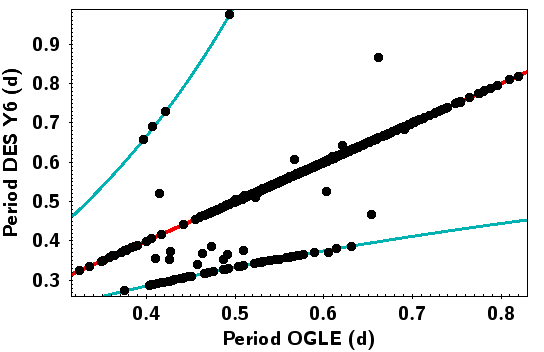}
\caption{\label{fig:bailey_ogle} 
{\it Left:} Period-amplitude diagram for DES Y6 RRab candidates. Overplotted are the Oosterhoff relations parameterized by \citet{Fabrizio:2019} and scaled to the $g$ band by \citet{Vivas:2020b}. 
{\it Right:} Comparison of the periods of 792 RRL in common with OGLE. The red line is a 1:1 relation (91\% of objects), while the cyan lines correspond to the $\pm 1$-day aliases.}
\end{figure*}

\subsection{The Magellanic Clouds} \label{sec:MC}

The variable star content of the Magellanic Clouds (MCs) has been extensively studied, with the OGLE-IV catalog of \citet{Soszynski19} containing 47,828 RRL in both MCs. Although the DES footprint only covers the outskirts of the MCs, a comparison with the OGLE catalog still yields a relatively large number of objects in common (792) and allows us to assess the recovery of light-curve properties of the DES Y6 RRab candidates. The OGLE catalog is ideal for this comparison because its light curves have $>100$ epochs (and consequently, period estimates and classifications are very robust) and it contains many types of variables (allowing us to investigate the contamination rate of our catalog). 

The variables in common between the two catalogs include 53 objects classified as RRc or RRd by OGLE, implying a $\roughly 7\%$ contamination by these types in our sample. This is consistent with estimates from our simulations (\figref{sim_results}) under the assumption that stellar populations are similar in the Milky Way and LMC halos, and  with the ``cloud'' seen in the period--amplitude diagram (left panel of \figref{bailey_ogle}) at short periods and low amplitudes.

720 (91\%) of the matches have an excellent agreement in period (right panel of \figref{bailey_ogle}). The median difference in period for objects on the 1:1 line in \figref{bailey_ogle} is only 0.001\%. Most period mismatches are due to 1-day aliasing.

We also search for matches between our catalog and other types of variables in OGLE, finding one object that they identified as an anomalous Cepheid (AC). The light curves of ACs and RRL are similar, and it is difficult to distinguish them in the field. ACs are rare, so the contamination of ACs is in the DES Y6 RRab catalog is expected to be low. We found no matches with the OGLE catalog of 40,204 eclipsing variables in the MCs \citep{Soszynski19}.

\begin{figure*}[!t]
  \center
  \setlength{\fboxsep}{-1pt}
  \fbox{\includegraphics[width=0.493\textwidth]{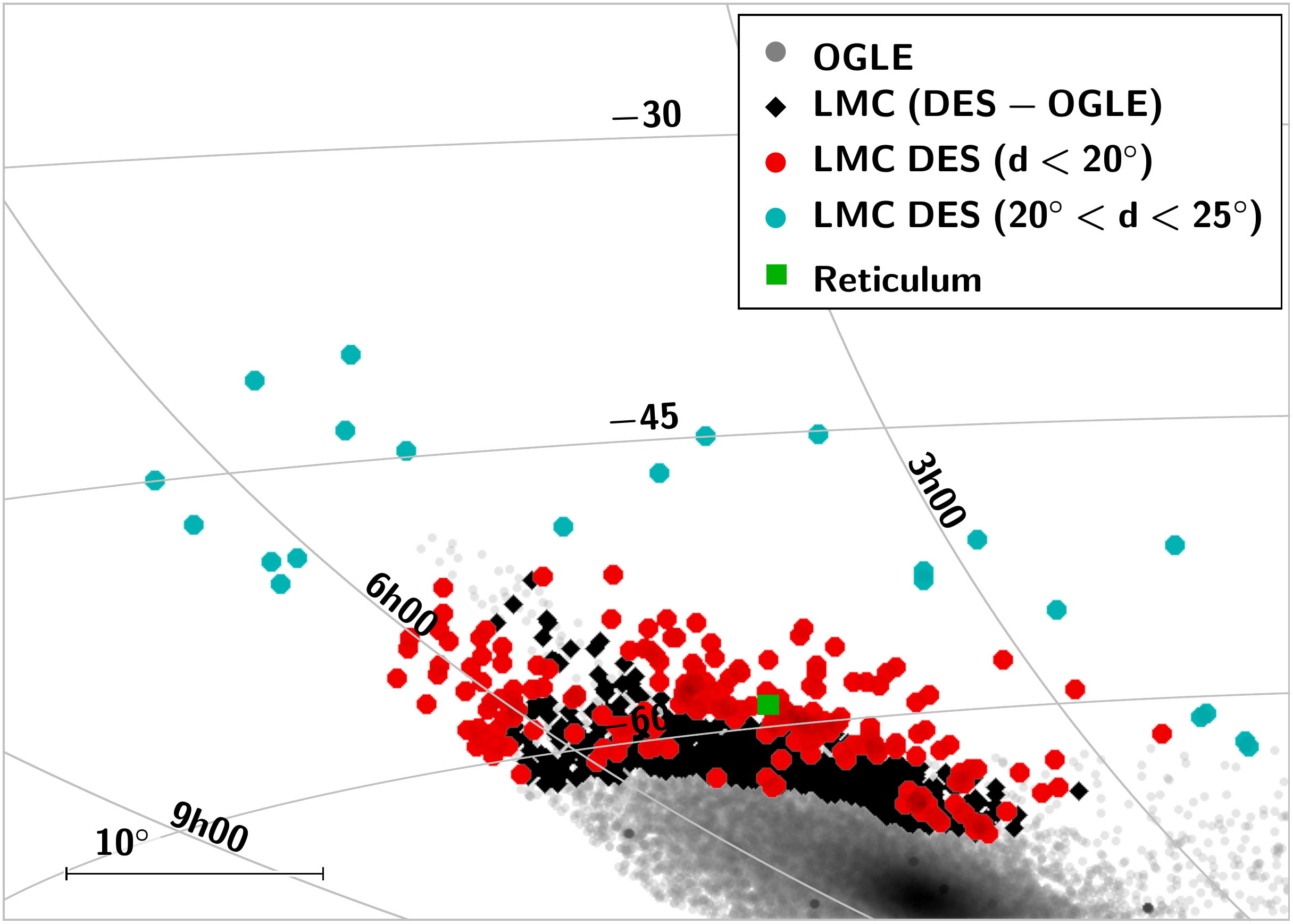}}
  \fbox{\includegraphics[width=0.485\textwidth]{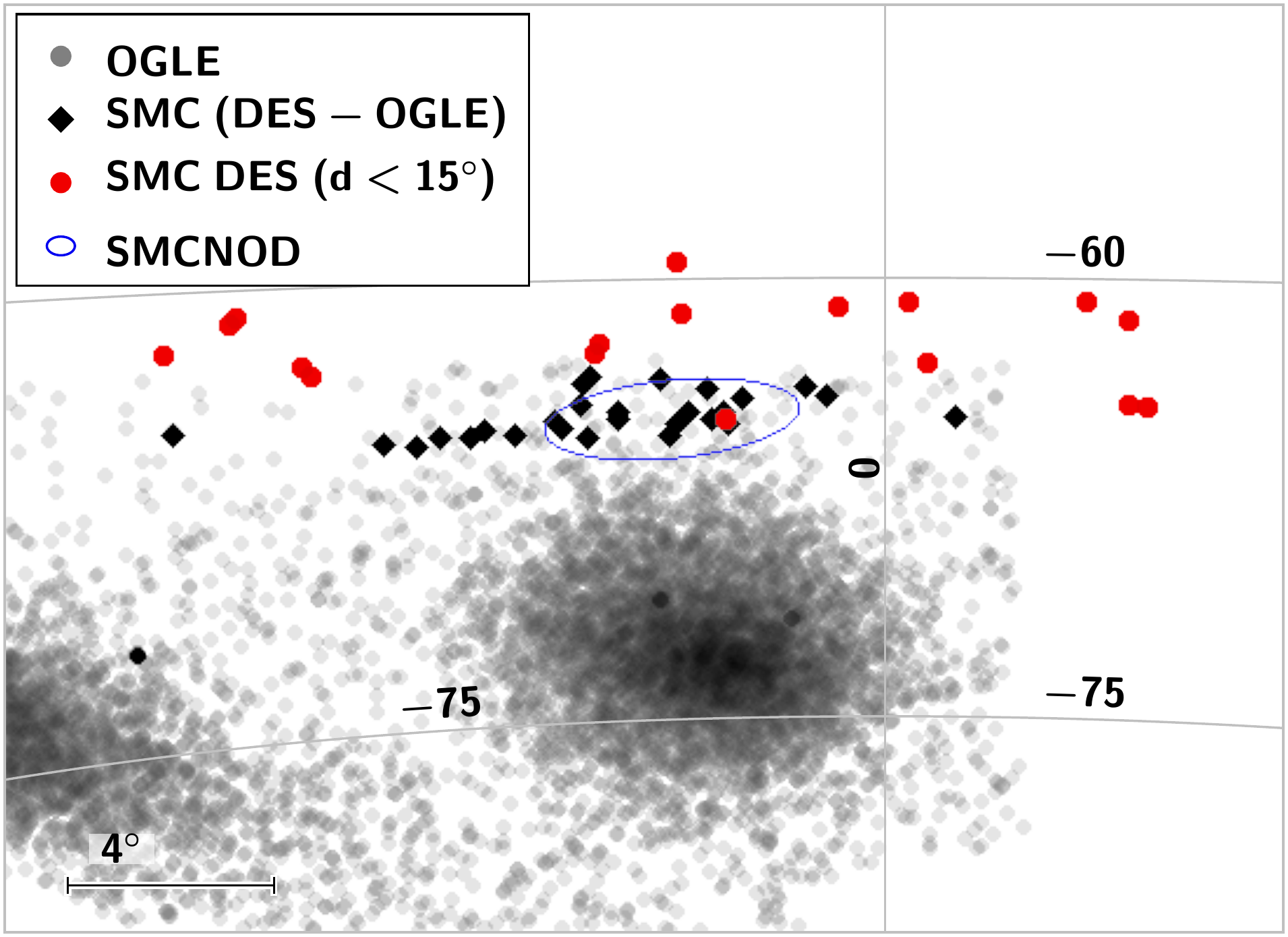}}
  \caption{Aitoff projection of the sky in equatorial coordinates showing the region around the LMC ({\it left}) and SMC ({\it right}). In both panels, the gray points show the density of OGLE RRL, the black diamonds show the DES RRab candidates matched with OGLE, while the red circles are DES RRab candidates that are not matched to OGLE but have a distance moduli in agreement with the distance of each galaxy. {\it Left:} Teal points show possible LMC member stars that have distance moduli matching the LMC ($17.9 < \mu <18.8$), but are separated from the LMC by $20\degr$--$25\degr$. The green square shows the location of the Reticulum globular cluster, which has 16 DES Y6 RRab candidates associated with it. {\it Right:} RRab candidates potentially associated with the SMC ($18.1 < \mu < 19.2$); the location of the SMCNOD \citep{Pieres:2017} is indicated with a blue ellipse.}
    \label{fig:MCSky}
\end{figure*}

We search for new RRab in the MCs using the DES Y6 coverage beyond the OGLE footprint. We start by selecting a range of distance moduli for each MC based on stars in common with OGLE. For the LMC, the distance modulus distribution showed a clear peak between $17.9 < \mu < 18.8$, containing 608 RRL. The SMC overlap with OGLE is significantly smaller, but there is still a clear peak in the distance modulus distribution at $18.1 < \mu < 19.2$, with 27 RRL matching with OGLE in that distance range. Matched OGLE-DES stars are shown as black diamonds in \figref{MCSky}. We then select DES Y6 RRab candidates that were not in OGLE but have distance moduli consistent with the selection defined above and have an angular separation of less than $25\degr$ and $15\degr$ from the centers of the LMC and SMC, respectively. These stars are potential new members of the MCs. Most of them are located outside the OGLE footprint, although there are also some stars that may have been missed by OGLE.

\begin{figure}[!bht]
    \includegraphics[width=0.45\textwidth]{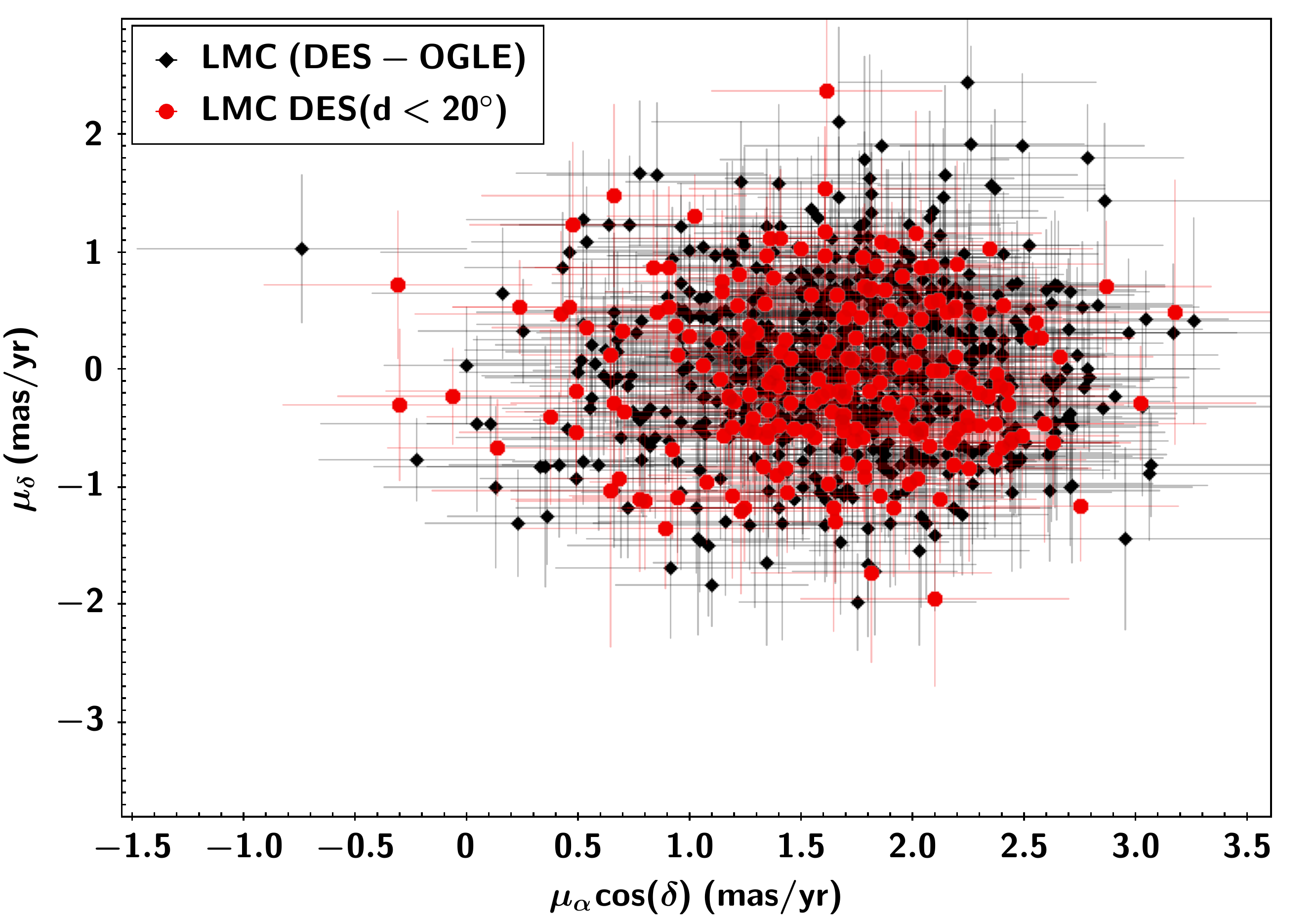}
    \includegraphics[width=0.45\textwidth]{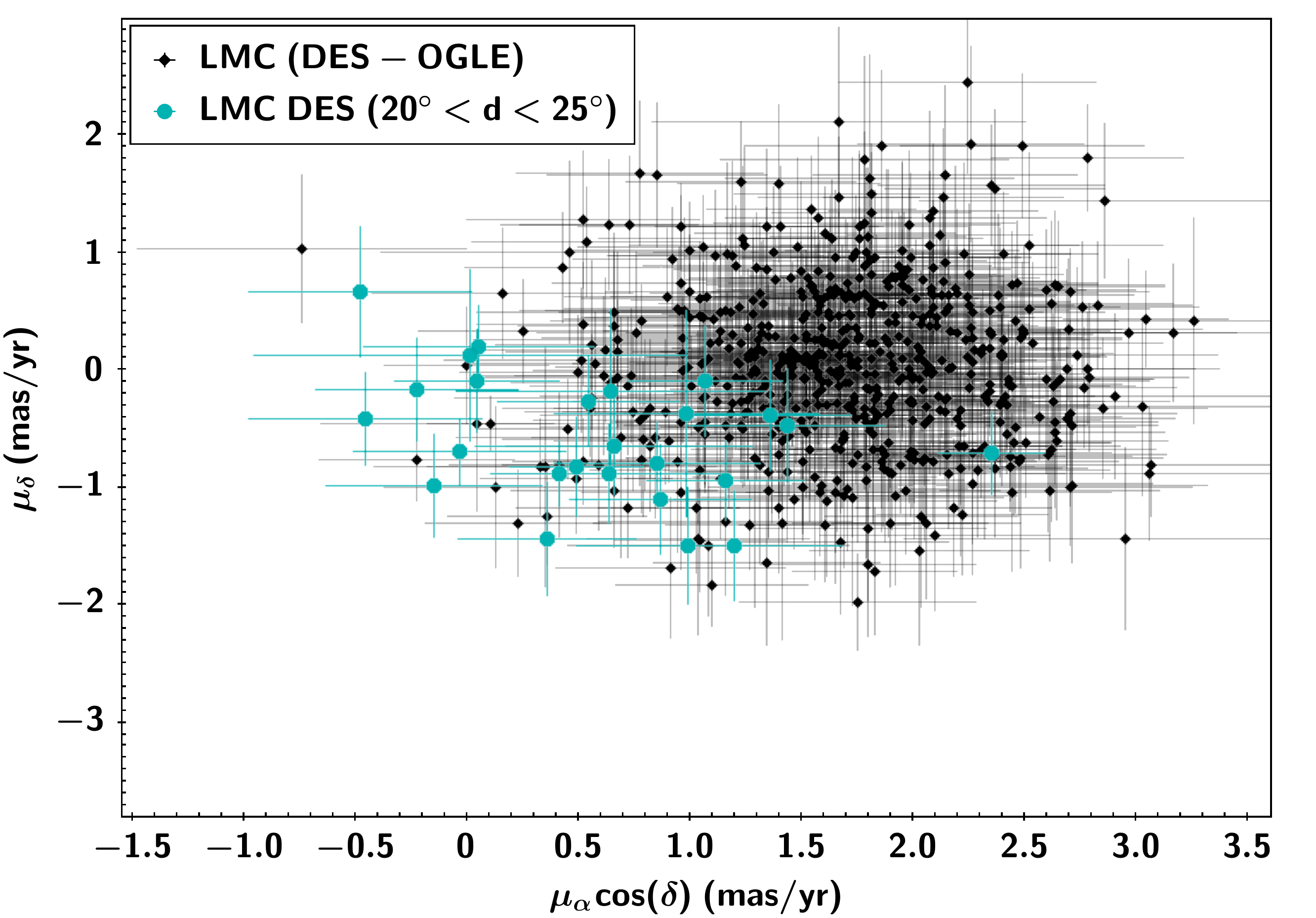} 
    \includegraphics[width=0.45\textwidth]{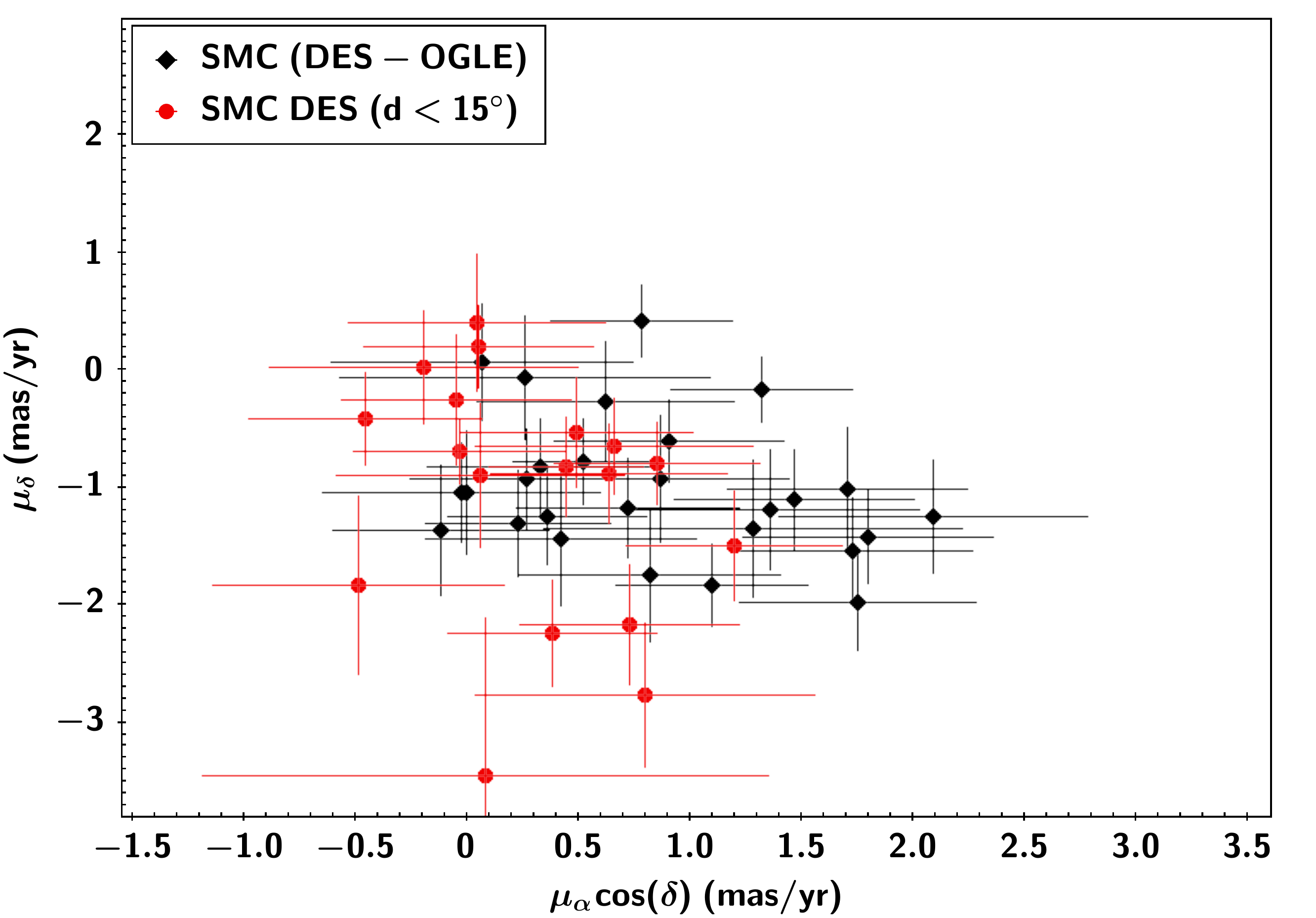}
    \caption{Gaia DR2 proper motions of DES Y6 RRab candidates around the LMC (top and middle panels) and in the SMC (bottom panel). In all panels, the black points are candidates that match with the OGLE catalog. For the LMC, we show separately the candidates closer to the center of the galaxy in red (top panel) and candidates with larger separation in teal (middle panel). For the SMC, we show new candidate members within $15\degr$ in red.}
    \label{fig:MC_PM}
\end{figure}

We use {\it Gaia} DR2 data \citep{Gaia:2018} to check whether our candidate MC members exhibit proper motions (PMs) consistent with those of the MCs. In \figref{MC_PM} we show the PMs of the stars matched to OGLE in each of the MCs and compare them with the new member candidates. For clarity, we show two plots for the LMC: one containing new candidate members within $20\degr$ of its center (top panel), and another for those lying between $20\degr$ and $25\degr$ (center panel). We find that new LMC member candidates in the former group have a distribution of PMs that is consistent with OGLE stars, while the latter do not. We find 202 LMC candidate RRL out to $20\degr$, excluding 16 members of the LMC globular cluster Reticulum (which we discuss separately in \secref{ultra}). Beyond this limit and out to $25\degr$, the density of LMC candidate RRL drops appreciably with only 25 possible members.

The old stellar population of the LMC extends to large angular separations, as shown in previous studies. \citet{Saha10} traced the LMC population with main-sequence turnoff stars out to $16\degr$ from its center. Using a similar technique, \citet{Nidever19} further extended the detection to $21\degr$. In addition, \citet{Belokurov:2016} used publicly available DES images  to find a lumpy distribution of Blue Horizontal Branch (BHB) stars extending out to $30\degr$, and possibly up to $50\degr$. We note in passing that some of our candidates match the location and distance of their ``S1'' substructure. Our strong detection of RRL in the LMC periphery confirms these previous results, and strengthens the case for an extended old ($\gtrsim 10\Gyr$) stellar population in the LMC. 

The southern declination limit of DES is $\delta = -65\fdg3$. Consequently, only the very external parts of the SMC (which is centered at $\delta \roughly -72\fdg8$) are within the footprint of the survey. Nonetheless, the OGLE catalog contains SMC RRL as far north as $\delta \sim -62\fdg7$, 27 of which are in common with our sample (\figref{MCSky}). We find 18 other RRL within the distance range of the SMC out to an angular separation of $15\deg$ (a region beyond the OGLE coverage). There is good agreement in the PMs of these new member candidates, as seen in the bottom panel of \figref{MC_PM}. 

An interesting feature associated with the SMC in this part of the sky is the so-called Small Magellanic Cloud Northern Over-Density \citep[SMCNOD,][]{Pieres:2017}, whose area is indicated by a blue ellipse in the left panel of \figref{MCSky}. \citet{Pieres:2017} estimated that the SMCNOD is primarily composed of an intermediate-age population of $\roughly 6\Gyr$, which is too young to contain RRL.  Indeed, the variable star content in the SMCNOD was examined by \citet{Prudil18} with an earlier release of the OGLE catalog \citep{Soszynski17}. Although they found eight OGLE RRab inside the SMCNOD area, they concluded their density was compatible with the expected SMC background at that distance. In contrast, we find almost twice as many (15 RRab candidates) in the same region. 

\subsection{Sculptor and Fornax} \label{sec:SculptorFornax}

In addition to the MCs, two other prominent overdensities of RRab candidates are within the DES footprint: the classical dwarf spheroidal (dSph) galaxies Sculptor $(\alpha,D = 15\fdg02, 86\kpc)$ and Fornax $(\alpha,D = 39\fdg96, 147 \kpc$). Both can be clearly seen in \figref{skyplots}.

\citet[][hereafter MV16]{Martinez-Vazquez:2016b} presented the most complete and extensive study of the variable star population in Sculptor, reporting 536 RRL, of which 289 were RRab. We search for potential Sculptor members by selecting all RRL in our catalog out to $1.5\times$ its tidal radius of $69\farcm1$ \citep{Munoz18} and with distance moduli within $19.67 \pm 0.75$.
We find 116 RRL within these limits, all but four matching the catalog of \citetalias{Martinez-Vazquez:2016b}. The lower-than-average completeness in this region ($39\%$ at $\mu\sim 19.5$) is due to crowding effects near the center of the galaxy.

\citetalias{Martinez-Vazquez:2016b} classified six of our RRL candidates as RRc/RRd. This represents a 5\% contamination, in agreement with the value derived from the MCs (\secref{MC}). Removing these misclassified stars, we find that the periods for the rest of our sample agree very well with those from  \citetalias{Martinez-Vazquez:2016b}, with a median difference of only 0.002\%. 

Three of the four new candidate members associated with this galaxy lie outside the footprint explored by \citetalias{Martinez-Vazquez:2016b}. One is located at $93\arcmin$, significantly beyond the tidal radius of Sculptor but with a compatible {\it Gaia} DR2 proper motion. The mean distance modulus of our Sculptor RRL is $\langle \mu \rangle = 19.49\pm0.09$ mag, while \citet{Martinez-Vazquez:2015} found a value of $19.62\pm 0.04$ mag. Although small, the difference can be attributed to the fact that our analysis assumes all RRL have the mean metallicity of the Milky Way halo, while the old stellar population in Sculptor has a wide range of metallicities extending down to $\feh \sim -2.4$ \citep{Martinez-Vazquez:2016a}.

\begin{figure}[!t]
    \includegraphics[width=0.45\textwidth]{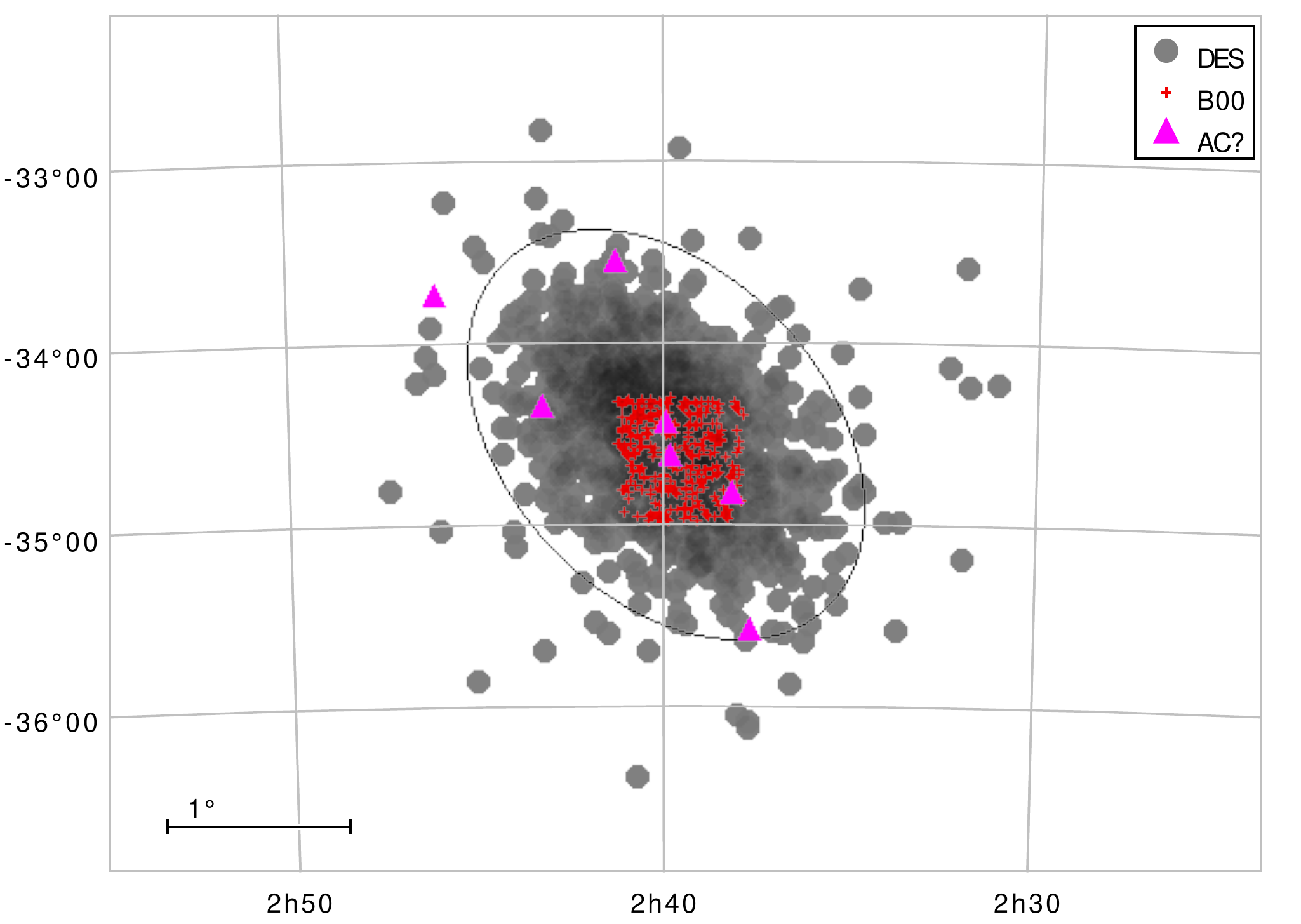}
    \caption{RRL from DES in the Fornax dSph galaxy. The black ellipse shows the tidal radius and structural parameters as derived by \citet{Wang19}. The red crosses indicate the RRL that match with \citet{Bersier:2002}. Magenta triangles are possible anomalous Cepheids in Fornax that were confused as RRL in the DES catalog.}
    \label{fig:Fornax}
\end{figure}

While more distant than the MCs and Sculptor, Fornax also displays a prominent overdensity of DES Y6 RRL candidates.  Again, we select candidate members as stars within $1.5\times$ its tidal radius of $77.5\arcmin$\citep{Wang19} and distance moduli within $20.82 \pm 0.75$, yielding 1385 RRL. The mean distance modulus of these stars is $20.67 \pm 0.08$, which is again slightly smaller than the best-available value of $20.82\pm 0.02$ \citep{Karczmarek17}, likely due to the higher metallicity of our template. 

Surprisingly, 51 of our Fornax RRL candidates are located beyond its tidal radius. These stars are uniformly distributed around the galaxy (\figref{Fornax}), reaching out to $114\arcmin$ from its center (note that our search radius only extended to $116\farcm2$). 
In a recent study based on data from DES Y3, \citet{Wang19} concluded that no significant extra-tidal disturbances are observed down to a surface brightness limit of $\sim 32.1$ mag arcsec$^{-2}$. 
Fornax is known to have several bursts of star formation and is dominated by a population of an age $\roughly 5 \Gyr$ \citep{deBoer12,Rusakov20}.
While this dominant population is too young to have produced the observed RRL, an old stellar population ($> 10 \Gyr$) is also present in Fornax.
This older stellar population has been found to be more spatially extended than the younger populations \citep{Wang19}. 
The detection of RRL outside the tidal radius suggests a very low surface brightness, old, extra-tidal population.

Fornax is known to be rich in RRL, but a complete census is not readily available yet. \citet{Bersier:2002} found 525 RRL in the central part of Fornax, although the quality of their light curves was not good enough to do a proper classification into RRab and RRc types. Based exclusively on their periods, they estimated that 396 of those stars may be RRab. We identified 275 of those stars in our catalog, shown in red in \figref{Fornax}, which suggests a 69\% recovery rate, higher than in Sculptor. 
A more complete search for RRL was most recently made by \citet{Fiorentino:2017}. In this work, they found 990 RRab and 436 RRc in a region of $54\arcmin \times 50\arcmin$ centered on the galaxy. Unfortunately, this catalog is not publicly available, preventing a direct comparison.  However, our results suggest that the total population of RRL in Fornax is not yet known, especially at large angular separations.

We notice that within the Fornax search area, there is a group of seven stars with $19.5 < g < 20.7$, $\roughly 1$ mag brighter than the Fornax RRL. It is unlikely that these are field halo stars, since RRL are rare at such large distances from the Galactic center. A more likely explanation for this group is that they are actually AC stars associated with Fornax. As discussed previously, the light curves of RRL and ACs are easily confused. In this case, these stars reside in the region of the Fornax color--magnitude diagram (CMD) that is expected for AC stars, and such objects are known to exist in Fornax \citep{Bersier:2002,Greco05}, although none of these stars match previously known variables. The spatial distribution of these candidates is shown in \figref{Fornax}.

\subsection{Cepheids in Local Group galaxies and beyond}\label{sec:cepheids}

The Local Group galaxies Phoenix, IC~1613, and Tucana are located within the DES footprint at approximate distances of 415\kpc, 755\kpc, and 887\kpc, respectively \citep{McConnachie12}. Both Phoenix and IC~1613 are spatially coincident with overdensities of objects in our catalog, while Tucana is not. 

In the case of Phoenix, we found three RRL candidates within $5\arcmin$ from the center of the galaxy, which has a half-light radius $r_h=3.76\arcmin$, with $22.3 < g < 22.7$. True RRL stars in this galaxy are expected to be $\roughly 1$~mag fainter.  Since it is unlikely to find distant halo field stars in the line of sight of Phoenix, we believe instead that these may be misclassified AC stars; such objects are known to exist in this galaxy \citep{Gallart:2004}. 

In the case of IC1613, we found a group of 11 RRL candidates with $22 < g < 23$ and within $3\,r_h$ from the center of the galaxy. Given the larger distance of IC1613, this range of observed magnitudes is appropriate for classical Cepheids, as even ACs will be too faint for our catalog in this galaxy. The sample of classical Cepheids in IC1613 by \citet{Bernard:2010} shows numerous stars in this magnitude range, a few of which have periods as short as 0.6 days. \citet{Udalski:2001} found 138 Cepheids within the central $14\farcm2 \times 14\farcm2$ region of IC1613.  We thus suspect some of the 11 objects in our catalog may be short-period classical Cepheids in this galaxy, even though we do not find any matches with previous catalogs. It is also possible that at these faint magnitudes crowding and the misclassification of background galaxies may lead to spurious detections.

Other nearby galaxies (beyond the Local Group) in the DES footprint  are ESO~410-G005, ESO~294-G010, NGC~55, NGC~300, and IC~5152. 
Since these galaxies have distance moduli of $\roughly 26.5$ mag \citep{McConnachie12}, it is unlikely that our analysis would  detect any variable stars. Indeed, no overdensities in the DES Y6 RRab catalog are associated with these objects.


\subsection{Ultra-Faint Dwarf Galaxies and Globular Clusters} 
\label{sec:ultra}

RRL in ultra-faint dwarf galaxies (UFDs) can provide independent estimates of the distances to these systems, which are particularly important given to their low surface brightnesses and sparsely populated CMDs. Recent censuses of RRL in UFDs can be found in \citet{Martinez-Vazquez:2019} and \citet{Vivas:2020}. We use the DES Y6 RRab candidate catalog to search for RRL in the vicinities of 19 UFDs in the DES footprint and find strong evidence of variables associated with Eridanus II, as well as  tentative identifications in Cetus III and Tucana IV. 

Eridanus II is among the most luminous ($M_V \sim -7.1$) and distant ($D \sim 366 \kpc$) UFDs \citep{Crnojevic:2016}. We search for RRL in our catalog within $10\,r_h=10\farcm1$ and find five stars with a very narrow range of distance moduli, $\mu_0=22.45\pm0.04$~mag. All variables are tightly concentrated within $2.4\, r_h=3\farcm 4$ and are confirmed to be RRL in this system by Mart\'inez-V\'azquez et al. (in prep.). Their mean period is 0.663 days, consistent with the Oo II group found in other UFDs \citep{Martinez-Vazquez:2019}. Given the low completeness of our survey at these faint magnitudes, the RRL population in this system should be significantly larger. There are two additional stars in our search area with $g=22.1-22.3$, or $\sim 0.8$ mag brighter than the RRL. These may be anomalous Cepheids in Eridanus II.

Cetus III is somewhat closer ($D \sim 251 \kpc$) and has an absolute magnitude of only $M_V\sim -2.5$ \citep{Homma:2018}. Little is known about this small ultrafaint dwarf and spectroscopic confirmation of its nature is not yet available. We found one RRab candidate located 1\arcmin from its center ($\sim 1\,r_h$) with $\mu=21.33$~mag, which would put the galaxy at 185 kpc from the Sun, somewhat closer than the aforementioned estimate. However, the difference is consistent with the bias introduced by our adoption of the mean halo metallicity for our RRL template and the lower metallicity expected for this UFD.
Further studies of Cetus III and this RRL candidate are needed to confirm their association.

We identify one RRL in Tucana IV with $g=18.78$, consistent with spectroscopically confirmed HB members of this galaxy \citep{Simon20}. The variable, however, is located at 56\arcmin from the center, equivalent to $6\,r_h$. Thus, the physical association is unclear although the {\it Gaia} DR2 proper motion is consistent. Another possibility is that this may be an extra-tidal star, similar to those seen in Tucana III, Eridanus III and Reticulum III \citep{Vivas:2020}.

We also note that we recover known RRL in other UFDs, including star V1 in Tucana II \citep{Vivas:2020}, V1 in Phoenix II, and V2 in Grus I \citep{Martinez-Vazquez:2019}. However, since more comprehensive analyses of these objects exist in the referenced publications, we do not examine them in detail here.

\begin{figure*}[t!]
    \includegraphics[width=\textwidth]{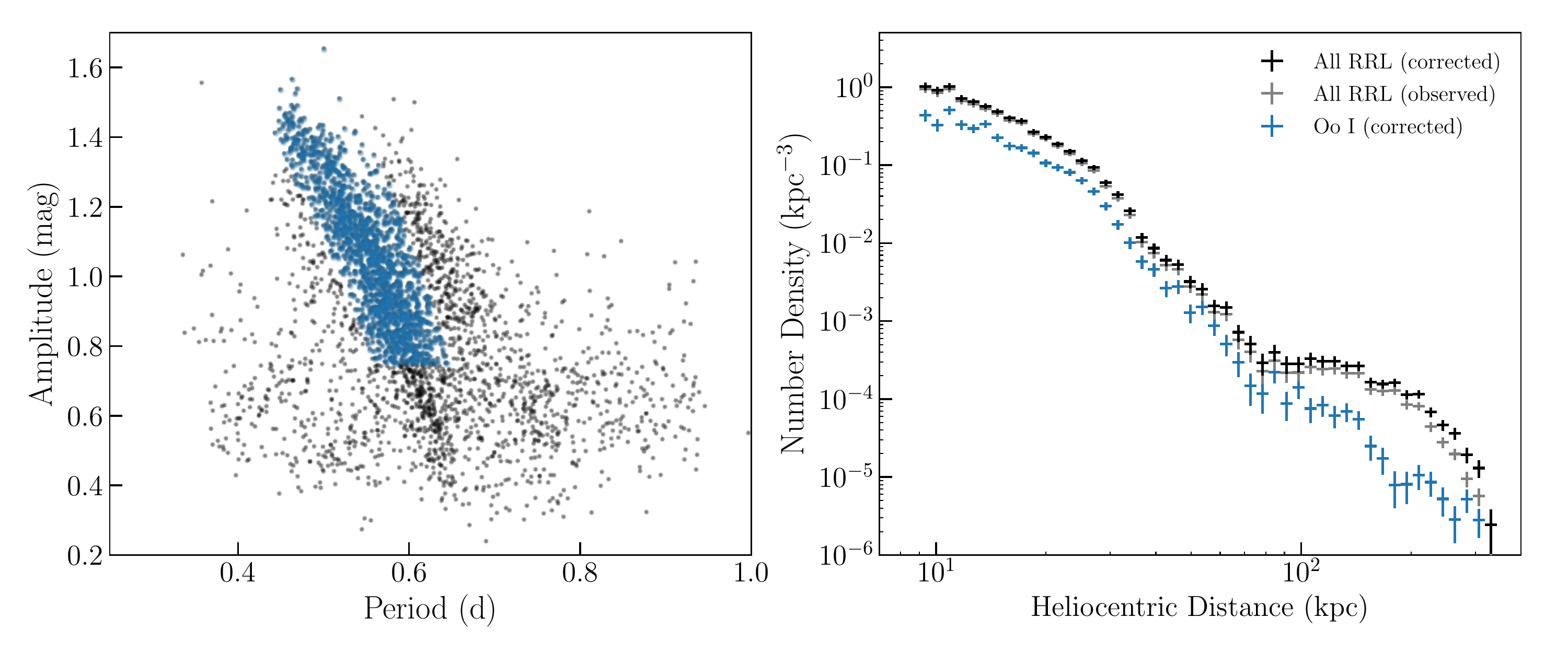}
    \caption{(Left) Period-amplitude diagram of DES Y6 RRab candidates (black) and high-confidence candidates associated with the Oosterhoff I sequence and possessing amplitude variations of $> 0.75\magn$ (blue). (Right) Density of RRab candidates as a function of heliocentric distance. The gray points show the observed distribution of all RRab candidates, while the black curve corrects for the detection efficiency estimated in \secref{sim}. The blue curve shows the density of high-confidence RRab associated with the Oo-I sequence in the left panel, corrected by detection efficiency (which is $>95\%$ out to 150 kpc).}
    \label{fig:field_hel}
\end{figure*}

Finally, there are several globular clusters within the DES footprint. RRL in Milky Way globular clusters closer than 10\kpc often saturate the DES images. Nonetheless, we recover star V21 in M2 \citep[NGC 7089,][2017 July version]{Clement01}, which is one of the 23 RRab known in that cluster. On the other hand, the LMC globular cluster Reticulum is a better target for the range of magnitudes of our catalog. We recover 14 out of its 22 known RRab \citep{Kuehn13}, a 63\% recovery rate. We find two additional RRab candidates spatially coincident with the cluster that have the appropriate magnitude and proper motions (from \Gaia DR2) to be members. The mean distance modulus of these 16 variables is $\mu=18.36\pm0.04$~mag, in agreement with the value obtained by \citet{Kuehn13} of $18.40 \pm 0.20$ mag. The good agreement in these estimates is due to the similar metallicity of the RRL in the cluster ($\feh\sim-1.6$) and the mean metallicity of the Milky Way halo adopted for our templates. Additionally, there is a third new possible member that has consistent magnitude and proper motion but it is located farther away, at $14\arcmin$ from the center of the cluster.

\subsection{Halo Density Profile} 
\label{sec:field}

The structure of the Milky Way stellar halo encapsulates information about the formation and evolution of our Galaxy \citep[e.g.,][]{Johnston:2002}. Several lines of observational evidence suggest that the halo density profile exhibits a break at Galactocentric distances of 20--35 \kpc, with star counts falling off more rapidly beyond this radius \citep[e.g.,][]{Watkins:2009,Sesar:2010,Deason:2011,Sesar:2011,Zinn:2014, Pila-Diez:2015,Xue:2015,Pieres:2020}. Such a broken-power-law profile may be produced through the accretion of a massive satellite galaxy \citep{Bullock:2005,Deason:2013,Deason:2018}, which corroborates recent claims of the \Gaia-Enceladus merger \citep{Belokurov:2018,Helmi:2018}. While quantitative estimates vary by analysis, studies across a wide range of stellar tracers find that shallower power-law slopes ($n_1 \sim -2$ to $-3.5$) are preferred in the inner region of the halo, while steeper values ($n_2 \sim -3.8$ to $-5.8$) are preferred at larger distances \citep[see][for a recent compilation]{Pila-Diez:2015}. RRL have provided an important probe of the stellar density profile, with evidence for the broken-power-law model first claimed by \citet{Saha:1985} and more recently by \citet{Zinn:2014} and \citet{Medina:2018}. The deep, wide-area catalog of DES Y6 RRab candidates offers an excellent opportunity to measure the density profile of the Milky Way stellar halo over a wide region of the southern sky.

We estimate the halo density profile by first removing RRab candidates around the LMC, SMC, Fornax, and Sculptor (see \appref{likelihood}), which yields a sample of $3{,}603$ candidate RRab. We group these stars into 51 bins of heliocentric distance from $9 < D < 400 \kpc$, and we calculate the number density by correcting for the detection efficiency of our catalog (\figref{field_hel}). We find that the density profile of RRab exhibits a break at a heliocentric distance of $D \sim 25 \kpc$, with an excess relative to a simple broken-power-law model at heliocentric distances of $80\kpc < D < 300 \kpc$. The excess of candidates at large distances is not explained by a change in detection efficiency, which is found to be decreasing monotonically with distance and is a fractionally smaller effect than the observed excess (\secref{sim}). To investigate this excess in more detail, we select a sub-population of high-confidence RRab candidates close to the Oosterhoff I sequence with a measured amplitude change of $> 0.75 \magn$. We find that the excess is greatly reduced in the high-confidence subpopulation, suggesting that contamination from faint sources may be responsible for the excess. To further investigate possible contamination, we cross-match RRab candidates in the S82 region with spectroscopically confirmed QSOs from SDSS DR7 \citep{Schneider:2010} and SDSS DR16 \citep{Lyke:2020}. We find that the QSO contamination of our catalog in the S82 region is $\roughly 2\%$; however, we find that all contaminating QSOs are fit with distance moduli $20 < \mu < 22$ ($100 \kpc < D < 250 \kpc$), leading to a contamination rate of $\roughly 16\%$ in the same distance range as the observed excess. 

While suggestive, this contamination is still far less than would be necessary to account for the observed excess. Furthermore, we note that the RRab candidates in this distance range are not uniformly distributed over the footprint, as would be expected from extragalactic contamination and the uniform recovery efficiency estimated from our RRL simulations. Rather, RRab candidates in this distance range are preferentially distributed at $\alpha \sim 0$, coincidentally close to where the Magellanic Stream crosses the DES footprint. 

Large-scale anisotropies in the halo RRL distribution have been claimed by \citet{Iorio:2018} using a combined catalog of \Gaia+2MASS RRL. \citet{Boubert:2019} provided supporting evidence for this structure using observations from the Catalina Surveys \citep[e.g.,][]{Torrealba:2015,Drake:2017} and the sample of RR Lyrae variables identified in PS1 \citep{Hernitschek:2016}. While this structure is significantly closer than the excess observed here (Galactocentric distance of $\roughly 20 \kpc$), \citet{Boubert:2019} claim a Magellanic origin for this overdensity, which could extend to greater distances where the infall track of the Magellanic Clouds crosses the DES footprint. 
Further investigation of these distant candidates is necessary to better understand possible anisotropies in the RRL distribution at distances $\gtrsim 80\kpc$.

Given the significant uncertainties in the contamination rate and possible anisotropies at heliocentric distances $\gtrsim 80\kpc$, we constrain our study of the Milky Way halo to smaller distances, where we estimate that our completeness is $\gtrsim 75\%$. To measure the Milky Way stellar halo density, we transform each of our RRab candidates into Galactocentric coordinates, $(x,y,z)$, assuming that the solar Galactic center distance is 8.178\kpc \citep{Abuter:2019}. We further calculate the elliptical Galactocentric radius, $r_e = \sqrt{x^2 + y^2 + (z/q)^2}$. Following \citet{Faccioli:2014}, we perform our fit assuming a fixed halo flattening of $q = 0.7$ \citep[e.g.,][]{Sesar:2011}. We group candidate RRab into 41 bins in elliptical Galactocentric radius from $9 \kpc < r_e < 100 \kpc$, and we fit the halo profile with an elliptical broken power law following the description of \citet{Pila-Diez:2015},
\begin{align}
    \rho(x,y,z) &= \rho_0
    \begin{cases}
      r_e^{n_1}, & r_e < R_0 \\ 
      r_e^{n_2} \cdot R_0^{n_1 - n_2}, & r_e \geq R_0, \\
    \end{cases}
    \label{eqn:pwl}
\end{align}
where $\rho_0$ is the density normalization, $n_1$ is the inner power-law index, $n_2$ is the outer power-law index, and $R_0$ is the break radius (in elliptical Galactocentric coordinates). We perform a binned Poisson maximum-likelihood fit for the observed counts in each bin, $k$, as a function of the model parameters, $\theta = \{\rho_0, R_0, n_1, n_2\}$. Our likelihood analysis is described in more detail in \appref{likelihood}. Within each bin, we correct the observed number of RRL for the detection completeness of our catalog estimated from simulated RRab described in \secref{sim}. We fit the model parameters using a Markov Chain Monte Carlo ensemble sampler \citep[\code{emcee};][] {Foreman-Mackey:2013}, and we report the median, 16th, and 84th percentiles of the marginalized posterior distributions of each parameter in \tabref{profile}. We perform this analysis for both the full DES Y6 RRab candidate sample and the high-confidence RRab associated with the Oosterhoff I sequence.

\begin{figure}[t!]
    \includegraphics[width=\columnwidth, trim=1cm 0 0.5cm 0, clip]{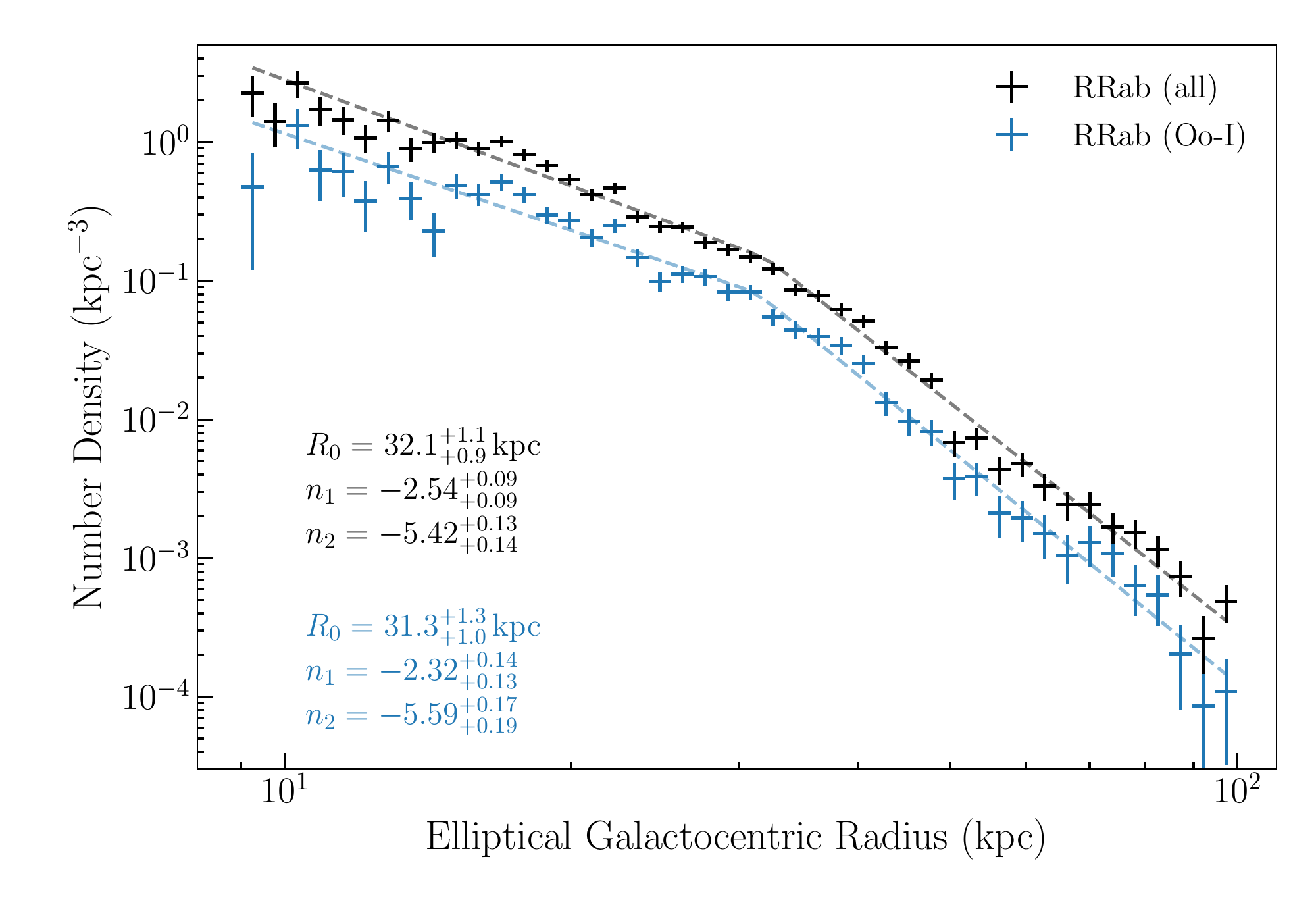}
    \caption{Density of RRab candidates binned in elliptical Galactocentric radius ($q=0.7$). The black markers show the full RRab candidate sample, while blue markers show high-confidence RRab associated with the Oosterhoff I sequence and having measured amplitude of variations of $> 0.75 \magn$. Errors represent the square root of the number of stars in each bin and the bin width. }
    \label{fig:field_gal}
\end{figure}

\begin{figure}[t!]
    \includegraphics[width=\columnwidth]{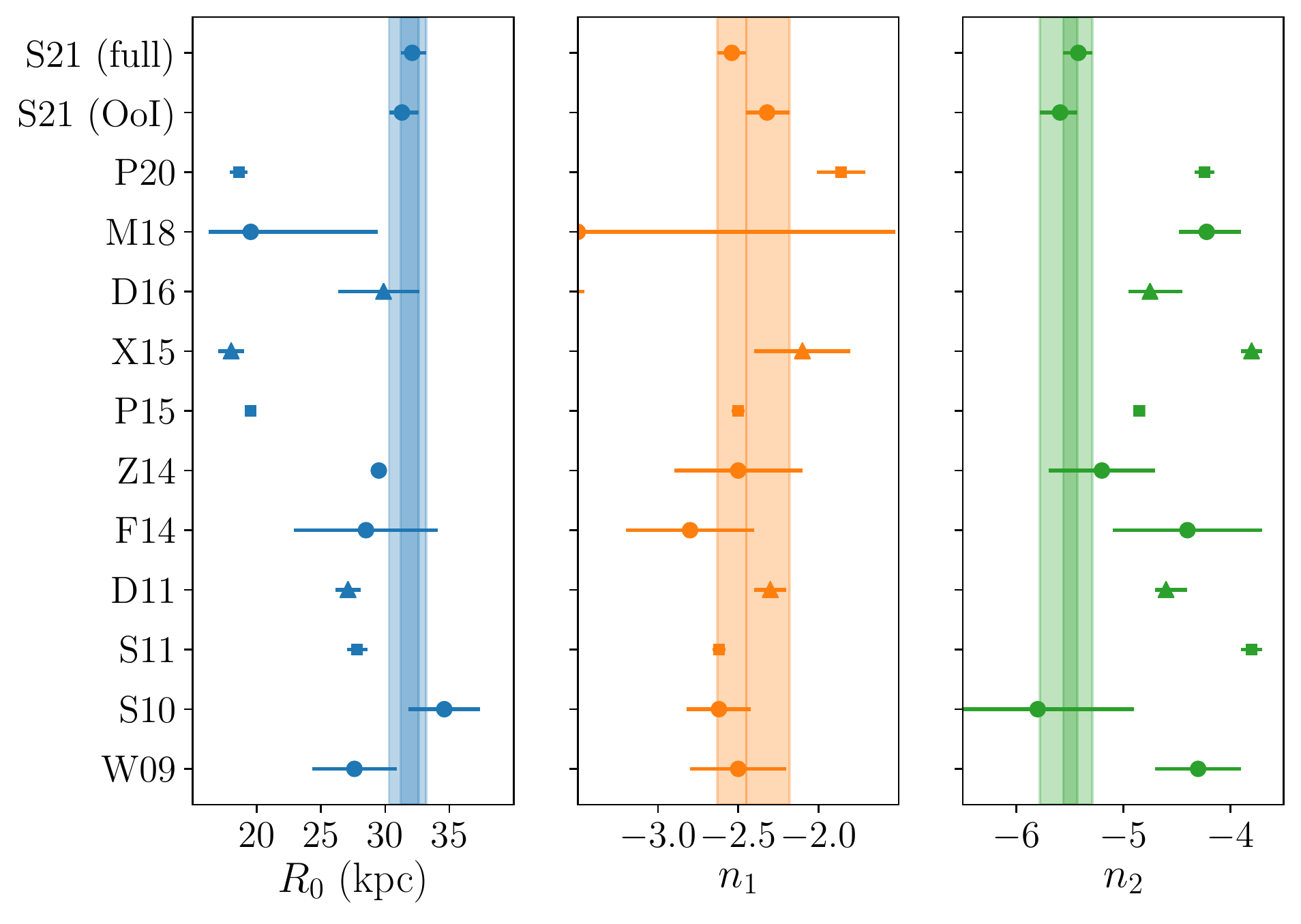}
    \caption{Comparison of best-fit broken-power-law halo parameters from this work (S21) and those collected from the literature. Circle markers represent measurements from RRL, squares represent measurements from main sequence turn-off stars, while triangles represent measurements from giant branch stars. Parameters are the break radius (left), the inner power-law index (center), and the outer power-law index (right). References are W09 \citep{Watkins:2009}, S10 \citep{Sesar:2010}, S11 \citep{Sesar:2011}, D11 \citep{Deason:2011}, F14 \citep{Faccioli:2014}, Z14 \citep{Zinn:2014}, P15 \citep{Pila-Diez:2015}, X15 \citep{Xue:2015}, D16 \citep{Das:2016}, M18 \citep{Medina:2018}, and P20 \citep{Pieres:2020}.}
    \label{fig:field_compare}
\end{figure}

In \figref{field_compare} we compare the best-fit parameters from our two RRab candidate samples to broken-power-law fits to the halo in the literature. We generally find that our best-fit break radius of $\roughly 32\kpc$ is slightly larger then many other analyses; however, this could be brought into better agreement through a smaller halo flattening. Our measured inner slope values are consistent within $2\sigma$ of each other and are broadly consistent with other values in the literature. Due to the saturation threshold of the DES images, fits for $n_1$ are largely driven by RRL with $r_e \gtrsim 14 \kpc$ and are highly correlated with the overall normalization parameter, $\rho_0$. Due to the large area and sensitivity of DES, we have several thousand RRab candidates in the range $30\kpc < r_e < 100\kpc$. This allows the DES data to provide tight constraints on the outer power-law slope, $n_2$. 
The value of $n_2 = \Nouter$ measured using DES RRL candidates is steeper than that measured by many other tracers and surveys \citep[e.g.,][]{Sesar:2011,Xue:2015,Pieres:2020}, but it is consistent with RRL measurements by \citet{Sesar:2010} and \citet{Zinn:2014}.

\begin{deluxetable}{lcccc}[t!]
  \tabletypesize{\footnotesize}
  \tablecaption{Broken-power-law halo density parameters. \label{tab:profile}}
  \tablehead{\colhead{Sample} & \colhead{$\rho_0$} & \colhead{$R_0$} & \colhead{$n_1$} & \colhead{$n_2$} \\[-1em]
    \colhead{} & \colhead{(kpc$^{-3}$)} & \colhead{(kpc)} & & }
  \decimals
  \startdata \\[-0.5em]
  Full  & \Norm   & \Rbreak	& \Ninner   & \Nouter \\
  Oo-I  & \NormOo & \RbreakOo & \NinnerOo & \NouterOo \\[+0.5em]
  \enddata
  \tablecomments{The halo density profile fit assumes a fixed halo flattening of $q = 0.7$.}
\end{deluxetable}


\begin{figure*}[t!]
  \center
  \includegraphics[width=0.9\textwidth]{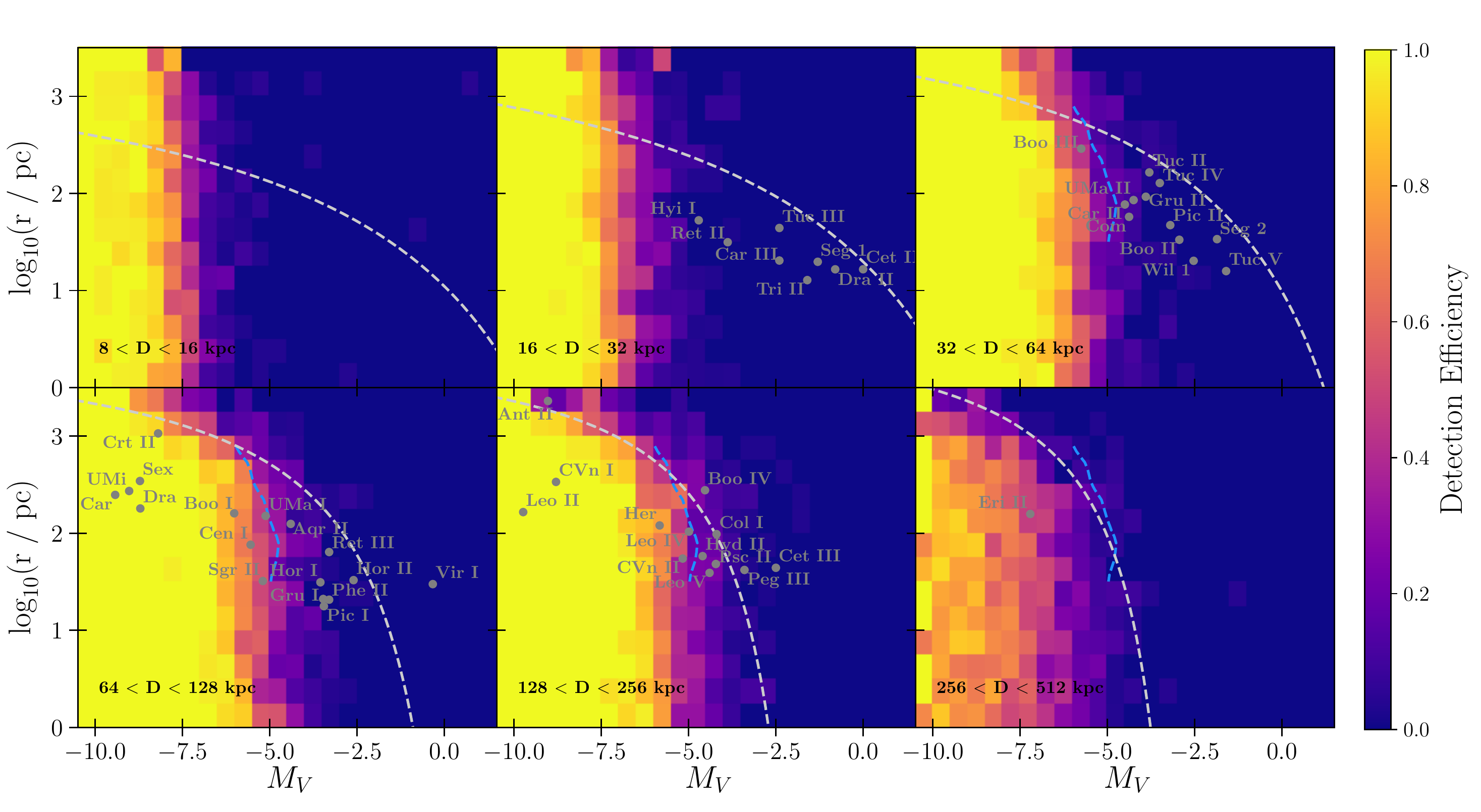}
  \caption{Detection efficiency of our search for Milky Way satellites using DES Y6 RRab candidates. Detection efficiency ranges from 0\% (blue) to 100\% (yellow) and is shown as a function of azimuthally averaged projected physical half-light radius and absolute $V$-band magnitude in different bins of heliocentric distance (logarithmically spaced from $8 \kpc$ to $512 \kpc$). The physical parameters of known satellites galaxies are indicated in gray. The light blue dashed line shows the 50\% detectability contour for RRL searches at distances $>50\kpc$ predicted by \citep{Baker:2015}, while the gray dashed line shows an analytic approximation to the 50\% detectability contour from isochrone matched filter searches \citep{Drlica-Wagner:2020}.}
    \label{fig:sensitivity}
\end{figure*}

\ \par

\subsection{Search for New Satellite Galaxies} \label{sec:satsearch}

Clusters of RRL can be used to identify faint satellite galaxies that may have evaded detection by other methods. In particular, \citet{Baker:2015} argued that the combination of three-dimensional information and the sparsity of halo RRL at distances $> 50 \kpc$ make RRL particularly useful for identifying Milky Way satellites with half-light radii $\rhalf > 500 \pc$ residing in the outer halo. 
The RRL catalogs from \Gaia DR2 have demonstrated the viability of this search technique through the discovery of the ultra-diffuse satellite Antlia II \citep{Torrealba:2019} and the detection of several candidate stellar streams \citep[e.g.,][]{Mateu:2018}.

We search for satellite galaxies coincident with each of the DES Y6 RRab candidates beyond the masked regions around the LMC, SMC, Fornax, and Sculptor. Our algorithm is based on a simple binned Poisson likelihood, which is qualitatively similar to searches for resolved stellar populations \citep{Bechtol:2015,Drlica-Wagner:2015}, but optimized to the detection of satellites with half-light radii $\rhalf \gtrsim 500\pc$.  At the location and distance of each RRab in our masked catalog, we define a search cylinder with a fixed radius of $2 \rhalf = 1 \kpc$. The depth of our search cylinder is calculated from the quadrature sum of $2 \rhalf$ and the systematic uncertainty on the measured distance to the RRL, $2 \sigma(\mu) = 0.2 \magn$. Each cylinder is expected to contain \CHECK{\roughly 90\%} of the RRL population of a satellite with $\rhalf = 500\kpc$ centered at the search location.

We determine the local expected density of field RRL, $\rho_f$, from a cylindrical annulus centered on our search location with inner and outer radii of $4 \rhalf$ and $8 \rhalf$, respectively. In cases where no RRL are contained within our background annulus, we assume the global background rate as estimated from the field density at that distance (\figref{field_hel}). We multiply the background density by the volume of our search cylinder to predict the expected number of field RRL within our search region, $\lambda$.

We calculate the significance of a putative satellite at each search location (i.e., at the location of each RRab candidate in our catalog) as the Poisson probability of detecting $k$ or more RRL given an expectation of $\lambda$,
\begin{align}
P(k|\lambda) = \frac{\lambda^k e^{-\lambda}}{k!}.
\label{eqn:poisson}
\end{align}
To select candidate satellites, we apply a significance threshold of \CHECK{$p < 3 \times 10^{-7}$}, which corresponds to a one-sided Gaussian significance of $5\sigma$. We also require that satellite candidates consist of at least three RRab candidates.

We quantify the sensitivity of our search using a suite of $10^5$ simulated satellite galaxies generated by \citet{Drlica-Wagner:2020}. These satellites span a range of stellar mass, heliocentric distance, size, ellipticity, and position angle \citep[see Table~1 of ][]{Drlica-Wagner:2020}. The simulated satellites are distributed uniformly over the DES footprint and uniformly in distance modulus. For each simulated satellite, we predict the expected number of RRL as a function of $M_V$ using a fit to the RRL population of Milky Way satellites provided in Eq.~4 of \citet{Martinez-Vazquez:2019},
\begin{align}
\log(N_{\rm RRL}) = -0.29 \times M_V - 0.80. 
\end{align}
To predict the number of RRab, we multiply the number of RRL by the fraction of RRab, $N_{\rm RRab} = f_{\rm ab} N_{\rm RRL}$, where $f_{\rm ab} = 0.71$ for Milky Way satellites \citep[i.e., Table 6 of ][]{2017ApJ...850..137M}. The expected number of RRab observed by DES is corrected for the detection efficiency of our search and fitting procedure, which depends on the distance of the simulated satellite. The spatial distribution of simulated RRab was drawn from an elliptical Plummer profile  \citep{Plummer:1911}. Distances were assigned from a Gaussian distribution centered on the distance of the simulated satellite matched to the 3D half-light radius of the satellite. An additional systematic Gaussian scatter in distance modulus, $\Delta(\mu) = 0.1 \magn$, was applied to each simulated RRab. We inject each simulated satellite into the DES Y6 RRab candidate catalog individually and attempt to recover it with our search algorithm. The resulting detection efficiency as a function of $M_V$ and physical half-light radius $\rhalf$ is shown in \figref{sensitivity}.

As expected, our search is more sensitive than isochrone-matched filter searches \citep[][dashed line]{Drlica-Wagner:2020} for satellites with large sizes. Our RRL satellite search is significantly less sensitive for compact satellites, due to the large assumed kernel ($2 \rhalf = 1 \kpc$). At small heliocentric distances, this large search kernel leads to an expected background contribution from halo RRL that is comparable with the RRL signal from a satellite with $M_V \sim -6$. At larger distances, the density of halo RRL decreases, and the sensitivity of our search increases until RRL detection efficiency starts to dominate. As a test, we re-run these simulations using a kernel matched to the true size of each simulated satellite, and we find a significant improvement in the sensitivity for small satellites. However, the isochrone-matched filter searches remain more sensitive for satellites with $M_V > -5$, due to the small number of RRab expected from these satellites. 

With the sensitivity of our search characterized on simulations, we apply our search to the DES Y6 candidate RRab catalog. We find three satellite candidates that pass our detection criteria of significance $> 5\sigma$ and $N_{\rm RRab} > 3$. The characteristics of each RRab candidate member associated with these satellite candidates are listed in \tabref{satellites}.

One of these candidates (SubId=1) is associated with the known satellite Eridanus II, located at a distance of 360 \kpc (\secref{ultra}). Based on the simulations described above (\figref{sensitivity}), we expect our search to be $\roughly 60\%$ efficient for satellites with the size, luminosity, and distance of Eridanus II. This could suggest that Eridanus II may have a larger-than-expected number of RRL.

\begin{deluxetable}{llcccc}[!t]
	\tabletypesize{\footnotesize}
	\tablecaption{Halo substructure candidates. \label{tab:satellites}}
	\tablehead{\colhead{SubID} & \colhead{Star ID} & \colhead{$\alpha$} & \colhead{$\delta$} & \colhead{$\mu$} & \colhead{$D$} \\[-0.8em]
	\colhead{} & \colhead{} & \multicolumn{2}{c}{deg, J2000} & \colhead{mag} & \colhead{kpc}} 
	\decimals
	\startdata
1 & 1410940364 & 56.10118 & -43.50492 & 22.5 & 316.2 \\
1 & 1410941177 & 56.07512 & -43.51257 & 22.4 & 301.8 \\
1 & 1410942171 & 56.09949 & -43.52116 & 22.4 & 305.1 \\
1 & 1410943700 & 56.07939 & -43.53486 & 22.5 & 314.1 \\
1 & 1410944738 & 56.01504 & -43.54419 & 22.4 & 307.6 \\
\hline
2 & 994558814 & 348.37430 & -55.77342 & 18.7 & 55.0 \\
2 & 1002871838 & 349.83519 & -55.33808 & 18.9 & 60.5 \\
2 & 1007027294 & 351.32392 & -55.76393 & 19.0 & 63.0 \\
\hline
3 & 1533696855 & 77.66305 & -38.90769 & 17.0 & 24.6 \\
3 & 1539669959 & 77.95832 & -38.64577 & 17.1 & 25.8 \\
3 & 1548634844 & 80.17935 & -39.58687 & 16.8 & 23.2 \\
3 & 1616940370 & 77.32652 & -37.11542 & 16.7 & 21.9 \\
\hline
	\enddata
\end{deluxetable}

\begin{figure}[!t]
  \center
  \includegraphics[width=0.48\textwidth]{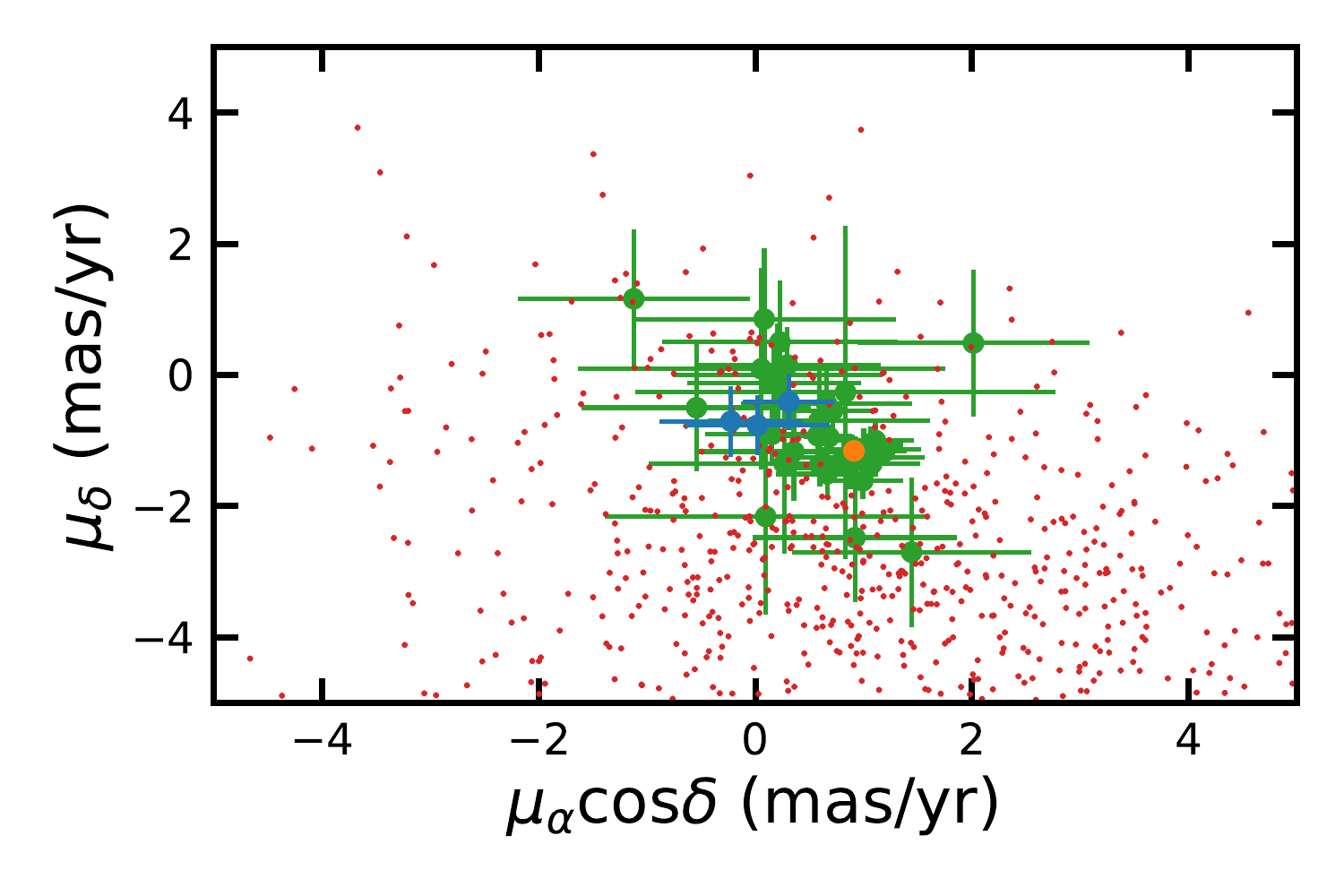}
  \caption{Proper motions of RRab candidate stars associated with satellite candidate 2 (\tabref{satellites}). 
  Stars associated with this satellite candidate are shown in blue, spectroscopically confirmed member stars and proper motion candidate members from the Tucana II satellite are shown in green, the systemic motion of Tucana II is shown in orange, and foreground stars are shown in red \citep{Pace:2019}.
  }
  \label{fig:pm}
\end{figure}

A second overdensity (SubId = 2) consists of three RRab candidates located at $(\alpha,\delta) \sim 349.8, -55.6$ at a distance of $59.5\kpc$. 
This overdensity is in the same region of the sky and at roughly the same distance as the Tucana II dwarf galaxy, $D = 57.5\kpc$; however, it is separated from Tucana II by $\roughly 5 \deg$ on the sky. 
Interestingly, the \Gaia DR2 proper motions of the RRab candidates in this overdensity are similar to the proper motions of confirmed members of Tucana II, albeit with large uncertainties (\figref{pm}). 
It has been suggested that the diffuse structure of Tucana II could be an indication of tidal disruption \citep[e.g.,][]{Bechtol:2015}. 
\citet{Belokurov:2016} showed evidence for extended stellar structure around Tucana II using BHB stars (i.e., their ``S2a'' cloud at $\mu \sim 18.8$), and \citet{Chiti:2020} used \Gaia proper motions combined with photometric metallicities to identify probable members $>1 \deg$ from Tucana II.
Further observations are required to confirm an association between this overdensity of RRab candidates and the Tucana II dwarf galaxy.

The third candidate (SubId =3) consists of four RRab candidates located at $(\alpha,\delta) \sim 78.3, -38.6$ at a distance of $23.8 \kpc$.  This is close in projection to the globular cluster NGC$\,$1851, but at a larger distance. \citet{Shipp:2018} found evidence for extended tidal features surrounding this cluster, but the proper motions of the stars associated with this candidate structure are not aligned with the motion of NGC 1851. We followed the procedure of \citet{Shipp:2018} to use the DES Y6 coadd object catalog to search for correlated overdensities of main-sequence turnoff stars associated with our candidate substructures using isochrone-filtered stellar density maps. However, we do not detect any previously unknown overdensities in turnoff stars, suggesting that radial velocities will be critical for confirming the second two candidate structures.

\section{Conclusion} \label{sec:conclusion}

We have used various statistical techniques to derive a catalog of candidate RRab detected in six years of deep, wide-area imaging from DES.  The DES Y6 data offers significant improvements on prior results using only three years of data \citep{paper1}, resulting in a catalog of \Nrrab RRab candidates. 
At the bright end, our catalog has significant overlap with surveys such as {\it Gaia} DR2, Pan-STARRS, and the Catalina surveys, providing strong evidence of the effectiveness of this method.  We publicly release the best-fit parameters and light curves for our RRL candidates. 

We find good agreement in the measured properties of our RRab candidates when matched against external catalogs from the MCs, Fornax, and Sculptor. In addition, we recover RRL detected in the ultra-faint dwarfs Tucana II, Phoenix II, and Grus I \citep{Martinez-Vazquez:2019}. DES extends significantly deeper than any of these surveys, allowing us to detect RRab candidates out to a distance of $\roughly 335\kpc$, about 1 mag deeper than \citet{paper1}. Indeed, we discover a group of five RRab candidates associated with the distant ultra-faint dwarf galaxy Eridanus II.  We also report tentative RRL members of the ultra-faint systems Cetus III \citep{Homma:2018} and Tucana IV \citep{Drlica-Wagner:2015}. We fit the stellar density profile of the Milky Way halo in the range of elliptical Galactocentric distances from $9 < r_e < 100\kpc$.  Assuming a halo flattening of $q = 0.7$, we find that the halo is well fit by a broken-power-law model with a break radius of $R_0 = \Rbreak$, an inner slope of $n_1 = \Ninner$, and an outer slope of $n_2 = \Nouter$. These values agree between analyses of full RRab candidate sample and a high-confidence sample of RRab candidates with large measured amplitude variations associated with the Oosterhoff I sequence. We further use our catalog of DES Y6 RRab candidates to search for halo substructures, with characteristic sizes of $\rhalf \sim 500\pc$. This search confidently detects the Eridanus II dwarf galaxy and two other candidate overdensities that are not confidently associated with known halo substructures.

RRL have long been recognized as a powerful probe of the Milky Way's outer stellar halo. However, it is only recently that surveys have been able to combine wide area coverage, deep imaging, and sufficient cadence to confidently identify RRL at distances $>100\kpc$. While DES was not optimized to search for RRL, it nonetheless provides an exceptional catalog of RRab candidates at distances beyond what is achievable by surveys on smaller telescopes (i.e., \Gaia, Catalina, and PS1). In the near future, the Vera C.\ Rubin Observatory Legacy Survey of Space and Time will provide hundreds of observations over the entire southern sky, promising to provide a high-quality catalog of RRL extending to the edge of the Milky Way halo.

\acknowledgments

K.M.S., L.M.M., and P.S.F. acknowledge support from the Mitchell Institute for Fundamental Physics and Astronomy at Texas A\&M University. This material is based upon work that was supported by the Fermilab Visiting Scholars Award Program of the Universities Research Association. This research has made use of the SIMBAD database, operated at CDS, Strasbourg, France.  A.B.P. acknowledges support from NSF grant AST-1813881. 

Funding for the DES Projects has been provided by the U.S. Department of Energy, the U.S. National Science Foundation, the Ministry of Science and Education of Spain, the Science and Technology Facilities Council of the United Kingdom, the Higher Education Funding Council for England, the National Center for Supercomputing Applications at the University of Illinois at Urbana-Champaign, the Kavli Institute of Cosmological Physics at the University of Chicago, the Center for Cosmology and Astro-Particle Physics at the Ohio State University, the Mitchell Institute for Fundamental Physics and Astronomy at Texas A\&M University, Financiadora de Estudos e Projetos, Funda{\c c}{\~a}o Carlos Chagas Filho de Amparo {\`a} Pesquisa do Estado do Rio de Janeiro, Conselho Nacional de Desenvolvimento Cient{\'i}fico e Tecnol{\'o}gico and the Minist{\'e}rio da Ci{\^e}ncia, Tecnologia e Inova{\c c}{\~a}o, the Deutsche Forschungsgemeinschaft and the Collaborating Institutions in the Dark Energy Survey. 

The Collaborating Institutions are Argonne National Laboratory, the University of California at Santa Cruz, the University of Cambridge, Centro de Investigaciones Energ{\'e}ticas, Medioambientales y Tecnol{\'o}gicas-Madrid, the University of Chicago, University College London, the DES-Brazil Consortium, the University of Edinburgh, the Eidgen{\"o}ssische Technische Hochschule (ETH) Z{\"u}rich, Fermi National Accelerator Laboratory, the University of Illinois at Urbana-Champaign, the Institut de Ci{\`e}ncies de l'Espai (IEEC/CSIC), the Institut de F{\'i}sica d'Altes Energies, Lawrence Berkeley National Laboratory, the Ludwig-Maximilians Universit{\"a}t M{\"u}nchen and the associated Excellence Cluster Universe, the University of Michigan, the NSF's NOIRLab, the University of Nottingham, The Ohio State University, the University of Pennsylvania, the University of Portsmouth, SLAC National Accelerator Laboratory, Stanford University, the University of Sussex, Texas A\&M University, and the OzDES Membership Consortium.

Based in part on observations at Cerro Tololo Inter-American Observatory at NSF's NOIRLab, which is operated by the Association of Universities for Research in Astronomy (AURA) under a cooperative agreement with the National Science Foundation.

The DES data management system is supported by the National Science Foundation under Grant Numbers AST-1138766 and AST-1536171. The DES participants from Spanish institutions are partially supported by MINECO under grants AYA2015-71825, ESP2015-66861, FPA2015-68048, SEV-2016-0588, SEV-2016-0597, and MDM-2015-0509, some of which include ERDF funds from the European Union. IFAE is partially funded by the CERCA program of the Generalitat de Catalunya. Research leading to these results has received funding from the European Research Council under the European Union's Seventh Framework Program (FP7/2007-2013) including ERC grant agreements 240672, 291329, and 306478. We  acknowledge support from the Australian Research Council Centre of Excellence for All-sky Astrophysics (CAASTRO), through project number CE110001020, and the Brazilian Instituto Nacional de Ci\^encia e Tecnologia (INCT) e-Universe (CNPq grant 465376/2014-2).

This manuscript has been authored by Fermi Research Alliance, LLC under Contract No. DE-AC02-07CH11359 with the U.S. Department of Energy, Office of Science, Office of High Energy Physics. The United States Government retains and the publisher, by accepting the article for publication, acknowledges that the United States Government retains a non-exclusive, paid-up, irrevocable, world-wide license to publish or reproduce the published form of this manuscript, or allow others to do so, for United States Government purposes.

Funding for the SDSS and SDSS-II has been provided by the Alfred P. Sloan Foundation, the Participating Institutions, the National Science Foundation, the U.S. Department of Energy, the National Aeronautics and Space Administration, the Japanese Monbukagakusho, the Max Planck Society, and the Higher Education Funding Council for England. The SDSS Web Site is \url{http://www.sdss.org/}.

The SDSS is managed by the Astrophysical Research Consortium for the Participating Institutions. The Participating Institutions are the American Museum of Natural History, Astrophysical Institute Potsdam, University of Basel, University of Cambridge, Case Western Reserve University, University of Chicago, Drexel University, Fermilab, the Institute for Advanced Study, the Japan Participation Group, Johns Hopkins University, the Joint Institute for Nuclear Astrophysics, the Kavli Institute for Particle Astrophysics and Cosmology, the Korean Scientist Group, the Chinese Academy of Sciences (LAMOST), Los Alamos National Laboratory, the Max-Planck-Institute for Astronomy (MPIA), the Max-Planck-Institute for Astrophysics (MPA), New Mexico State University, Ohio State University, University of Pittsburgh, University of Portsmouth, Princeton University, the United States Naval Observatory, and the University of Washington.

Based on data products from observations made with ESO Telescopes at the La Silla Paranal Observatory under programme IDs 177.A-3016, 177.A-3017 and 177.A-3018, and on data products produced by Target/OmegaCEN, INAF-OACN, INAF-OAPD and the KiDS production team, on behalf of the KiDS consortium. OmegaCEN and the KiDS production team acknowledge support by NOVA and NWO-M grants. Members of INAF-OAPD and INAF-OACN also acknowledge the support from the Department of Physics \& Astronomy of the University of Padova, and of the Department of Physics of Univ. Federico II (Naples).

%

\vspace{5mm}
\facilities{Blanco (DECam), Gaia}


\software{\code{astropy} \citep{Astropy:2018},  
  \code{healpix} \citep{Gorski:2005},
  \code{matplotlib} \citep{Hunter:2007},
  \code{numpy} \citep{numpy:2011},
  \code{SourceExtractor} \citep{Bertin:1996}
  \code{scikit-learn} \citep{Pedegrosa:2012},
  \code{scipy} \citep{scipy:2001}
}




\appendix
\section{Variability Statistics}\label{app:varmetrics}

Here, we describe the variability statistics used to select objects for template fitting. A glossary of terms is provided in \tabref{stats}. For each object, we multiplied the reported photometric uncertainties by scaling factors based on the best-fit values from Equation \ref{eq:quadfit}. We use these rescaled errors in the calculation of all variability statistics. To avoid confusion, we will refer to the error-weighted mean as $\overline{m}$ and the median as med($m$). 

In addition to the reduced median chi-squared, \medchisq (Equation \ref{eqn:medchisq}), we calculate the reduced chi-squared from the mean magnitude in each band, $\chi^2_{\nu,b}$, and its common logarithm 
\begin{equation}
   \mathrm{log}_{10}(\chi^{2}_{\nu,b})= \mathrm{log}_{10} \left(\frac{1}{N_{b}-1}\sum_{1}^{N_{b}}\frac{(m_{i,b}-\overline{m_{b}})^{2}}{\sigma_{i,b}^{2}}\right).
\end{equation}
\noindent As these quantities use the rescaled errors determined in \secref{error_rescale}, the non-varying sources have a distribution centered around zero, with positive outliers denoting true variable objects.

We also measure the range in magnitude, which we call $\Delta(\mathrm{mag})_b$, in each single-band light curve to relay information about the amplitude of an object's variation. The uncertainties on the maximum and minimum magnitudes used to calculate this quantity are recorded as well.

\begin{deluxetable*}{lllr}
	\tabletypesize{\footnotesize}
	\tablecaption{Glossary of Terms}
	\tablehead{\colhead{Abbreviation} & \colhead{Full Name} & \colhead{Bands Used} & \colhead{Ref.}}
	\decimals
	\startdata
        IQR$_b$                & Interquartile range for band $b$    & $g,r,i,z$          & 1     \\
        $J_b$                & Stetson's J statistic for band $b$    & $g,r,i,z,(grizY)$  & 2    \\
        MAD$_b$                & Median absolute deviation for band $b$    & $g,r,i,z,(grizY)$  & 1     \\
        NAPD$_b$               & N absolute pairwise distances for band $b$ & $g,r,i,z,(grizY)$  & 1     \\
        $\sigma^{2}_{{\rm NXS},b}$ & Normalized excess variance for band $b$ & $g,r,i,z,(grizY)$  & 3,4 \\
        $Q_{n,b}$            & $k^{\rm th}$ value of absolute pairwise distances for band $b$       & $g,r,i,z$          & 5    \\
        $\Delta({\rm mag})_b$ & Magnitude range for band $b$  & $g,r,i,z$          &      \\
        $\chi^{2}_{\nu,b}$   &  Reduced chi-squared statistic for band $b$   & $g,r,i,z,(grizY)$  &  \\
        $\tilde{\chi}^{2}_{\nu,b}$   &  Reduced median chi-squared statistic for band $b$  & $g,r,i,z,(grizY)$  &  \\
        RoMS$_b$             & Robust median statistic for band $b$  & $g,r,i,z,(grizY)$  & 6    \\
        $S_{n,b}$            & Median value of absolute pairwise distances for band $b$                           & $g,r,i,z$          & 5    \\
        W\_Range$_b$         & Weighted magnitude range for band $b$     & $g,r,i,z$          & \ref{app:varmetrics} \\
        log$_{10}(\chi_{\nu,b}^{2})_{\rm th}$ & Common logarithm of the reduced chi-squared above threshold for band $b$ & $g,r,i,z$  & \ref{app:varmetrics}     \\
        rss$_{\nu,j}$        & Residual sum of squares from template curves. $j\in[0,2]$ denotes the rank order minima of this function & $(grizY)$ & S19 \\
        $\mu$                & Distance modulus of the best-fitting template & --- & S19 \\
        $A_{g}$              & $g$-band amplitude for the best-fitting template & --- & S19 \\
        amp$_{j}$            & $g$-band amplitude for the $j^{\rm th}$ best-fitting template, $j \in [0,2]$ & --- & S19 \\
        $P$ & Best-fitting template period in units of days & --- & S19 \\ 
        $\phi$ & Template estimated phase offset in units of days & --- & S19 \\
        D(Oo-I)$_{j}$    & Distance from the Oosterhoff I sequence for $j^{\rm th}$ best-fit period and $g$-band amplitude, $j \in [0,2]$ & --- & S19 \\
        D(Oo-II)$_{j}$   & Distance from the Oosterhoff II sequence for $j^{\rm th}$ best-fit period and $g$-band amplitude, $j \in [0,2]$ & --- & S19 \\
        D(Oo-int)$_{j}$  & Distance from the intermediate Oosterhoff sequence for $j^{\rm th}$ best-fit period and $g$-band amplitude, $j \in [0,2]$ & --- &  S19 \\
        RF$i$\_score  & Output score from the $i^{\rm th}$ random forest classifier, $i\in[1,2]$ &  \\
	\enddata
        \tablerefs{(1) \citealt{Sokolovsky:2017}, (2) \citealt{Stetson:1996}, (3) \citealt{Nandra:1997}, (4) \citealt{Simm:2015}, (5) \citealt{Rousseeuw:1993}, (6) \citealt{Enoch:2003}, (\ref{app:varmetrics}) \appref{varmetrics}, (S19) \citealt{paper1}.}
    \label{tab:stats}
\end{deluxetable*}

Another proxy for variability amplitude is the normalized excess variance ($\sigma_{\rm NXS}^{2}$). This statistic was first defined by \citet{Nandra:1997} as 

\begin{equation}
    \sigma_{\rm NXS}^{2} = \frac{1}{N\overline{m} ^{2}}\smashoperator[r]{\sum_{i=1}^{N}}[(m_{i}-\overline{m})^{2}-\sigma_{i}^{2}].
\end{equation}

\noindent Although this metric is commonly used for X-ray analyses of active galactic nuclei, it has successfully been deployed on the sparsely sampled Pan-STARRS 3$\pi$ optical light curves \citep{Simm:2015}.

We measure the overall scatter of a light curve using the median absolute deviation (MAD). 
\begin{equation}
    \mathrm{MAD} = \mathrm{med}\left(\ |m_{i} - \mathrm{med}(m)|\ \right)
\end{equation}

MAD is slightly more stable in the presence of outliers than the standard deviation as it does not amplify the effects of an outlier by squaring it. This metric should be sensitive to repeated variations; however, a real RRL with observations sampled at close to the same phase value will not appear variable in MAD.

The robust median statistic (RoMS) is a more robust analog of $\chi^2_{red}$, which is less sensitive to bias in the presence of non-Gaussian uncertainties. This metric was first defined in \citet{Enoch:2003} as
\begin{equation}
    \mathrm{RoMS} = \frac{1}{N-1}\smashoperator[r]{\sum_{i=1}^{N}}\frac{|m_{i}-\mathrm{med}(m)|}{\sigma_{i}}
\end{equation}
\noindent In a single band, RoMS tends toward values of one for non-varying sources.

Alternative measures of deviation $S_{n}$ and $Q_{n}$ were first proposed by \citet{Rousseeuw:1993}. These metrics seek to measure the midpoint of a dataset like MAD, but do not rely on a central reference value and are thus better estimators for asymmetric distributions. To account for the scatter caused by the large uncertainties of faint observations, we will apply these statistics to the pairwise distances between all individual observations divided by their combined uncertainties. Here, $S_{n}$ measures the median of the median of these weighted pairwise differences and is defined as
\begin{equation}
    S_{n} = 1.1926\ \mathrm{med}_{i}\left( \mathrm{med}_{j}\left(\ \frac{|m_{i} - m_{j}|}{\sqrt{\sigma_{m_{i}}^{2} + \sigma_{m_{j}}^2}} \ \right)\right)
\end{equation}

\noindent where $m_{i}$ and $m_{j}$ are separate observations. 
A similar metric $Q_{n}$ is defined as
\begin{equation}
    Q_{n} = 2.2219\left(\ \frac{|m_{i}-m_{j}|}{\sqrt{\sigma_{m_{i}}^{2} + \sigma_{m_{j}}^2}};\ i<j\ \right)_{k}.
\end{equation}
Note that $Q_{n}$ records the \textit{k}th value in the weighted absolute pairwise differences between all of the observations, and $k$ is the binomial coefficient $\binom{h}{2}$ with $h=\mathrm{floor}(n/2) +1$ for $n$ observations. Thus, $Q_{n}$ records roughly the midpoint of these values.   
Another way to measure this is to calculate quantiles for the differences in separate observations. We use the 90\% quantile values of these error-weighted $N$ absolute pairwise distances (NAPD) as a proxy for the weighted range that is less sensitive to outliers. 

Stetson's $J$ variability index \citep{Stetson:1996} has been widely used to identify pulsating variables such as Cepheids and RRL. It builds upon the Welch--Stetson variability index \textit{I} \citep{Welch:1993}, which measures the correlation between $n$ subsequent pairs of observations. Stetson's $J$ index measures this correlation using single observations as well as pairs, so we can apply it to both the single-band and multiband light curves:
\begin{equation}
    J = \frac{\smashoperator[r]{\sum_{k=1}^{n}}w_{k}\ \mathrm{sgn}(P_{k})\ \sqrt{|P_{k}|}}{\smashoperator[r]{\sum_{k=1}^{n}}w_{k}}
\end{equation}

\noindent where sgn is the sign function and
\begin{equation}
    P_{k} = 
    \begin{cases}
        \mathrm{pair} & \left(\sqrt{\frac{n_{v}}{n_{v}-1}}\frac{v_{i}-\overline{v}}{\sigma_{v_{i}}}\right)\left(\sqrt{\frac{n_{b}}{n_{b}-1}}\frac{b_{i}-\overline{b}}{\sigma_{b_{i}}}\right) \\[10pt]
        \mathrm{single} & \frac{n_{v}}{n_{v}-1}\left(\frac{v_{i}-\overline{v}}{\sigma_{v_{i}}}\right)^{2} - 1 
    \end{cases}
\end{equation}

\noindent for pairs of observations in bands \textit{v} and \textit{b} and for single observations. For $\overline{v}$ and $\overline{b}$, we calculate the weighted mean for the observations in that band. Sources with purely noisy light curves have values close to zero. 

The DES observing strategy often yields sequences of two to three exposures taken of the same field in different filters, with the filters chosen according to the seeing, Moon phase, and number of missing observations in that field (see Figure 3 in \citealt{Diehl:2016}). Thus, to take advantage of this, we group together any observations taken within an hour. Any group with three observations $abc$ within the same hour was treated as three pairs $ab$, $bc$, $ac$, each weighted with a factor of $2/3$ as prescribed in \citet{Stetson:1996}. All other groups with one or two observations are weighted equally with a factor of 1. For this measurement, we excluded the generally noisier observations in $Y$.

\begin{figure}[!t]
  \includegraphics[width=0.5\textwidth]{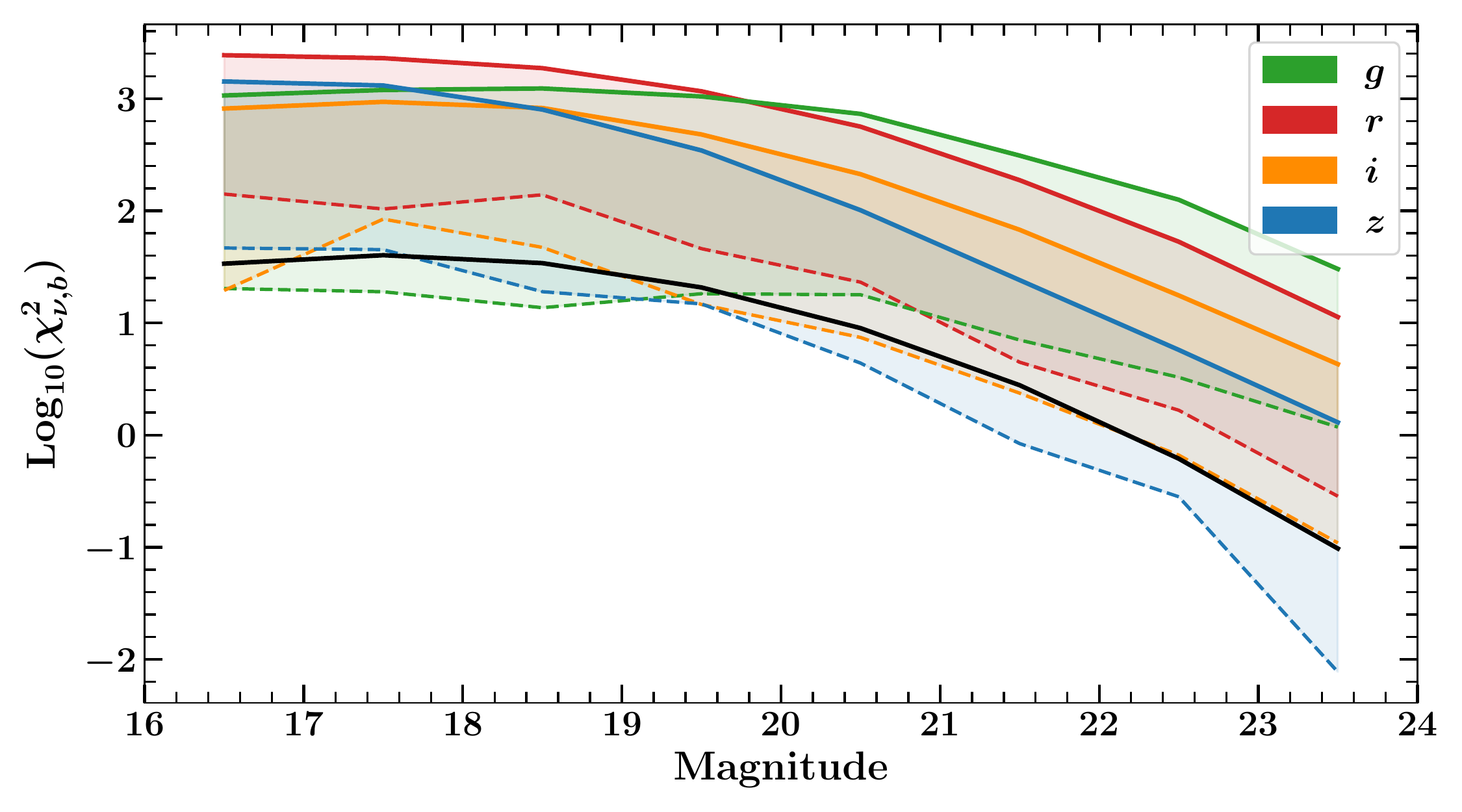}
  \caption{Distribution of $\log_{10}(\chi_{\nu,b}^{2})$ for simulated RRab light curves in each band binned by magnitude. The mean $\log_{10}(\chi_{\nu,b}^{2})$ in each band are plotted with solid curves, while the first percentile values for $\log_{10}(\chi_{\nu,b}^{2})$ in each band are plotted with dashed curves. The black curve shows the best-fit quadratic curve to the first percentile value of the $i$-band curve, which was used to transform the $\log_{10}(\chi_{\nu,b}^{2})$ features for target selection.}
  \label{fig:weighted_chi}
\end{figure}

We defined two new variability metrics intended to avoid overly penalizing faint objects, which can manifest measurement variability beyond that expected from their statistical uncertainties.
The first of these metrics was calculated from the difference between the minimum and maximum measured magnitude in each band, $\magn_{b}$, weighted by the photometric uncertainties on these values added in quadrature:
\begin{equation}
    \mathrm{W\_Range}_{b}= \frac{\magn_{b, \max} - \magn_{b,\min}}{[\sigma^2_{\magn_{b,\max}}+\sigma^2_{\magn_{b,\min}}]^{1/2}}.
    \label{eq:w_mag_range}
\end{equation}

Fainter stars have larger photometric errors and are harder to identify as variable, as reflected in smaller values of $\log(\chi_{\nu,b}^{2})$ (\figref{weighted_chi}).
To avoid rejecting faint variable objects while retaining high purity for bright variables, we define a threshold on $\log(\chi_{\nu,b}^{2})$ that changes with magnitude.
To derive this threshold, we binned the simulated RRab by their weighted-average magnitudes in bins of 1 magnitude and fit a quadratic curve to the 1\% percentile value of the $\log_{10}(\chi_{\nu,i}^{2})$, shown by the black curve in \figref{weighted_chi}, which follows the form
\begin{equation}
    \begin{aligned}
    \log_{10}(\chi_{\nu,b}^{2})_{\rm th} = \log_{10}(\chi_{\nu,b}^{2}) - (a_1 \mathrm{mag}_b - a_2)^2 - a_3,
    \end{aligned}
    \label{eq:chi2_thresh}
\end{equation}
\noindent where mag$_b$ is the weighted-average PSF magnitude of the individual measurements (WAVG\_MAG\_PSF) in band $b$.
The best-fit values of the constants were found to be $a_1=-0.07305$, $a_2=17.5165$, and $a_3=1.6036$.
The magnitude independent quantity, $\log_{10}(\chi_{\nu,b}^{2})_{\rm th}$, can be thought of as broadening the $\log_{10}(\chi_{\nu,b}^{2})$ criteria to retain high efficiency for faint variable sources.
RRL have positive values or values near zero across their entire magnitude range, while non-variable objects have more negative values.

We also define a metric for discriminating RRL based on their locations in period--amplitude space. 
We define a distance metric for the separation between the template-fit period and amplitude of objects with the Oosterhoff I, Oosterhoff II, and Osterhoff intermediate sequences as parameterized by \citet{Fabrizio:2019} and scaled to the $g$-band amplitude using $A_g/A_V = 1.29$ as determined by \citet{Vivas:2020b}. 
We calculate the distance between each object and each Oosterhoff sequence from the logarithm of the period in days and $g$-band amplitude in magnitudes using a rectangular approximation.
These distances are denoted D(Oo-I)$_{j}$, D(Oo-II)$_{j}$, and D(Oo-int)$_{j}$, where $j \in [0,2]$ indicates the $j$th best template fit.

\section{Performance Curves and Top Features for Random Forest Classifiers}\label{app:roc_features}

\begin{figure*}
  \includegraphics[width=1\textwidth]{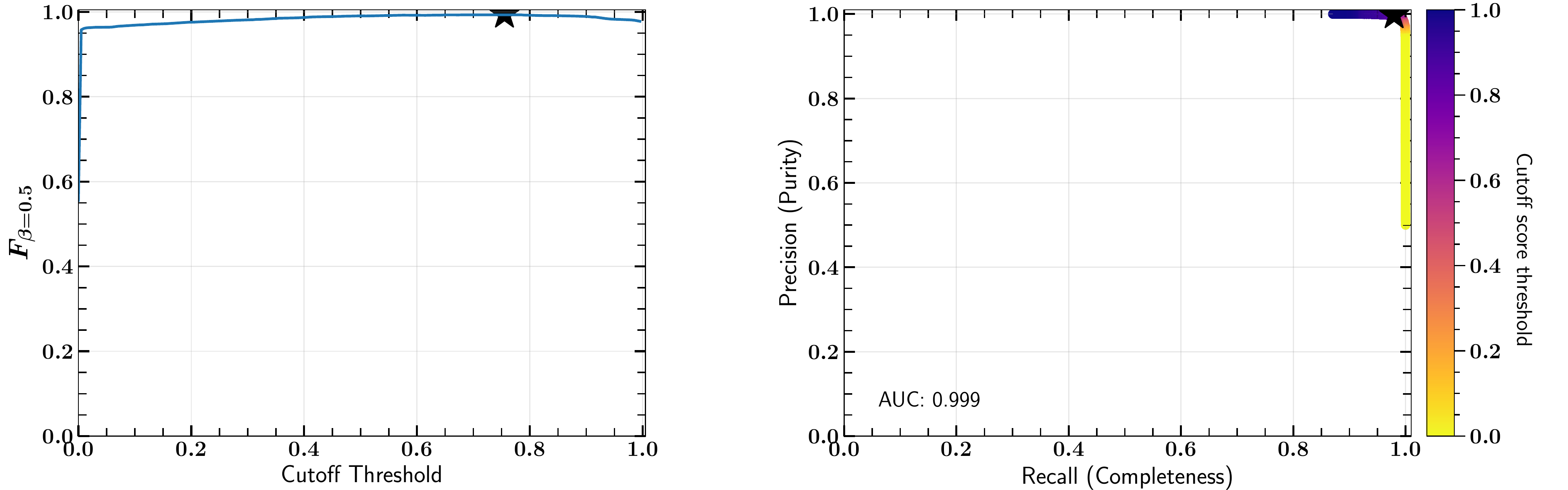}
  \caption{Performance curves for the initial variability classifier. \textit{Left:} $F_{\beta}$ scores over all possible cutoff score choices. The black star denotes where $F_{\beta}$ is maximized. \textit{Right:} ROC curve of the RF classifier trained on cross-matched RRab and calibration stars. The black star shows the location of the preferred cutoff score.}\label{fig:rf1_stds_roc} 
\end{figure*}

\begin{figure*}
  \includegraphics[width=1\textwidth]{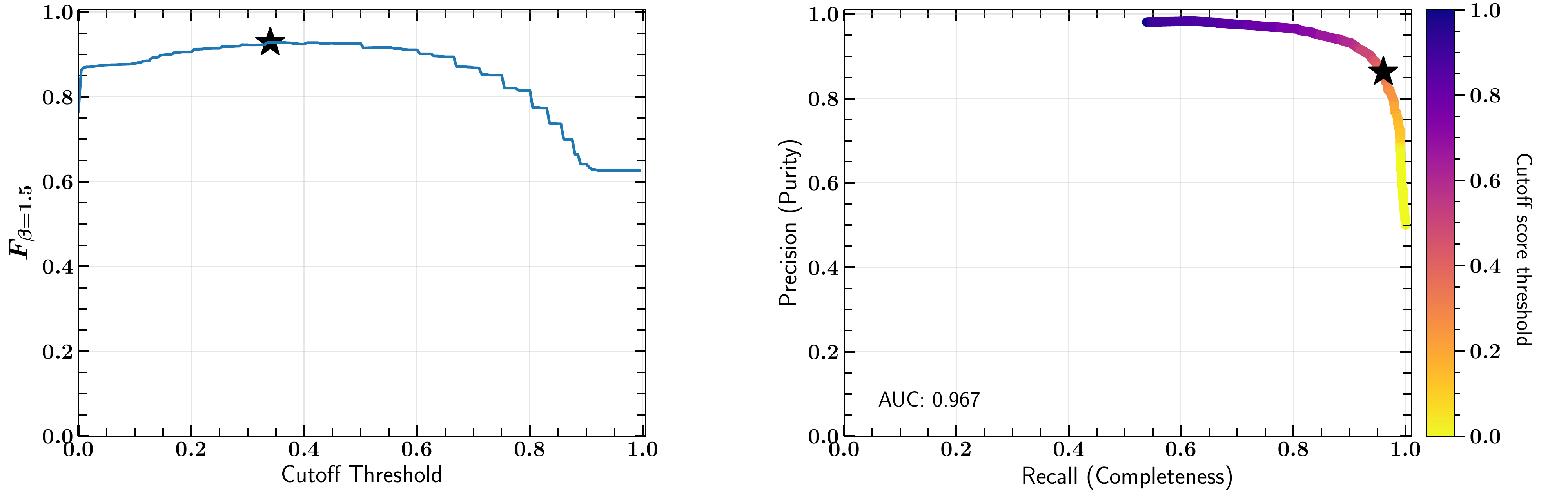}
  \caption{Performance curves for the second-stage classifier.}
  \label{fig:rf2_roc}
\end{figure*}

\begin{figure*}
  \includegraphics[width=1\textwidth]{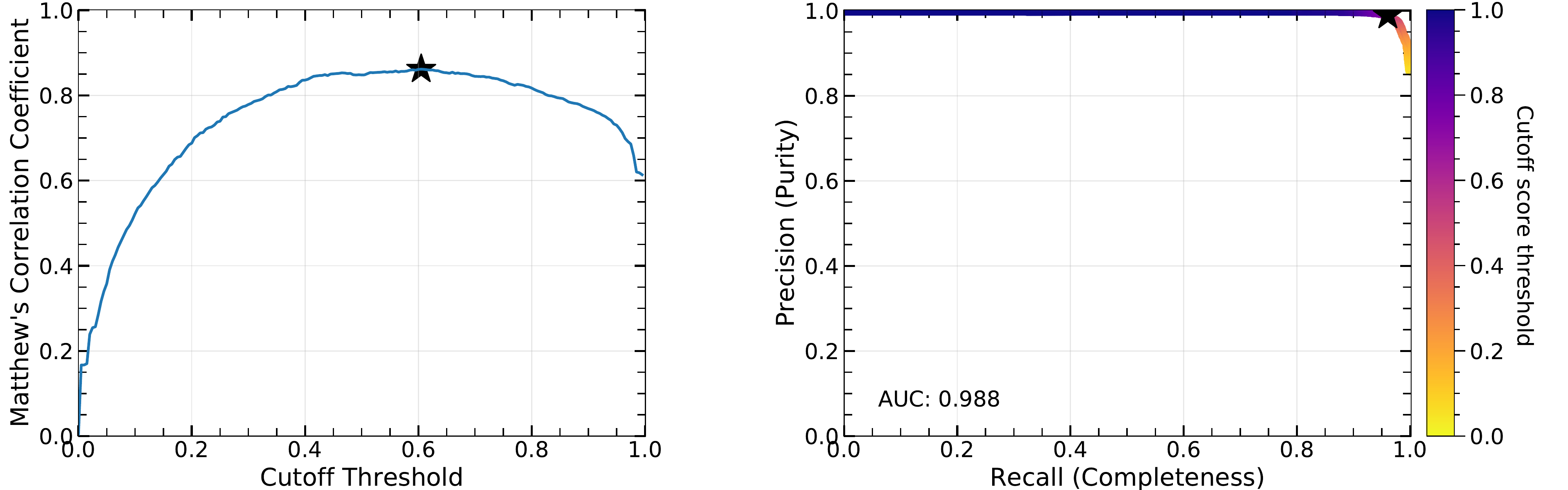}
  \caption{Performance curves for the third-stage classifier after template fitting.}
  \label{fig:rf3_roc}
\end{figure*}

Tables~\ref{tab:rf1_features}-\ref{tab:rf3_features} present the top 10 features for the initial variable selection, second-stage variability classifier, and light-curve selection, respectively. The performance curves for these selections are shown in Figs.~\ref{fig:rf1_stds_roc}-\ref{fig:rf3_roc}.

\begin{deluxetable}{ll}[!h]\label{tab:rf1_features}
	\tabletypesize{\footnotesize}
	\tablecaption{Top 10 features for initial variability selection}
	\tablehead{\colhead{Feature Name} & \colhead{Importance}}
	\decimals
	\startdata
        RoMS$_{g}$                       &  0.1637 \\
        $J_{grizY}$                      &  0.1604 \\
        log$_{10}(\chi^{2}_{\nu,i})_{\rm th}$       &  0.1357 \\
        W\_Range$_{g}$                   &  0.1055 \\
        log$_{10}(\chi^{2}_{\nu,g})_{\rm th}$    &  0.0842 \\
        RoMS$_{r}$                       &  0.0815 \\
        $J_{g}$                          &  0.0581 \\
        log$_{10}(\chi^{2}_{\nu,r})_{\rm th}$    &  0.0571 \\
        $J_{r}$                          &  0.0544 \\
        $\sigma^{2}_{{\rm NXS},g}$       &  0.0292 \\
	\enddata
\end{deluxetable}

\begin{deluxetable}{ll}[!b]\label{tab:rf2_features}
	\tabletypesize{\footnotesize}
	\tablecaption{Top 10 features for second variability classifier}
	\tablehead{\colhead{Feature Name} & \colhead{Importance}}
	\decimals
	\startdata
        RoMS$_{g}$                            &  0.1977 \\
        $J_{g}$                               &  0.1097 \\
        W\_Range$_{g}$                        &  0.0769 \\
        NAPD$_{90,g}$                         &  0.0562 \\
        $\chi^{2}_{\nu,g}$/$\chi^{2}_{\nu,r}$ &  0.0485 \\
        RoMS$_{r}$                            &  0.0358 \\
        NAPD$_{90,i}$                         &  0.0562 \\
        log$_{10}(\chi^{2}_{\nu,r})_{\rm th}$ &  0.0340 \\
        RoMS$_{z}$                            &  0.0312 \\
        log$_{10}(\chi^{2}_{\nu,g})_{\rm th}$ &  0.0321 \\
	\enddata
\end{deluxetable}

\begin{deluxetable}{ll}\label{tab:rf3_features}
	\tabletypesize{\footnotesize}
	\tablecaption{Top 10 features for candidate light-curve selection}
	\tablehead{\colhead{Feature Name} & \colhead{Importance}}
	\decimals
	\startdata
        $\frac{\mathrm{rss}_{\nu,1} - \mathrm{rss}_{\nu,0}}{\mathrm{rss}_{\nu,0}}$  &  0.1603 \\
        $\frac{\mathrm{rss}_{\nu,2} - \mathrm{rss}_{\nu,0}}{\mathrm{rss}_{\nu,0}}$  &  0.1559 \\
        $\mathrm{rss}_{\nu,1} - \mathrm{rss}_{\nu,0}$                               &  0.0660 \\
        rss$_{\nu,0}$                                                               &  0.0644 \\
        $\frac{\mathrm{amp}_{0}}{\mathrm{rss}_{\nu,0}}$                             &  0.0543 \\
        $\mathrm{rss}_{\nu,2} - \mathrm{rss}_{\nu,0}$                               &  0.0448 \\
        $\frac{1}{2}(\mathrm{RF1\_score} + \mathrm{RF2\_score})$                      &  0.0282 \\
        D(Oo-int)$_{0}$                                                             &  0.0267 \\
        RF1\_score                                                                  &  0.0262 \\
        amp$_{0}$                                                                   &  0.0241 \\
	\enddata
\end{deluxetable}

\vfill\pagebreak\newpage

\vfill\pagebreak\newpage

\section{Halo Profile Fit}
\label{app:likelihood}

We fit the halo profile using a standard binned Poisson maximum-likelihood approach. However, as with many maximum-likelihood analyses, the calculation of the predicted number of counts from our model is sufficiently complex that it merits a devoted discussion.

We define our likelihood, $\mathcal{L}$, as the product over bins, $i$, of the Poisson likelihood for observing $k$ RRab candidates given a model prediction of $\lambda$ counts:
\begin{equation}
\mathcal{L} = \prod_i \frac{\lambda_i^{k_i} e^{-\lambda_i}}{k_i!}.
\end{equation}
It is more computationally feasible to work with the logarithm of the likelihood, and thus we define
\begin{equation}
\log \mathcal{L} = \sum_i { k_i \log(\lambda_i) - \lambda_i } - \log(k_i!)
\end{equation}
where the term $-\log(k!)$ does not depend on the model parameters and can be discarded.
The model-predicted number of counts in a bin, $\lambda_i$, is a function of the elliptical Galactocentric radius of the bin, $r_{e,i}$, and the model parameters $\theta$, of the broken-power-law density model described in \eqnref{pwl}: 
\begin{equation}
\lambda_i(r_{e,i},\theta) = \int \rho(r_{e,i},\theta) f(D) {\rm d}V
\label{eqn:model}
\end{equation}
where $f(D)$ is the efficiency of detecting RRab at the heliocentric distance, $D$, of each volume element.
This formulation naturally incorporates the geometric effects of the solar offset from the Galactic center, which are most noticeable for distances of $D \lesssim 15 \kpc$ (i.e., the heliocentric distance cut translates to a cut in $r_e$).
Numerically, we perform the integration in \eqnref{model} over \code{HEALPix} pixels of area $\roughly 0.84 \deg^2$ ($\code{nside}=64$), incorporating the coverage fraction of each pixel, which is estimated at scales of $\roughly 166 \asec^2$ ($\code{nside} = 16384$).

\begin{figure*}[t!]
  \centering
  \includegraphics[width=0.6\textwidth]{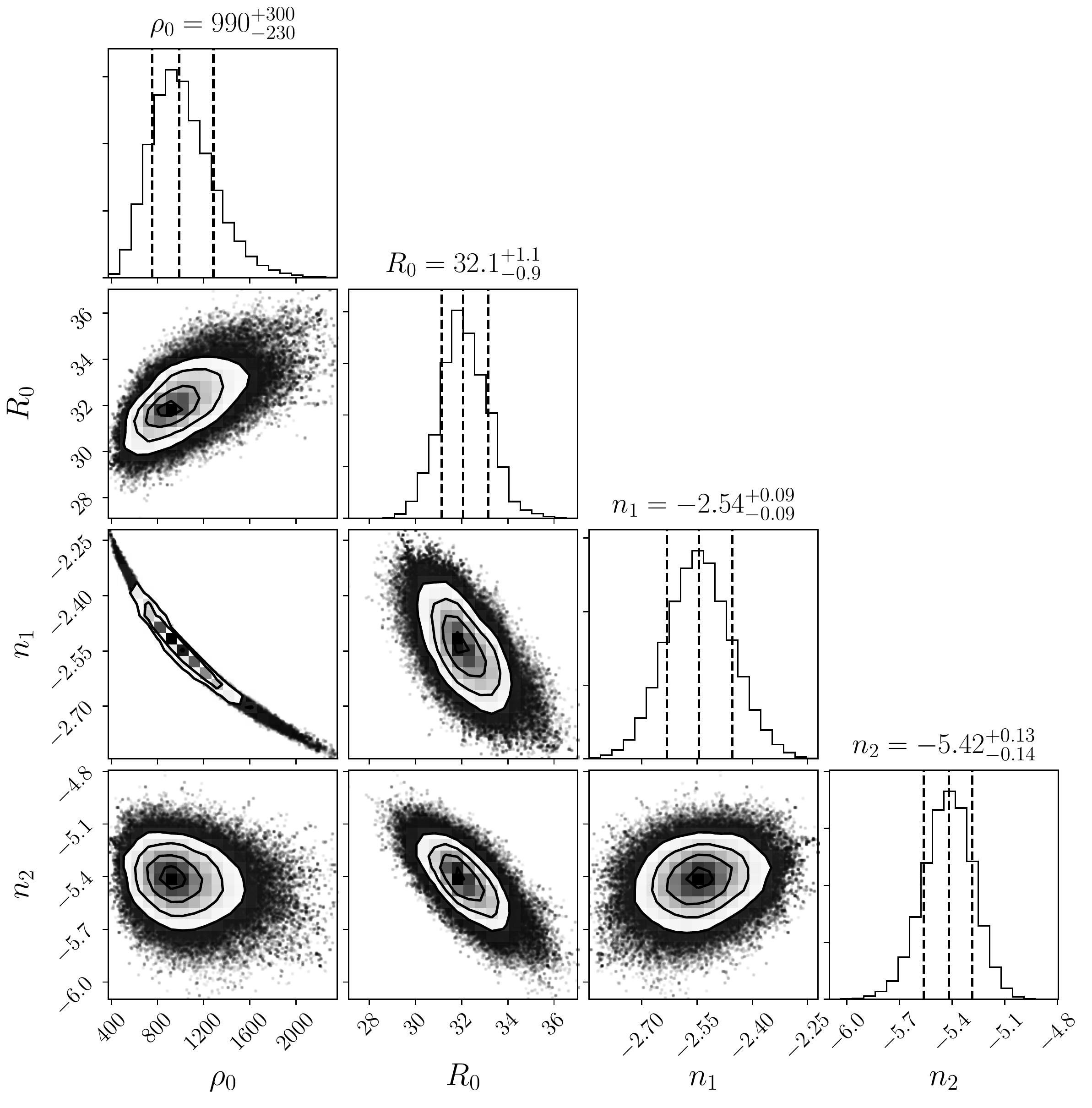}
  \includegraphics[width=0.6\textwidth]{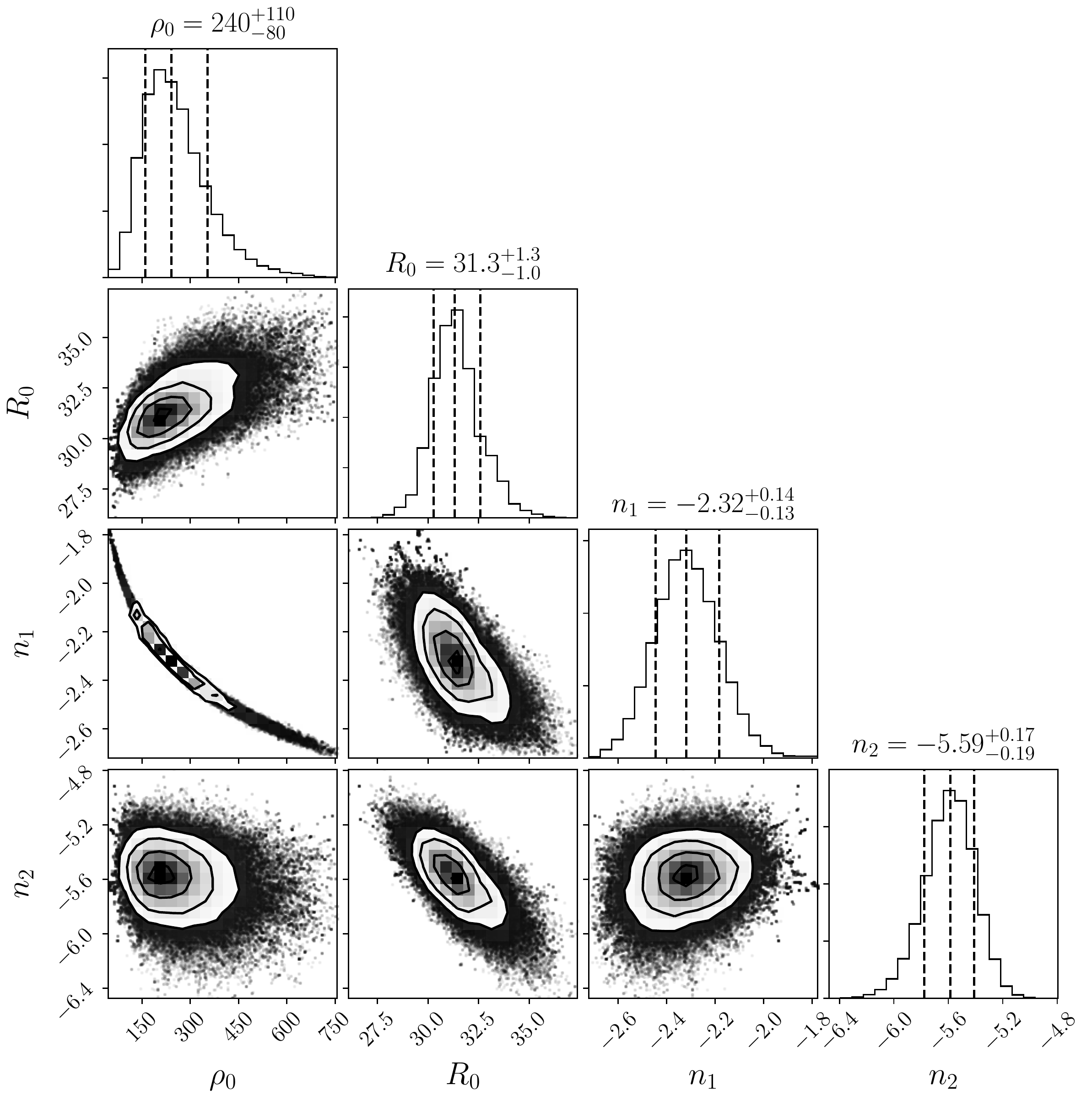}
  \caption{Posterior probability distributions for the parameters of the broken-power-law fits to the full RRab candidate sample (top) and the high-confidence sample associated with the Oosterhoff I sequence (\secref{field}).}
  \label{fig:posteriors}
\end{figure*}

With our likelihood thus defined, we seek to sample the posterior probability distribution as defined by Bayes' theorem.
We explore the posterior probability distribution with Markov Chain Monte Carlo using the affine-invariant ensemble sampler \code{emcee} \citep{Foreman-Mackey:2013}. We exclude the regions listed in \tabref{masks} and  assign uniform priors to each of the model parameters following the range described in \tabref{priors}.
We sample the posterior using 100 walkers with 5000 samples each, discarding the first 1000 samples as burn-in. 
The resulting posterior distributions from the RRab and Oo-I subselection can be found in \figref{posteriors}.
The best-fit parameters are assigned from the median of the posterior, and the errors are derived from the 16th and 84th percentiles of the posterior distribution.

As can be seen in \figref{posteriors}, there is significant correlation between the normalization parameter, $\rho_0$, and the inner power-law slope, $n_1$. We note that in this regime, the analysis is especially sensitive to the geometric corrections described above for calculating the model-predicted counts (i.e., Equation \ref{eqn:model}).

\begin{deluxetable}{lcc}\label{tab:masks}
	\tabletypesize{\footnotesize}
	\tablecaption{Masked regions for halo RRL study.\label{tab:mask}}
	\tablehead{ \colhead{Galaxy} & \colhead{($\alpha,\delta$)} & \colhead{Mask Radius} \\[-1em]
	\colhead{} & \multicolumn{1}{c}{(deg, J2000)} & \colhead{(deg)}}
	\decimals
	\startdata
        LMC          & (80.8938,-69.7561)   &  $27.1 $ \\
	SMC	     & (13.1867,-72.8286)   &  $12.6 $ \\
	Fornax	     & (39.9583,-34.4997)   &  $2.0  $ \\
	Sculptor     & (15.0183,-33.7186)   &  $1.0  $ 
	\enddata
\end{deluxetable}

\begin{deluxetable}{lll}
  \label{tab:priors}
  \tabletypesize{\footnotesize}
  \tablecaption{Priors on the broken-power-law model parameters.}
  \tablehead{\colhead{Parameter Name} & \colhead{Prior} & \colhead{Range}}
  \startdata
  Normalization, $\rho_0$ (kpc$^{-3}$)  & uniform   & $[0,10^6]$ \\
  Break radius, $R_0$ (kpc)             & uniform   & $[10,50]$ \\
  Inner slope, $n_1$                    & uniform   & $[-3.5,-1.0]$ \\
  Outer slope, $n_2$                    & uniform   & $[-3.0,-6.5]$ \\
  \enddata
  \tablecomments{The halo density profile fit assumes a fixed halo flattening of $q = 0.7$.}
\end{deluxetable}

\clearpage

\bibliography{main.bib}



\end{document}

%% file: commands.tex
\usepackage{soul} 
\usepackage{amsmath}
\usepackage{amssymb}
\usepackage{xspace}
\usepackage{xifthen}

\newcommand{\CHECK}[1]{{#1}}
\newcommand{\COMMENT}[3]{{}}

\mathchardef\mhyphen="2D

\newcommand{\roughly}{\ensuremath{ {\sim}\,} }

\newcommand{\less}{\ensuremath{ {<}\,} }
\newlength{\dhatheight}

\newcommand{\code}[1]{\texttt{#1}\xspace}

\newcommand{\unit}[1]{\ensuremath{\mathrm{\,#1}}\xspace}
\newcommand{\Gyr}{\unit{Gyr}}

\newcommand{\degree}{\ensuremath{{}^{\circ}}\xspace}

\newcommand{\asec}{\unit{arcsec}}

\newcommand{\pc}{\unit{pc}}
\newcommand{\kpc}{\unit{kpc}}
\newcommand{\second}{\unit{s}}

\newcommand{\magn}{\unit{mag}}
\newcommand{\mmag}{\unit{mmag}}
\newcommand{\e}{\unit{e^{-}}}

\newcommand{\secref}[1]{\S\ref{sec:#1}}
\newcommand{\appref}[1]{Appendix~\ref{app:#1}}
\newcommand{\tabref}[1]{Table~\ref{tab:#1}}
\newcommand{\figref}[1]{Figure~\ref{fig:#1}}
\newcommand{\eqnref}[1]{Equation~\eqref{eqn:#1}}

\newcommand{\bandvar}[2][]{%
  \ifthenelse{\isempty{#1}}{\var{#2}}{\var{#2\_#1}}%
}

\newcommand{\feh}{{\ensuremath{\rm [Fe/H]}}\xspace}

\newcommand{\SExtractor}{\code{Source{\allowbreak}Extractor}}

\newcommand{\HEALPix}{\code{HEALPix}}
\newcommand{\healpix}{\HEALPix}
\newcommand{\nside}{\code{nside}}

\newcommand{\var}[1]{\ensuremath{\texttt{\MakeUppercase{#1}}}\xspace}
\newcommand{\Gaia}{{\it Gaia}\xspace}

\newcommand{\medchisq}{\ensuremath{\tilde{\chi}_{\nu,b}^{2}}\xspace}

\newcommand{\rhalf}{\ensuremath{ r_{\rm h} }\xspace}

\providecommand\physrep{\ref@jnl{Phys.~Rep.}}%
\providecommand\apjs{\ref@jnl{ApJS}}%
\providecommand{\jcap}{\ref@jnl{JCAP}}%

\newcommand{\Nrrab}{\CHECK{6,971}\xspace}

\newcommand{\Norm}{\CHECK{\ensuremath{990^{+300}_{-230}}}\xspace}
\newcommand{\Rbreak}{\CHECK{\ensuremath{32.1^{+1.1}_{-0.9}}}\xspace}
\newcommand{\Ninner}{\CHECK{\ensuremath{-2.54^{+0.09}_{-0.09}}}\xspace}
\newcommand{\Nouter}{\CHECK{\ensuremath{-5.42^{+0.13}_{-0.14}}}\xspace}

\newcommand{\NormOo}{\CHECK{\ensuremath{240^{+110}_{-80}}}\xspace}
\newcommand{\RbreakOo}{\CHECK{\ensuremath{31.3^{+1.3}_{-1.0}}}\xspace}
\newcommand{\NinnerOo}{\CHECK{\ensuremath{-2.32^{+0.14}_{-0.13}}}\xspace}
\newcommand{\NouterOo}{\CHECK{\ensuremath{-5.59^{+0.17}_{-0.19}}}\xspace}

\defcitealias{DR1:2018}{DES Collaboration (2018)}



%% file: authors.tex

\author[0000-0002-4624-2772]{K.~M.~Stringer}
\affiliation{George P. and Cynthia Woods Mitchell Institute for Fundamental Physics and Astronomy, Department of Physics and Astronomy, Texas A\&M University, College Station, TX 77843,  USA}
\author[0000-0001-8251-933X]{A.~Drlica-Wagner}
\affiliation{Fermi National Accelerator Laboratory, P. O. Box 500, Batavia, IL 60510, USA}
\affiliation{Kavli Institute for Cosmological Physics, University of Chicago, Chicago, IL 60637, USA}
\affiliation{Department of Astronomy and Astrophysics, University of Chicago, Chicago, IL 60637, USA}
\author[0000-0002-1775-4859]{L.~Macri}
\affiliation{George P. and Cynthia Woods Mitchell Institute for Fundamental Physics and Astronomy, Department of Physics and Astronomy, Texas A\&M University, College Station, TX 77843,  USA}
\author[0000-0002-9144-7726]{C.~E.~Mart{\'\i}nez-V{\'a}zquez}
\affiliation{Cerro Tololo Inter-American Observatory, NSF's National Optical-Infrared Astronomy Research Laboratory, \\ Casilla 603, La Serena, Chile}
\author[0000-0003-4341-6172]{A.~K.~Vivas}
\affiliation{Cerro Tololo Inter-American Observatory, NSF's National Optical-Infrared Astronomy Research Laboratory, \\ Casilla 603, La Serena, Chile}
\author[0000-0001-6957-1627]{P.~Ferguson}
\affiliation{George P. and Cynthia Woods Mitchell Institute for Fundamental Physics and Astronomy, Department of Physics and Astronomy, Texas A\&M University, College Station, TX 77843,  USA}
\author[0000-0002-6021-8760]{A.~B.~Pace}
\affiliation{Department of Physics, Carnegie Mellon University, Pittsburgh, Pennsylvania 15312, USA}
\author[0000-0002-7123-8943]{A.~R.~Walker}
\affiliation{Cerro Tololo Inter-American Observatory, NSF's National Optical-Infrared Astronomy Research Laboratory, \\ Casilla 603, La Serena, Chile}
\author[0000-0002-7357-0317]{E.~Neilsen}
\affiliation{Fermi National Accelerator Laboratory, P. O. Box 500, Batavia, IL 60510, USA}
\author[0000-0001-6584-6144]{K.~Tavangar}
\affiliation{Department of Astronomy and Astrophysics, University of Chicago, Chicago, IL 60637, USA}
\affiliation{Kavli Institute for Cosmological Physics, University of Chicago, Chicago, IL 60637, USA}
\author[0000-0003-0072-6736]{W.~Wester}
\affiliation{Fermi National Accelerator Laboratory, P. O. Box 500, Batavia, IL 60510, USA}
\author{T.~M.~C.~Abbott}
\affiliation{Cerro Tololo Inter-American Observatory, NSF's National Optical-Infrared Astronomy Research Laboratory, \\ Casilla 603, La Serena, Chile}
\author{M.~Aguena}
\affiliation{Departamento de F\'isica Matem\'atica, Instituto de F\'isica, Universidade de S\~ao Paulo, CP 66318, S\~ao Paulo, SP, 05314-970, Brazil}
\affiliation{Laborat\'orio Interinstitucional de e-Astronomia - LIneA, Rua Gal. Jos\'e Cristino 77, Rio de Janeiro, RJ - 20921-400, Brazil}
\author[0000-0002-7069-7857]{S.~Allam}
\affiliation{Fermi National Accelerator Laboratory, P. O. Box 500, Batavia, IL 60510, USA}
\author{D.~Bacon}
\affiliation{Institute of Cosmology and Gravitation, University of Portsmouth, Portsmouth, PO1 3FX, UK}
\author{K.~Bechtol}
\affiliation{Physics Department, 2320 Chamberlin Hall, University of Wisconsin-Madison, 1150 University Avenue Madison, WI  53706-1390}
\author{E.~Bertin}
\affiliation{CNRS, UMR 7095, Institut d'Astrophysique de Paris, F-75014, Paris, France}
\affiliation{Sorbonne Universit\'es, UPMC Univ Paris 06, UMR 7095, Institut d'Astrophysique de Paris, F-75014, Paris, France}
\author[0000-0002-8458-5047]{D.~Brooks}
\affiliation{Department of Physics \& Astronomy, University College London, Gower Street, London, WC1E 6BT, UK}
\author{D.~L.~Burke}
\affiliation{Kavli Institute for Particle Astrophysics \& Cosmology, P. O. Box 2450, Stanford University, Stanford, CA 94305, USA}
\affiliation{SLAC National Accelerator Laboratory, Menlo Park, CA 94025, USA}
\author[0000-0003-3044-5150]{A.~Carnero~Rosell}
\affiliation{Instituto de Astrofisica de Canarias, E-38205 La Laguna, Tenerife, Spain}
\affiliation{Universidad de La Laguna, Dpto. Astrof\'isica, E-38206 La Laguna, Tenerife, Spain}
\author[0000-0002-4802-3194]{M.~Carrasco~Kind}
\affiliation{Department of Astronomy, University of Illinois at Urbana-Champaign, 1002 W. Green Street, Urbana, IL 61801, USA}
\affiliation{National Center for Supercomputing Applications, 1205 West Clark St., Urbana, IL 61801, USA}
\author[0000-0002-3130-0204]{J.~Carretero}
\affiliation{Institut de F\'{\i}sica d'Altes Energies (IFAE), The Barcelona Institute of Science and Technology, Campus UAB,\\ 08193 Bellaterra (Barcelona) Spain}
\author{M.~Costanzi}
\affiliation{INAF-Osservatorio Astronomico di Trieste, via G. B. Tiepolo 11, I-34143 Trieste, Italy}
\affiliation{Institute for Fundamental Physics of the Universe, Via Beirut 2, 34014 Trieste, Italy}
\author[0000-0002-9745-6228]{M.~Crocce}
\affiliation{Institut d'Estudis Espacials de Catalunya (IEEC), 08034 Barcelona, Spain}
\affiliation{Institute of Space Sciences (ICE, CSIC),  Campus UAB,\\ Carrer de Can Magrans, s/n,  08193 Barcelona, Spain}
\author{L.~N.~da Costa}
\affiliation{Laborat\'orio Interinstitucional de e-Astronomia - LIneA, Rua Gal. Jos\'e Cristino 77, Rio de Janeiro, RJ - 20921-400, Brazil}
\affiliation{Observat\'orio Nacional, Rua Gal. Jos\'e Cristino 77, Rio de Janeiro, RJ - 20921-400, Brazil}
\author{M.~E.~S.~Pereira}
\affiliation{Department of Physics, University of Michigan, Ann Arbor, MI 48109, USA}
\author[0000-0001-8318-6813]{J.~De~Vicente}
\affiliation{Centro de Investigaciones Energ\'eticas, Medioambientales y Tecnol\'ogicas (CIEMAT), Madrid, Spain}
\author[0000-0002-0466-3288]{S.~Desai}
\affiliation{Department of Physics, IIT Hyderabad, Kandi, Telangana 502285, India}
\author[0000-0002-8357-7467]{H.~T.~Diehl}
\affiliation{Fermi National Accelerator Laboratory, P. O. Box 500, Batavia, IL 60510, USA}
\author{P.~Doel}
\affiliation{Department of Physics \& Astronomy, University College London, Gower Street, London, WC1E 6BT, UK}
\author{I.~Ferrero}
\affiliation{Institute of Theoretical Astrophysics, University of Oslo. P.O. Box 1029 Blindern, NO-0315 Oslo, Norway}
\author[0000-0002-9370-8360]{J.~Garc\'ia-Bellido}
\affiliation{Instituto de Fisica Teorica UAM/CSIC, Universidad Autonoma de Madrid, 28049 Madrid, Spain}
\author[0000-0001-9632-0815]{E.~Gaztanaga}
\affiliation{Institut d'Estudis Espacials de Catalunya (IEEC), 08034 Barcelona, Spain}
\affiliation{Institute of Space Sciences (ICE, CSIC),  Campus UAB,\\ Carrer de Can Magrans, s/n,  08193 Barcelona, Spain}
\author[0000-0001-6942-2736]{D.~W.~Gerdes}
\affiliation{Department of Astronomy, University of Michigan, Ann Arbor, MI 48109, USA}
\affiliation{Department of Physics, University of Michigan, Ann Arbor, MI 48109, USA}
\author[0000-0003-3270-7644]{D.~Gruen}
\affiliation{Department of Physics, Stanford University, 382 Via Pueblo Mall, Stanford, CA 94305, USA}
\affiliation{Kavli Institute for Particle Astrophysics \& Cosmology, P. O. Box 2450, Stanford University, Stanford, CA 94305, USA}
\affiliation{SLAC National Accelerator Laboratory, Menlo Park, CA 94025, USA}
\author{R.~A.~Gruendl}
\affiliation{Department of Astronomy, University of Illinois at Urbana-Champaign, 1002 W. Green Street, Urbana, IL 61801, USA}
\affiliation{National Center for Supercomputing Applications, 1205 West Clark St., Urbana, IL 61801, USA}
\author[0000-0003-3023-8362]{J.~Gschwend}
\affiliation{Laborat\'orio Interinstitucional de e-Astronomia - LIneA, Rua Gal. Jos\'e Cristino 77, Rio de Janeiro, RJ - 20921-400, Brazil}
\affiliation{Observat\'orio Nacional, Rua Gal. Jos\'e Cristino 77, Rio de Janeiro, RJ - 20921-400, Brazil}
\author[0000-0003-0825-0517]{G.~Gutierrez}
\affiliation{Fermi National Accelerator Laboratory, P. O. Box 500, Batavia, IL 60510, USA}
\author{S.~R.~Hinton}
\affiliation{School of Mathematics and Physics, University of Queensland,  Brisbane, QLD 4072, Australia}
\author{D.~L.~Hollowood}
\affiliation{Santa Cruz Institute for Particle Physics, Santa Cruz, CA 95064, USA}
\author[0000-0002-6550-2023]{K.~Honscheid}
\affiliation{Center for Cosmology and Astro-Particle Physics, The Ohio State University, Columbus, OH 43210, USA}
\affiliation{Department of Physics, The Ohio State University, Columbus, OH 43210, USA}
\author[0000-0002-2571-1357]{B.~Hoyle}
\affiliation{Faculty of Physics, Ludwig-Maximilians-Universit\"at, Scheinerstr. 1, 81679 Munich, Germany}
\affiliation{Max Planck Institute for Extraterrestrial Physics, Giessenbachstrasse, 85748 Garching, Germany}
\affiliation{Universit\"ats-Sternwarte, Fakult\"at f\"ur Physik, Ludwig-Maximilians Universit\"at M\"unchen, Scheinerstr. 1, 81679 M\"unchen, Germany}
\author[0000-0001-5160-4486]{D.~J.~James}
\affiliation{Center for Astrophysics $\vert$ Harvard \& Smithsonian, 60 Garden Street, Cambridge, MA 02138, USA}
\author[0000-0003-0120-0808]{K.~Kuehn}
\affiliation{Australian Astronomical Optics, Macquarie University, North Ryde, NSW 2113, Australia}
\affiliation{Lowell Observatory, 1400 Mars Hill Rd, Flagstaff, AZ 86001, USA}
\author[0000-0003-2511-0946]{N.~Kuropatkin}
\affiliation{Fermi National Accelerator Laboratory, P. O. Box 500, Batavia, IL 60510, USA}
\author[0000-0002-9110-6163]{T.~S.~Li}
\affiliation{Department of Astrophysical Sciences, Princeton University, Peyton Hall, Princeton, NJ 08544, USA}
\affiliation{Observatories of the Carnegie Institution for Science, 813 Santa Barbara St., Pasadena, CA 91101, USA}
\author[0000-0001-9856-9307]{M.~A.~G.~Maia}
\affiliation{Laborat\'orio Interinstitucional de e-Astronomia - LIneA, Rua Gal. Jos\'e Cristino 77, Rio de Janeiro, RJ - 20921-400, Brazil}
\affiliation{Observat\'orio Nacional, Rua Gal. Jos\'e Cristino 77, Rio de Janeiro, RJ - 20921-400, Brazil}
\author[0000-0003-0710-9474]{J.~L.~Marshall}
\affiliation{George P. and Cynthia Woods Mitchell Institute for Fundamental Physics and Astronomy, Department of Physics and Astronomy, Texas A\&M University, College Station, TX 77843,  USA}
\author[0000-0002-1372-2534]{F.~Menanteau}
\affiliation{Department of Astronomy, University of Illinois at Urbana-Champaign, 1002 W. Green Street, Urbana, IL 61801, USA}
\affiliation{National Center for Supercomputing Applications, 1205 West Clark St., Urbana, IL 61801, USA}
\author[0000-0002-6610-4836]{R.~Miquel}
\affiliation{Instituci\'o Catalana de Recerca i Estudis Avan\c{c}ats, E-08010 Barcelona, Spain}
\affiliation{Institut de F\'{\i}sica d'Altes Energies (IFAE), The Barcelona Institute of Science and Technology, Campus UAB,\\ 08193 Bellaterra (Barcelona) Spain}
\author{R.~Morgan}
\affiliation{Physics Department, 2320 Chamberlin Hall, University of Wisconsin-Madison, 1150 University Avenue Madison, WI  53706-1390}
\author[0000-0003-2120-1154]{R.~L.~C.~Ogando}
\affiliation{Laborat\'orio Interinstitucional de e-Astronomia - LIneA, Rua Gal. Jos\'e Cristino 77, Rio de Janeiro, RJ - 20921-400, Brazil}
\affiliation{Observat\'orio Nacional, Rua Gal. Jos\'e Cristino 77, Rio de Janeiro, RJ - 20921-400, Brazil}
\author[0000-0002-6011-0530]{A.~Palmese}
\affiliation{Fermi National Accelerator Laboratory, P. O. Box 500, Batavia, IL 60510, USA}
\affiliation{Kavli Institute for Cosmological Physics, University of Chicago, Chicago, IL 60637, USA}
\author{F.~Paz-Chinch\'{o}n}
\affiliation{Institute of Astronomy, University of Cambridge, Madingley Road, Cambridge CB3 0HA, UK}
\affiliation{National Center for Supercomputing Applications, 1205 West Clark St., Urbana, IL 61801, USA}
\author[0000-0002-2598-0514]{A.~A.~Plazas}
\affiliation{Department of Astrophysical Sciences, Princeton University, Peyton Hall, Princeton, NJ 08544, USA}
\author[0000-0001-5326-3486]{A.~Roodman}
\affiliation{Kavli Institute for Particle Astrophysics \& Cosmology, P. O. Box 2450, Stanford University, Stanford, CA 94305, USA}
\affiliation{SLAC National Accelerator Laboratory, Menlo Park, CA 94025, USA}
\author[0000-0002-9646-8198]{E.~Sanchez}
\affiliation{Centro de Investigaciones Energ\'eticas, Medioambientales y Tecnol\'ogicas (CIEMAT), Madrid, Spain}
\author[0000-0001-9504-2059]{M.~Schubnell}
\affiliation{Department of Physics, University of Michigan, Ann Arbor, MI 48109, USA}
\author{S.~Serrano}
\affiliation{Institut d'Estudis Espacials de Catalunya (IEEC), 08034 Barcelona, Spain}
\affiliation{Institute of Space Sciences (ICE, CSIC),  Campus UAB,\\ Carrer de Can Magrans, s/n,  08193 Barcelona, Spain}
\author[0000-0002-1831-1953]{I.~Sevilla-Noarbe}
\affiliation{Centro de Investigaciones Energ\'eticas, Medioambientales y Tecnol\'ogicas (CIEMAT), Madrid, Spain}
\author[0000-0002-3321-1432]{M.~Smith}
\affiliation{School of Physics and Astronomy, University of Southampton,  Southampton, SO17 1BJ, UK}
\author[0000-0001-6082-8529]{M.~Soares-Santos}
\affiliation{Department of Physics, University of Michigan, Ann Arbor, MI 48109, USA}
\author[0000-0002-7047-9358]{E.~Suchyta}
\affiliation{Computer Science and Mathematics Division, Oak Ridge National Laboratory, Oak Ridge, TN 37831}
\author[0000-0003-1704-0781]{G.~Tarle}
\affiliation{Department of Physics, University of Michigan, Ann Arbor, MI 48109, USA}
\author{D.~Thomas}
\affiliation{Institute of Cosmology and Gravitation, University of Portsmouth, Portsmouth, PO1 3FX, UK}
\author[0000-0001-7836-2261]{C.~To}
\affiliation{Department of Physics, Stanford University, 382 Via Pueblo Mall, Stanford, CA 94305, USA}
\affiliation{Kavli Institute for Particle Astrophysics \& Cosmology, P. O. Box 2450, Stanford University, Stanford, CA 94305, USA}
\affiliation{SLAC National Accelerator Laboratory, Menlo Park, CA 94025, USA}
\author{T.~N.~Varga}
\affiliation{Max Planck Institute for Extraterrestrial Physics, Giessenbachstrasse, 85748 Garching, Germany}
\affiliation{Universit\"ats-Sternwarte, Fakult\"at f\"ur Physik, Ludwig-Maximilians Universit\"at M\"unchen, Scheinerstr. 1, 81679 M\"unchen, Germany}
\author[0000-0002-3908-7313]{R.D.~Wilkinson}
\affiliation{Department of Physics and Astronomy, Pevensey Building, University of Sussex, Brighton, BN1 9QH, UK}
\author[0000-0001-5969-4631]{Y.~Zhang}
\affiliation{Fermi National Accelerator Laboratory, P. O. Box 500, Batavia, IL 60510, USA}

\collaboration{69}{(DES Collaboration)}